\documentstyle[seceq,epsf,wrapft]{ptptex}

\makeatletter
\def\rmm{\@ifundefined{DeclareOldFontCommand}{\rm}{
 \mathrm}}
\def\@bgnmark{<}
\def\@endmark{>}
\def\WKht{.85}
\def\WKsep{.4}
\def\WKrule{.03}
\newdimen\@tempdimc
\newdimen\@tempdimd
\newdimen\bgnp@sition
\newdimen\endp@sition
\newdimen\bgnh@ight
\newdimen\endh@ight
\newdimen\h@ight 
\newdimen\w@dth  
\def\SEPbgn#1<1#2#3>1#4#5\@@{\xdef\@MAE{#1}\xdef\@BGNCHAR{#2}
\xdef\@NAKA{#3}\xdef\@ENDCHAR{#4}\xdef\@USIRO{#5}}
\def\wicksymbol#1#2#3#4#5{
 \@tempdima=#3 \advance\@tempdima-#1%
 \@tempdimc=#5\h@ight \@tempdimb=\@tempdimc \advance\@tempdimb-\w@dth%
 \@tempdimd=#2 \advance\@tempdimd-1.587ex
 \hskip#1%
 \vrule height \@tempdimc width\w@dth depth-\@tempdimd \kern-\w@dth%
 \vrule height \@tempdimc width\@tempdima depth-\@tempdimb\kern-\w@dth%
 \@tempdimd=#4 \advance\@tempdimd-1.587ex 
 \vrule height \@tempdimc width\w@dth depth-\@tempdimd}
\def\wick#1#2{%
 \h@ight=\WKht ex \w@dth=\WKrule em%
 \def\@list{}
 \def\sqrt{\radical"270370}%
 \SEPbgn#2\@@ 
 \setbox0=\hbox{$\displaystyle \@BGNCHAR$}
 \@tempdima\wd0 \bgnh@ight\ht0 
 \settowidth{\@tempdimb}{$\displaystyle \@MAE$}
 \divide\@tempdima by2 \advance\@tempdima by \@tempdimb
 \bgnp@sition\@tempdima
 \xdef\@list{\@MAE\@BGNCHAR\@NAKA}
 \setbox0=\hbox{$\displaystyle \@ENDCHAR$}
 \@tempdima\wd0 \endh@ight\ht0 
 \settowidth{\@tempdimb}{$\displaystyle \@list$}
 \divide\@tempdima by2 \advance\@tempdima by \@tempdimb
 \endp@sition\@tempdima
 \xdef\@list{\@MAE\@BGNCHAR\@NAKA\@ENDCHAR\@USIRO}
 \mathop{\vbox{\m@th\ialign{##\crcr\noalign{\kern\WKsep ex}%
 $\m@th \setbox0=\hbox{$\wicksymbol{\bgnp@sition}{\bgnh@ight}%
 {\endp@sition}{\endh@ight}{#1}$} \dp0\z@ \wd0\z@ \box0$ %
 \crcr\noalign{\kern\WKsep ex\nointerlineskip}%
 \setbox0=\hbox{$\displaystyle\@list$}\ht0=1.587ex%
 \box0\crcr}}}\limits}
\makeatother
\def\aa{x_u}
\def\bb{x_\ro}
\def\cc{x_\rc}
\newcommand{\nn}{\nonumber\\}
\newcommand{\QB}{Q_{\rm B}}
\newcommand{\tQB}{{\tilde Q}_{\rm B}}
\newcommand{\half}{\frac{1}{2}}
\newcommand{\abs}[1]{\left| #1 \right|}
\newcommand{\bra}[1]{\langle#1 |}
\newcommand{\ket}[1]{\left| #1 \right>}
\newcommand{\bidx}[1]{{\textstyle {\atop #1}}\!}
\newcommand{\ro}{{\propto}}
\newcommand{\rc}{\infty}
\newcommand{\Nmn}[2]{\bar{N}^{#1}_{#2}}
\renewcommand{\overline}{\bar}
\def\Amn#1#2{A^{#1}_{#2}}
\def\Czzero#1{C_{\setbox0=\hbox{\tiny$(#1)$}%
\ht0=3pt\hbox{\scriptsize$z$}_0^{\box0}}}

\def\VEV#1{\left<#1\right>}

\def\half{\hbox{\large ${1\over2}$}}
\def\uoprj{\Pi}
\def\o{{\rmm o}}
\def\c{{\rmm c}}
\def\rH{\rmm H}
\def\wbar{{\bar w}}
\def\zbar{{\bar z}}
\def\bbar{{\bar b}}
\def\cbar{{\bar c}}

\def\kPhi{\ket{\Phi}}
\def\kphi{\ket{\phi}}
\def\kpsi{\ket{\psi}}
\def\kPsi{\ket{\Psi}}

\def\czm{c_0^-}

\def\czp{c_0^+}

\def\bzm{b_0^-}
\def\bzp{b_0^+}

\def\bzp{b_0^+}
\def\calO{{\cal O}}

\def\calP{{\cal P}}
\def\calN{{\cal N}}

\def\Lzero#1{L_0^{(#1)}}

\def\uoP#1{\Pi^{(#1)}}
\def\dZdW{\left(\frac{dZ_1}{dw_1}\frac{d\hat{Z}_1}{d\bar{w}_1}
     \frac{dZ_2}{dw_2}\frac{d\hat{Z}_2}{d\bar{w}_2}\right)^{-1}}
\def\zzs{z_0^{(s)}}
\def\bPP#1{(b_0^-{\cal P}\Pi)^{(#1)}}
\def\PP{({\cal P}\Pi)^{(r)}}
\def\odr{O}
\newcommand{\pderop}[1]{\frac{\partial}{\partial{#1}}}

\newcommand{\Nnm}[1]{\overline{N}^{#1}_{nm}}
\newcommand{\nm}{\frac{(2n-1)!!}{(2n)!!}\frac{(2m-1)!!}{(2m)!!}}
\newcommand{\N}[2]{\overline{N}^{#1}_{#2}}
\newcommand{\cdott}{\hspace{1pt}{\cdot}\hspace{1pt}}

\preprintnumber[4cm]{
UT-795\\
KUNS-1478\\ HE(TH)~97/18\\ 
hep-th/9711100\\
\today}

\title{
Unoriented Open-Closed String Field Theory
}

\author{
Taichiro {\sc Kugo}\footnote{E-mail address:
kugo@gauge.scphys.kyoto-u.ac.jp} 
and Tomohiko {\sc Takahashi}$^{\dagger,}$\footnote{
JSPS Research Fellow. E-mail address: tomo@hep-th.phys.s.u-tokyo.ac.jp}
}

\inst{
Department of Physics, Kyoto University, Kyoto 606-01
\\
$^\dagger$Department of Physics,
University of Tokyo, Tokyo 113
}

\recdate{
November 14, 1997}

\abst{
The string field theory for unoriented open-closed string mixed system 
is constructed up to quadratic order based on the joining-splitting type
vertices. The gauge invariance with closed string transformation parameter 
is proved. The infinity cancellation mechanism between disk and 
projective plane amplitudes plays an essential role for the gauge 
invariance of the theory.}

\begin{document}

\maketitle

\section{Introduction}
The discovery of D-branes\cite{rf:pol} has brought us with a quite new 
insight into string theories and various duality properties there, and 
actually triggered the recent rapid 
developments in the fields. Now the string theory seems no longer the 
theory of string alone, but the theory for a totality of branes with 
various dimensional extension. 

The most remarkable one among the recent developments would be matrix 
theory formulations for such a `string theory',\cite{rf:BFSS,rf:IKKT}
in which the various 
dimensional branes seems to be described unifiedly by a very simple 
looking system of matrix variables. It may also suggest a possibility 
for a quite new formulation for the `second-quantized theory' of many 
body systems in place of the conventional field theory. There are, 
however, many subtle problems in the limiting procedure of 
$N\rightarrow\infty$ and  it is still lacking some key idea for the matrix 
theory to become a truly powerful `second quantized theory' by which 
the conventional field theory can be a replaced. We still have something 
to learn from the conventional field theory.

Despite these developments the dynamics of the D-branes does not seem 
fully made clear. At this stage it would be necessary to try various 
approaches to reveal the string dynamics, namely, of various branes. 
There already appears some trials for studying the string system with 
D-branes in terms of string field theories. One approach by Hashimoto 
and Hata\cite{rf:HashiHata} is to introduce the D-brane explicitly in 
the closed string field theory as a source term using the boundary 
state. Another approach by Zwiebach,\cite{rf:zwie} who 
worked in the non-polynomial string field theory framework, 
is based on oriented open-closed string field theory, in which D-branes 
were not directly introduced but it is described by the end points of 
the open string by performing the T-duality transformation.

The purpose of this paper is to construct the string field theory for 
unoriented open-closed string mixed system with joining-splitting type 
vertices and to reveal the gauge 
symmetry structure.  This is of course in the hope to study the 
D-brane dynamics eventually. 
In open-closed string theories, the unoriented Type I theory with 
gauge group $SO(32)$ is the only known consistent one. Unless the gauge
group is $SO(32)$, the anomaly dose not cancel, and the infinities 
coming from the dilaton tadpole do appear so that the supersymmetric 
vacuum becomes unstable.\cite{rf:GreenSchwarz,rf:ItoyamaMoxhay} \  In 
bosonic case also, the infinities of the dilaton tadpole are known to be
canceled if and only if the gauge group is $SO(2^{13}=8192)$ in the 
unoriented theory.\cite{rf:DougGrin} \ Therefore, we can naturally expect 
that this infinity cancellation mechanism also work to guarantee the 
gauge invariance of the open-closed string field theory.

Since the bosonic string theories always suffer from the tachyon problem, 
we should ideally treat supersymmetric string theories. But, at the
present stage, there are 
still difficult problems for constructing supersymmetric string field 
theories.\cite{rf:wendt,rf:KugoTerao} \  So we here have to content ourselves with 
the construction of bosonic string field theory for the unoriented 
open-closed mixed system with gauge group SO($n$). 
Moreover, in this open-closed mixed system, there is an anomaly for 
the gauge transformation with open-string field parameter, which is 
similar to more familiar Lorentz anomaly in the light-cone gauge string 
field theory.\cite{rf:ST1,rf:ST2,rf:KikkawaSawada} \ 
Since we need a bit complicated loop amplitude calculations 
to treat it, we defer the discussion of the open-string parameter gauge 
transformation to the forthcoming paper.\cite{rf:KT} \  
In this paper, we treat only the gauge transformation with closed
string field parameter  
restricting the action to the quadratic part in the string fields.

\begin{wrapfigure}[15]{r}{\halftext}
    \epsfxsize= 6.5cm   
    \centerline{\epsfbox{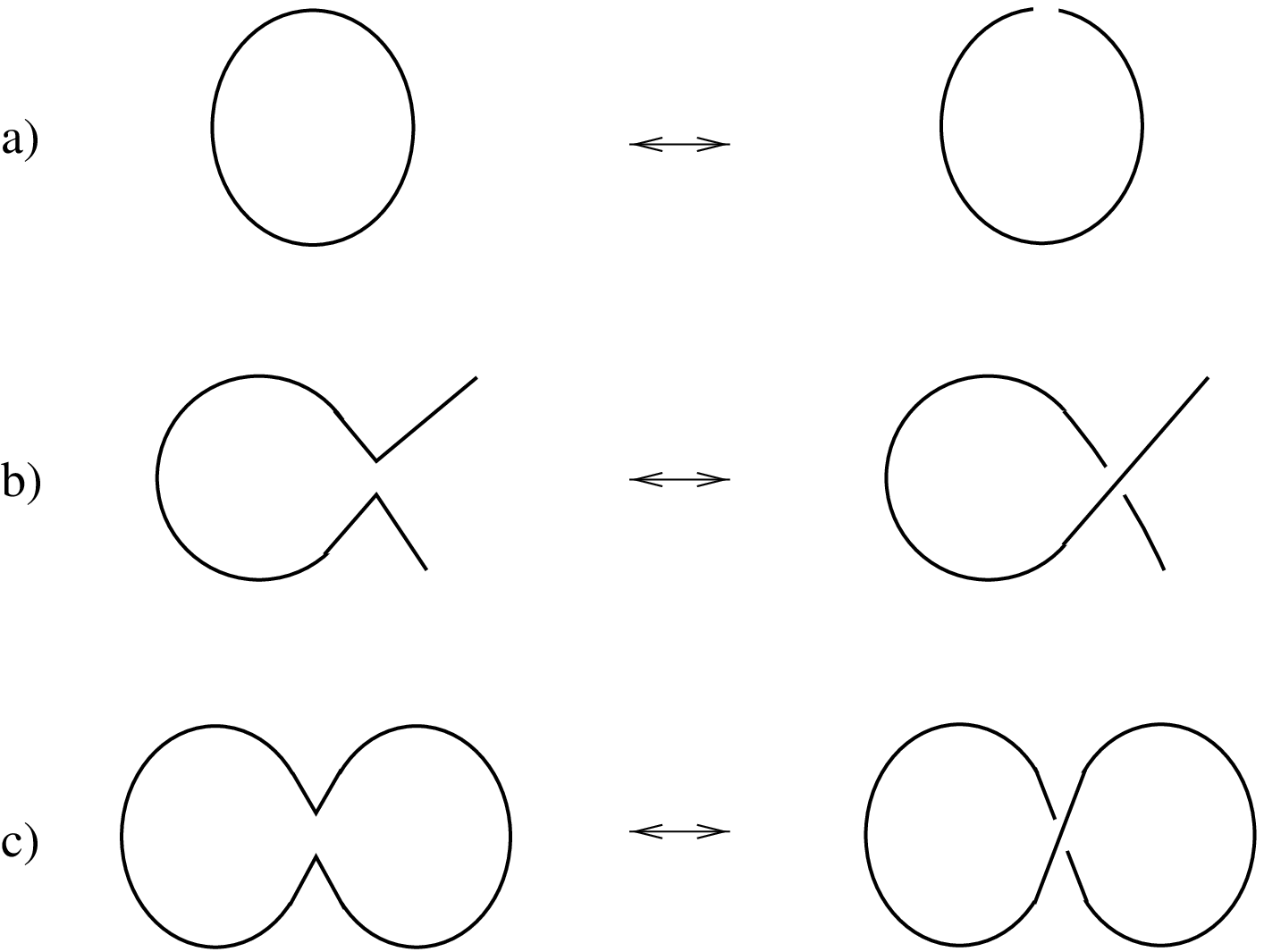}}
\vspace{1.5ex}
 \caption{Interaction vertices a) $U$, b) $V_\ro$ and c) $V_\rc$.}
 \label{fig:vertex}
\end{wrapfigure}

In the unoriented open-closed string field theory, there are three type 
of vertices for quadratic interactions as drawn in Fig.~\ref{fig:vertex}. 
One of them, $U$, corresponds to the transition from open to closed string, 
and vice versa. The others, $V_\ro$ and $V_\rc$, are the vertices in 
which an open or closed string intersects itself and rearranges. So, we 
can write the quadratic part of the action in the form:
\begin{eqnarray}
\label{eq:action}
S&=& -\half\Psi{\cdott}\tQB\Psi 
   -\half\Phi{\cdott}\tQB b_0^-\Phi \nn
  && \qquad +\aa g\,\Psi{\cdott} U\Phi 
  +\bb\,\frac{g^2}{2}\,\Psi{\cdott} V_\ro\Psi 
  +\cc\,\frac{g^2}{2}\,\Phi{\cdott} V_\rc\Phi,
\end{eqnarray}
where $\Psi$ and $\Phi$ denote open  and closed string field,
respectively, and $\aa$, $\bb$ and $\cc$ are coupling constants 
(relative to the open 3-vertex coupling constant $g$) 
for the open-closed transition $U$, open intersection $V_\ro$ and closed
intersection $V_\rc$ interactions, respectively. The details of the 
definition will be explained in the text. 

We consider the following form of 
gauge transformation with a closed string field parameter $\Lambda^\c$:
\begin{eqnarray}
\label{eq:gauge}
  \delta(b_0^-\Phi) &=& -\tQB b_0^- \Lambda^\c
     +\cc g^2 V_\rc\,\Lambda^\c, \nn
  \delta\Psi&=& \aa g U\Lambda^\c,
\end{eqnarray}
where $\Lambda^\c$ denotes a functional of closed string 
coordinate. Under this transformation, the action of 
Eq.~(\ref{eq:action}) is transformed, roughly writing, as
\begin{eqnarray}
\label{eq:dS}
  \delta S &=& \Phi\cdot\tQB^2 b_0^-\Lambda^\c
   +\aa g\,\Psi\cdot[\tQB,\,U]\Lambda^\c 
   +g^2\,\Phi\cdot\left(\aa^2\,UU-\cc\,\{\tQB,\,{V}_\rc\} 
        \right)b_0^-\Lambda^\c
      \nn
&&   +\aa g^3\,\Psi\cdot\left(-\cc\,{U}V_\rc
        +\bb\,V_\ro U\right)\Lambda^\c
   -\cc g^4\,\Phi\cdot{V}_\rc V_\rc\Lambda^\c. 
\end{eqnarray}
In order to realize the gauge invariance, 
each term of Eq.~(\ref{eq:dS}) must vanish order by order.
Moreover, in $O(g^2)$ terms in Eq.~(\ref{eq:dS}), the first $UU$ term 
corresponds to a singular world sheet of a disk and the second $\{\tQB,\,{V}_\rc\}$ term to real projective
($RP^2$) plane, each with two external states. 
They both contain infinities due to the dilaton and tachyon contributions 
in the configuration drawn in Fig.~\ref{fig:rappa} below. 
Thus, the unoriented open-closed string field 
theory can be gauge invariant if and only if the infinity cancellation
occurs between the disk and $RP^2$ amplitudes.
Conversely speaking, in the case of {\it oriented} open-closed string 
field theory, there is no $RP^2$ contribution since it comes from the 
$V_\rc$ vertex characteristic to the unoriented string, and therefore 
it requires other novel infinity cancellation mechanism
\cite{rf:CLNY} in order for 
the oriented string theory to be gauge invariant. This point is also the 
problem for the Lorentz invariance in the light-cone gauge string
field theory of oriented open-closed strings, which seems to have not
been noticed by other authors up to
now.\cite{rf:ST1,rf:ST2,rf:KikkawaSawada}

This paper is organized as follows. 
In sect.~2, we fix our conventions and notations for string fields 
as well as string coordinates as conformal fields. 
In sects.~3 and 4, we give precise definitions for the above 
vertices and action. In sect.~5, we prove the gauge invariance of the 
quadratic action for unoriented open-closed string field theory and 
determine relations between various coupling constants to satisfy the 
gauge invariance. 
Then, in sect.~6, we calculate two closed tachyon amplitudes 
corresponding to the disk and $RP^2$ from string field theory action, 
and see that the infinity cancellation occurs if and only if the 
gauge invariance condition is met. 
Sect.~7 is devoted to the summary and some more discussions on the 
gauge symmetry. Appendix A is added for clarifying the relations 
between three different expressions for the vertices by LeClair,
Peskin and Preitschopf (LPP),\cite{rf:LPP} Kunitomo and Suehiro
(KS),\cite{rf:KS} and Hata, Itoh, Kugo, Kunitomo and Ogawa
(HIKKO),\cite{rf:HIKKO1,rf:HIKKO2} which is necessary for the
gauge-invariance proof in Sect.~5. Appendix B gives the
explicit formula for the Neumann coefficients of open-closed
transition vertex.

\section{String Fields}
Let us recall the mode expansion of 
string coordinates $\phi \equiv(X^\mu,\ c, \ b)$ and fix our 
convention. In a unit disk $\abs{w_r}\leq1$, the string 
coordinates are expanded as
\begin{eqnarray}
  \label{modeW}
  &&X^\mu(w_r) = x^\mu-i\,\alpha_0^\mu\ln w_r
          +i\sum_{n\neq 0}\frac{1}{n}\alpha^\mu_n w_r^{-n}, \nn
  &&c(w_r) = \sum_n c_n w_r^{-n+1}, \qquad 
    b(w_r) = \sum_n b_n w_r^{-n-2}, \nn
 &&[x^\mu,\,\alpha^\nu_0]=i\eta^{\mu\nu},\qquad 
   \eta^{\mu\nu}={\rm diag}(-1,+1,\cdots,+1),
\end{eqnarray}
where $r$ denotes $r$-th string
and $l_s=\sqrt{2\alpha'}$ is set equal to 1. There is also an
anti-holomorphic counterpart  
$\bar \phi(\bar w_r)
=(\bar X^\mu(\bar w_r),\ \bar c(\bar w_r),\ \bar b(\bar w_r))$ 
for closed string, which lives in the unit disk with anti-holomorphic 
coordinate $\bar w_r$, $|\bar w_r|\leq1$. For any conformal mapping 
$z\rightarrow z'$, the string coordinates $\phi(z)$ are mapped as 
follows according to their conformal weights 
$d_\phi= (0,\ -1,\ +2)$ for $\phi=X^\mu, c, b$:
\begin{equation}
\phi(z) \rightarrow 
\phi'(z') = \left({dz\over dz'}\right)^{d_\phi}\phi(z).
\end{equation}
The unit disk can of course be mapped to a semi-infinite cylinder 
with more familiar world-sheet coordinate 
$\rho_r=\tau_r+i\sigma_r$, $\tau_r\leq0$ 
by a conformal mapping $w_r=e^{\rho_r}=e^{\tau_r+i\,\sigma_r}$, 
and the string coordinates on the $\rho_r$ plane are therefore 
given by
\begin{eqnarray}
  \label{modeRho}
  &&X^\mu(\rho_r) = x^\mu-i\,\alpha_0^\mu\rho_r
          +i\sum_{n\neq 0}\frac{1}{n}\alpha^\mu_n e^{-n\rho_r}, \nn
  &&c(\rho_r) = \sum_n c_n e^{-n\rho_r}, \qquad 
    b(\rho_r) = \sum_n b_n e^{-n\rho_r}.
\end{eqnarray}

For open-string, the $\sigma_r$ originally runs only over 
$0\leq\sigma_r\leq\pi$, but it is extended in these equations to 
run over $0\leq\sigma_r\leq2\pi$ just as for the closed string. 
Because of $2\pi$ periodicity of $\sigma_r$, we shall conveniently 
take also the region $-\pi\leq\sigma_r\leq\pi$ as well as 
$0\leq\sigma_r\leq2\pi$ as the fundamental region of $\sigma_r$ 
both for open and closed strings, depending on the situations.
Note also that the real string coordinate
$X^\mu(\sigma,\tau)$ is given by
\begin{eqnarray}
\label{eq:realcoord}
X^\mu(\sigma,\tau)=\half\big(X^\mu(\rho)+ \bar X^\mu(\bar\rho)\big)
\end{eqnarray}
with the understanding that $\bar X^\mu=X^\mu$ for open string case.

The open string field $\Psi$ and closed one $\Phi$ are denoted in our
notation by
\begin{eqnarray}
        \kPsi&=&\kphi+c_0\kpsi, \nn
        \kPhi&=&\czm\left(\kphi+\czp\kpsi\right)
                +\left(\ket\chi+\czp\ket\eta\right).
\label{eq:component}
\end{eqnarray}
It should be kept in mind that the open string field is 
$n\times n$ matrix valued; $\ket{\Psi}_{ij}$.
As for the ghost zero-modes for the closed string, we use the notation:
\begin{eqnarray}
&&\czp\equiv\half(c_0+\bar c_0), \quad \czm = c_0-\bar c_0, \nn 
&&\bzp\equiv b_0+\bar b_0, \quad \bzm\equiv\half(b_0-\bar b_0).
\end{eqnarray}
The physical fields are contained in the $\kphi$ component.
We take the $\kphi$ component to be Grassmann even, so that 
the closed string field $\kPhi$ is Grassmann {\em odd} while the open 
string field $\kPsi$ is {\em even}. 
We define the Fock vacuum $\ket{1} = (\bra{1})^\dagger$ by
\begin{eqnarray}
\alpha_n^\mu\ket{1} &=& 0 \quad {\rm for}\ \ n\geq1, \nn 
c_n \ket{1} &=& 0 \quad {\rm for}\ \ n\geq1, \nn 
b_n \ket{1} &=& 0 \quad {\rm for}\ \ n\geq0,
\end{eqnarray}
and we keep to use the coordinate (or momentum) representation for the
zero modes $p^\mu$, $x^\mu$. $\ket{1,1}=(\bra{1,1})^\dagger$
is the Fock vacuum for the closed string possessing the holomorphic
and anti-holomorphic oscillator freedoms. Concerning the ghost mode
part, in particular, the Fock vacuum is related with the conformal
vacuum $\ket{0}$ by 
\begin{equation}
\ket{1} = c_1 \ket{0}, \quad \bra{1} = \bra{0}c_{-1}, \quad 
\hbox{so that} \quad \bra{1}c_0\ket{1}=\bra{0}c_{-1}c_0c_1\ket{0}=1. 
\end{equation}
The label 1 of the Fock vacuum $\ket1$, thus, implies the ghost number 
relative to the conformal vacuum. As for the order of the holomorphic and 
anti-holomorphic ghost zero-modes for the closed string case, we take a 
convention 
\begin{equation}
\bra{1,1} \czm\czp \ket{1,1} = 
\bra{1,1} c_0 \bar c_0 \ket{1,1} = 1.
\end{equation}

We here define reflectors,\cite{rf:kugozwie} which convert the ket
representation to the bra representation, for the open and closed strings,
respectively: 
\begin{eqnarray}
\bra{R^\o(1,2)}&=&\delta(1,2)\,\bidx{1}\bra{1}\bidx2\bra{1}
({c_0}^{(1)}\!+{c_0}^{(2)})\exp{(E^\o_{12})}, \nn
\bra{R^\c(1^\c,2^\c)}&=&\delta(1,2)\,\bidx{1}\bra{1,1}
\bidx{2}\bra{1,1}
(c_0^{+(1)}\!+c_0^{+(2)})
(c_0^{-(1)}\!+c_0^{-(2)})
\exp{(E^\c_{12})}, \hspace{2em}
\end{eqnarray}
with the exponents given by
\begin{eqnarray}
\label{eq:refexp}
E^\o_{12} &=& \sum_{n\geq1}(-)^{n+1}\left(\frac{1}{n}
{\alpha_n^\mu}^{(1)}{\alpha_n}_\mu^{(2)}
+{c_n}^{(1)}{b_n}^{(2)}
-{b_n}^{(1)}{c_n}^{(2)}\right), \nn
E^\c_{12} &=& -\sum_{n\geq1}\left(\frac{1}{n}
{\alpha_n^\mu}^{(1)}{\alpha_n}_\mu^{(2)}
+{c_n}^{(1)}{b_n}^{(2)}
-{b_n}^{(1)}{c_n}^{(2)}\right)+{\rm a.h.},
\end{eqnarray}
and the delta functions by
\begin{eqnarray}
  \delta(1,2,\cdots,n)=(2\pi)^{d}\,\delta^{d}(
\sum^n_{r=1}p_r^\mu)
\end{eqnarray}
with $d=26$. We understood that the delta function 
$\delta(1,2)$ in $\bra{R^\o(1,2)}$ case also implies the matrix 
reflection $\delta(i_1,j_1;i_2,j_2)\equiv\delta_{i_1,j_2}\delta_{i_2,j_1}$.
We also define ket reflectors $\ket{R^{\o,\c}}$ as the
inverse of the bra reflectors $\bra{R^{\o,\c}}$ in the sense that
\begin{eqnarray}
&&\bra{R^\o(1,3)}\ket{R^\o(3,2)}\ket{\Psi}_1 =\ket\Psi_2, 
\quad {\rm for}\ \ \forall\Psi,  \nn
&&\bra{R^\o(1^\c,3^\c)}\ket{R^\o(3^\c,2^\c)}
\ket{\Phi}_{1^\c} =\ket\Phi_{2^\c}
\quad {\rm for}\ \ \forall\Phi,
\end{eqnarray}
where the inner-product of the bra and ket states implies,
in addition to the usual product for oscillators, integration 
over the zero-modes, $\int d^{d}p/(2\pi)^{d}$ (as well as the trace 
operation over the matrix for the open string case). The ket reflectors 
are easily seen to be the minus of the hermitian conjugates of the bra 
reflectors:
\begin{equation}
\ket{R^\o(1,2)}=-\left(\bra{R^\o(2,1)}\right)^\dagger, \qquad 
\ket{R^\c(1^\c,2^\c)}=-\left(\bra{R^\c(2^\c,1^\c)}\right)^\dagger.
\end{equation} 
We note the symmetry property of the reflectors:
\begin{eqnarray}
&&\bra{R^\o(2,1)}\ket\Psi_1\ket{\Psi'}_2 = 
- (-1)^{|\Psi| |\Psi'|} \bra{R^\o(1,2)}\ket{\Psi'}_2\ket\Psi_1, \nn
&&\bra{R^\c(2^\c,1^\c)}\ket\Phi_1\ket{\Phi'}_2 = (-1)^{|\Phi| |\Phi'|}
\bra{R^\o(1^\c,2^\c)}\ket{\Phi'}_2\ket\Phi_1,
\end{eqnarray}
where $|\Psi|$ is 0 (1) if $\ket\Psi$ is Grassmann even (odd).
The  minus sign for the former comes from the interchange of 
the bra Fock vacua: 
$\bidx{2}\bra{1}\bidx1\bra{1} = -\bidx{1}\bra{1}\bidx2\bra{1}$, which 
effectively contains the ghost zero-mode integration 
$\int dc_0^{(2)}dc_0^{(1)} = -\int dc_0^{(1)}dc_0^{(2)}$. 

In defining string vertices below, we assign string length $\alpha$
to each participating string. Basically we adopt in this paper 
the so-called $\alpha=p^+$ HIKKO theory,\cite{rf:kugozwie} in which
we identify the string length $\alpha$ as follows:
\begin{eqnarray}
\alpha=\cases{
2p^+ & for open string \cr
p^+ & for closed string. \cr}
\end{eqnarray}
So the $\alpha$ parameter is not an independent freedom unlike
in HIKKO theory. (The following discussion in this paper will, however,
be equally valid even when we instead adopt the standpoint of 
`covaliantized light-cone string 
field theory'\cite{rf:Siegel,rf:NW,rf:Uehara,rf:Kugo}
provided that we make a suitable
re-interpretation like replacing the zero mode variable $p^\mu$ by 
an OSp($d+2|2$) vector.) Since only the ratio of the $\alpha$
parameters is relevant to the vertices, the string length parameter 
$\alpha$ can be fixed to be $\pm2$ for open and $\pm1$ for closed
strings, respectively, in this paper where only the quadratic vertices 
are treated.

\section{Vertices}

We construct BRS invariant vertices corresponding to three types of 
interactions following the method of LeClair, Peskin and 
Preitschopf (LPP).\cite{rf:LPP} \

To define BRS invariant vertices a la LPP, we have to specify the way 
how to connect the strings: a single $\rho$-plane is constructed by 
gluing the $\rho_r$-planes (or, unit disks $w_r$,) of individual 
strings. In the case of the open-closed transition vertex $U$, the 
coordinate $\rho$ is identified region-wise with the coordinate 
of each string, as depicted in Fig.~\ref{fig:op-cl}:
\begin{figure}[b]
    \epsfxsize=7cm
    \centerline{\epsfbox{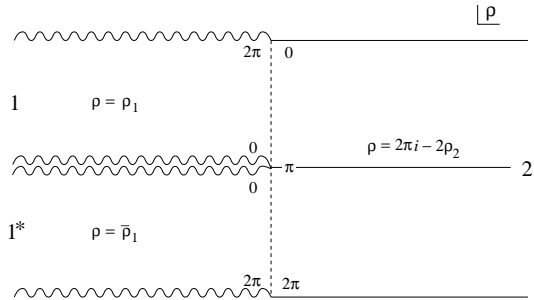}}
    \caption{The $\rho$ plane representing the connection of
open-closed transition vertex $U$.}
\label{fig:op-cl}
\end{figure}
%
\begin{equation}
\label{eq:rho o-c}
\rho= \tau+i\sigma= 
\cases{ 
\rho_1=\tau_1+i\sigma_1
        & for $\tau\leq0$,\ \ $0\leq\sigma\leq2\pi$ : region 1 \cr 
\bar\rho_1=\tau_1-i\sigma_1
        & for $\tau\leq0$,\ \ $-2\pi\leq\sigma\leq0$ : region $1^*$ \cr 
-2\rho_2 + 2\pi i & \cr
\quad = -2(\tau_2+i\sigma_2) + 2\pi i
        & for $\tau\geq0$, \ \ $-2\pi\leq\sigma\leq2\pi$ : region 2.
\cr}
\end{equation}
Note that, as explained before, we have assigned string length
parameter $\alpha_1=1$ for incoming closed string 1 and $\alpha_2=-2$
for outgoing open string 2. The regions 1 and $1^*$ stand for the
holomorphic and anti-holomorphic parts of the closed string 1, and the
region 2 for the open string 2. With this gluing, a single holomorphic
function $\phi(\rho)$ is defined on the $\rho$ plane:
\begin{equation}
\phi(\rho)=
\cases{
 \phi^{(1)}(\rho)
                &in region 1 \cr
 \bar{\phi}^{(1)}(\rho)
                &in region $1^*$ \cr
 (-2)^{-d_\phi}\,\phi^{(2)}(-\rho/2+\pi i)
                &in region 2. \cr}
 \label{eq:connect-oc}
\end{equation}
Therefore, considering the overlapping line $\tau=0$ and 
Eq.~(\ref{eq:connect-oc}),
we have the following connection equations on
the LPP vertex for the open-closed transition,  $\bra{u(2,1^\c)}$:
\begin{eqnarray}
&&\bra{u(2,1^\c)}\bigl[\phi^{(1)}(i\sigma)
   -(-2)^{-d_\phi}\phi^{(2)}(i\pi-i \sigma/2)\bigr]=0, 
    \nn
&&\bra{u(2,1^\c)}\bigl[\bar{\phi}^{(1)}(-i\sigma)
   -(-2)^{-d_\phi}\phi^{(2)}(i\pi+i \sigma/2)\bigr]=0,
\label{eq:OCconnect}
\end{eqnarray}
for $0\leq\sigma\leq2\pi$.
Note that the real coordinate 
$X^\mu(\sigma,\tau)$ 
at $\tau=0$ is given by 
$X^\mu(\sigma,0)= \half\big( X^\mu(i\sigma) + \bar
X^\mu(-i\sigma)\big)$.
So $X^{(1)}(\sigma,0)=X^{(2)}(\pi-\sigma/2,0)$ is realized 
on the vertex $\bra{u(2,1^\c)}$ as desired.

\begin{figure}[b]
 \parbox{\halftext}{
    \epsfxsize=6.5cm
    \centerline{\epsfbox{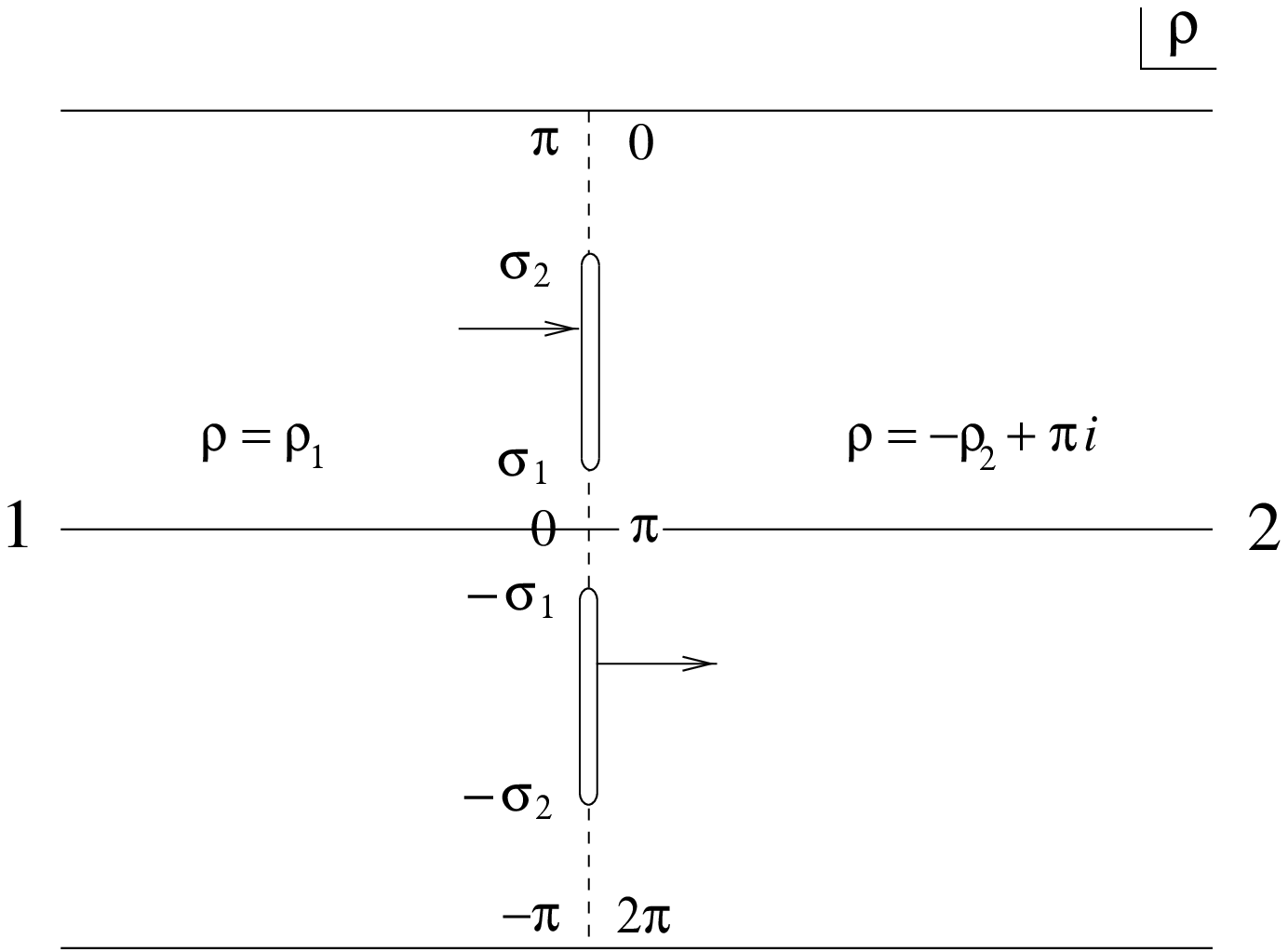}}
    \caption{The $\rho$ plane for open intersection vertex ${V_\ro}$.}
    \label{fig:op-op}}
 \hspace{7mm}
 \parbox{\halftext}{
    \vspace{4ex}\hspace{-2mm}
    \epsfxsize= 7cm   
    \centerline{\epsfbox{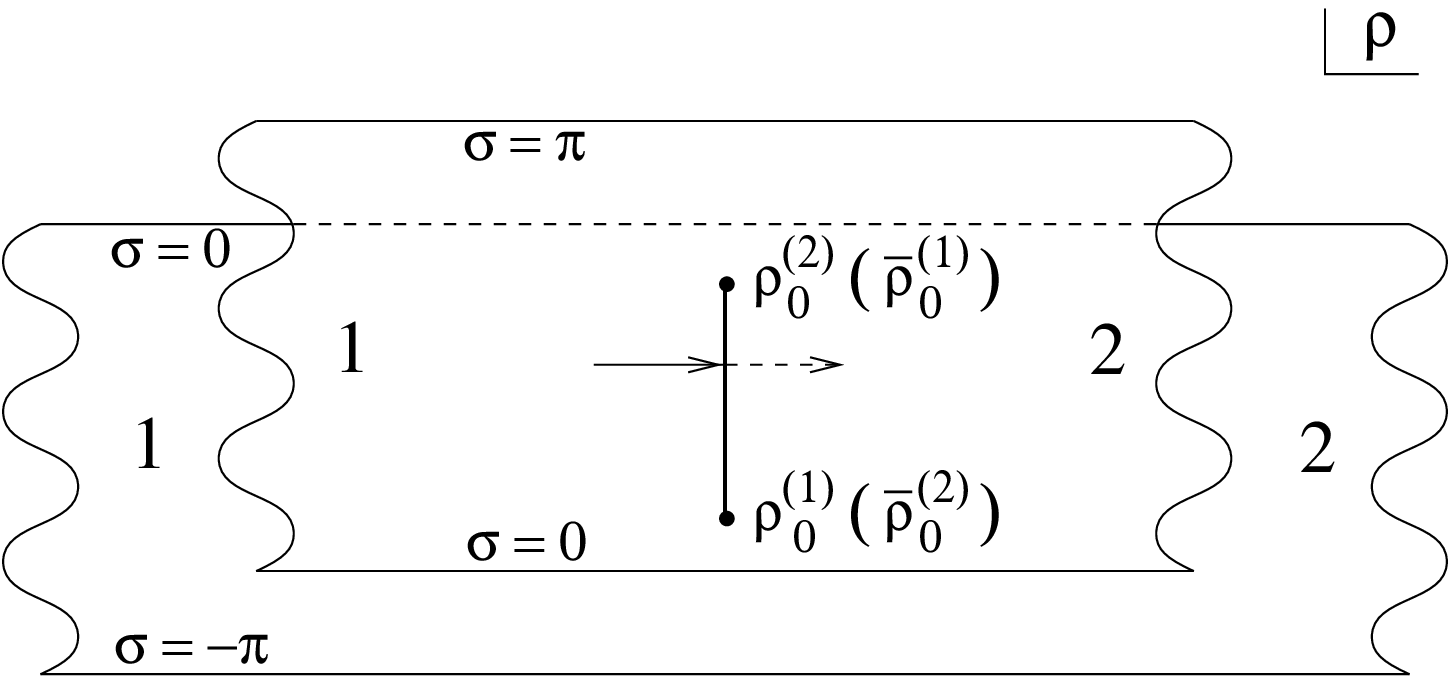}}
    \vspace{3ex}
    \caption{Redrawing of the $\rho$ plane for the vertex ${V_\ro}$
    with the cut overlapped. $(\bar\rho_0^{(i)})$ denotes the branch points 
on the second sheet.}
    \label{fig:op-opnew}}
\end{figure}

The $\rho$ plane for the open intersection interaction $V_\ro$ is a bit 
more complicated. Prepare the $\rho$ plane first as follows:
\begin{equation}
  \rho=
\cases{
 \rho_1=\tau_1+i\,\sigma_1
                &for $\tau\leq0$,\ \ $-\pi\leq\sigma\leq\pi$ : region 1
\cr
 -\rho_2+\pi i=-(\tau_2+i\,\sigma_2)+\pi i
                &for $\tau\geq0$,\ \ $-\pi\leq\sigma\leq\pi$ : region
2.\cr}
\end{equation}
Then, as shown in  Fig.~\ref{fig:op-op}, we make two cuts running from 
$\sigma=\sigma_1$ to $\sigma=\sigma_2$ and from $\sigma=-\sigma_2$ to
$\sigma=-\sigma_1$ at $\tau=0$. 
And the left hand side points on the one cut are identified with the 
right hand side points on the other cut; namely, the point 
$\sigma\in[\sigma_1,\,\sigma_2]$ at $\tau=0$ in the region 1 (2) is
identified with the point 
$\sigma-(\sigma_1+\sigma_2)\in[-\sigma_2,\,-\sigma_1]$ in the region 2 (1).  
With this identification through the cuts, the desired $\rho$ plane 
is defined, as drawn in Fig.~\ref{fig:op-opnew}. 
The single holomorphic field on the $\rho$ plane is defined as usual by
\begin{eqnarray}
  \phi(\rho)=
\cases{
\phi^{(1)}(\rho) 
                &in region 1\cr
(-1)^{-d_\phi}\,\phi^{(2)}(-\rho+\pi i) 
                &in region 2.\cr}
\label{eq:connect-oo}
\end{eqnarray}
Considering the cut structure of the Riemann surface of $\rho$ and
Eq.~(\ref{eq:connect-oo}), the open intersection vertex, $\bra{v_\ro}$,
satisfies the following connection condition for $0\leq\sigma\leq\pi$,
\begin{eqnarray}
 &&\bra{v_\ro(1,2)}\bigl[\phi^{(1)}(i\sigma)
          -(-1)^{-d_\phi}\,\phi^{(2)}(i\pi-i\sigma)\bigr]=0
\qquad (\sigma\leq\sigma_1, \ \ \sigma\geq\sigma_2), \nn
 && \bra{v_\ro(1,2)}\bigl[\phi^{(1)}(i\sigma)
          -(-1)^{-d_\phi}\,\phi^{(2)}(-i\sigma_1-i\sigma_2+i\sigma 
)\bigr]=0
\ (\sigma_1\leq\sigma\leq\sigma_2),\hspace{2em}
\end{eqnarray}
and, for $-\pi\leq\sigma\leq0$,
\begin{eqnarray}
 && \bra{v_\ro(1,2)}\bigl[\phi^{(1)}(i\sigma)
          -(-1)^{-d_\phi}\,\phi^{(2)}(-i\pi-i\sigma)\bigr]=0
\qquad (\sigma\geq-\sigma_1, \ \ \sigma\leq-\sigma_2), \nn
 && \bra{v_\ro(1,2)}\bigl[\phi^{(1)}(i\sigma)
          -(-1)^{-d_\phi}\,\phi^{(2)}(i\sigma_1+i\sigma_2+i\sigma 
)\bigr]=0
\ (-\sigma_2\leq\sigma\leq-\sigma_1).\hspace{2em}
\end{eqnarray}
These conditions imply that a open string is twisted 
in the region of $\sigma_1<\abs{\sigma}<\sigma_2$.

\begin{figure}[b]
    \vspace*{-1ex}
 \parbox{\halftext}{
    \epsfxsize=6.5cm
    \centerline{\epsfbox{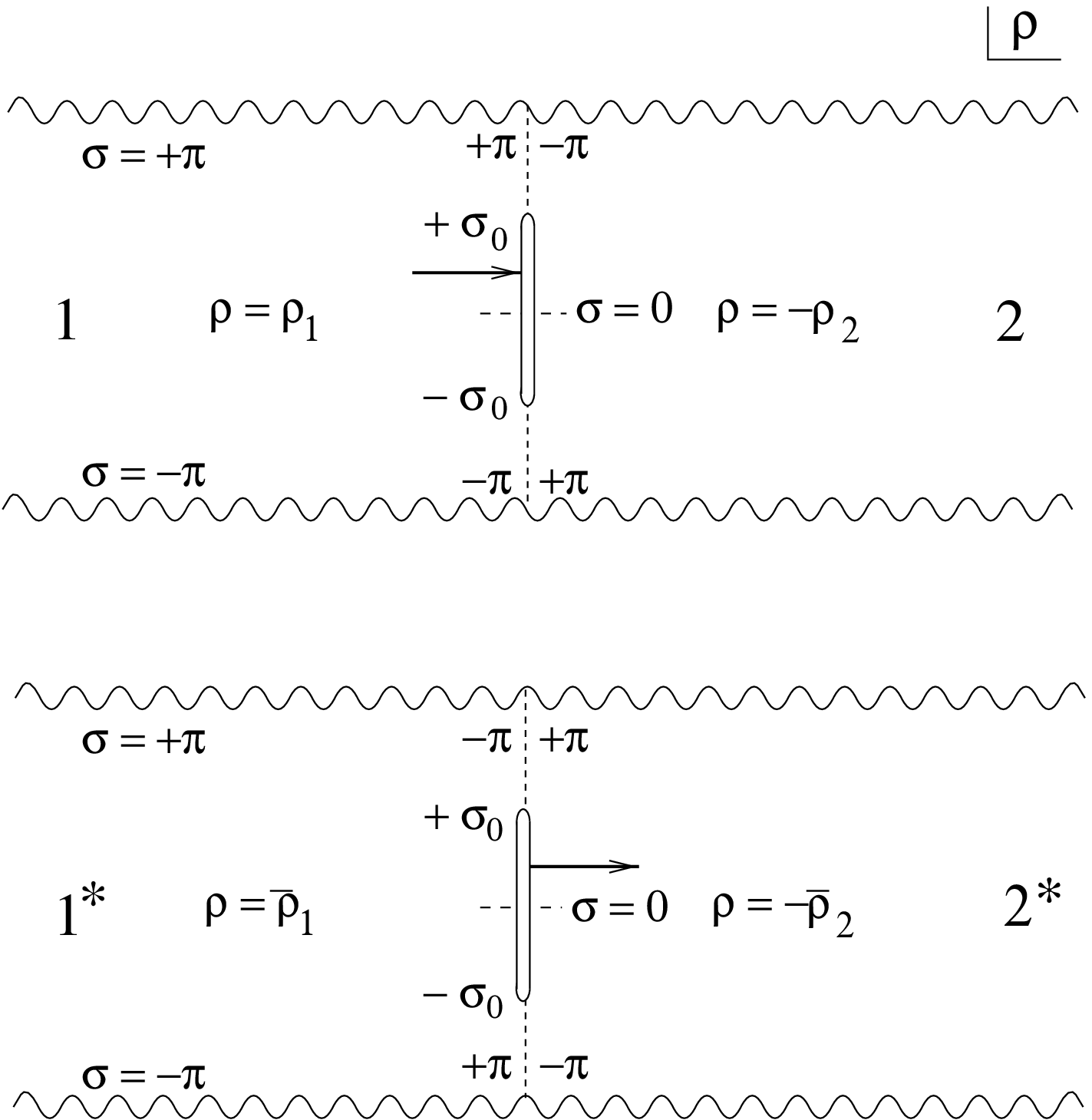}}
    \caption{Two sheets of $\rho$ planes for vertex ${V_\ro}$. }
    \label{fig:cl-cl}}
 \hspace{7mm}
 \parbox{\halftext}{
    \vspace{7ex}\hspace{-2mm}
    \epsfxsize= 7cm
    \centerline{\epsfbox{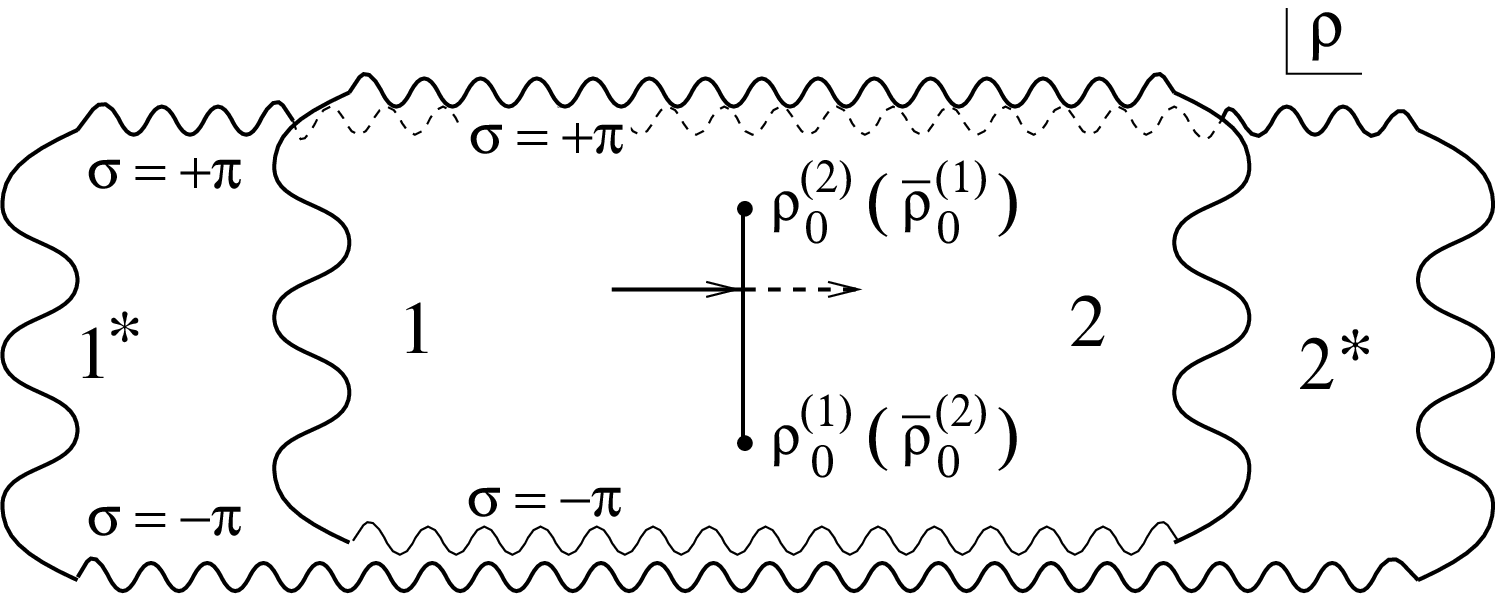}}
    \vspace{6.2ex}
    \caption{The $\rho$ plane for the closed intersection vertex 
${V_\ro}$
    drawn in such a way that the cut is overlapped. 
    Again $(\bar\rho_0^{(i)})$ denotes the coordinates of the branch 
    points on the second sheet.} 
    \label{fig:cl-clnew}}
\end{figure}

In the case of the closed intersection vertex $V_\rc$, we first prepare 
two sheets of $\rho$ planes, one for the holomorphic and another for the
anti-holomorphic parts, separately:
\begin{eqnarray}
&&\hspace{-1em}\hbox{first sheet}\nn
  &&\rho=
\cases{
 \rho_1=\tau_1+i\,\sigma_1
                &for $\tau\leq0$,\ \ $-\pi\leq\sigma\leq\pi$ : region 1
\cr
 -\rho_2=-\tau_2-i\,\sigma_2
                &for $\tau\geq0$,\ \ $-\pi\leq\sigma\leq\pi$ : region 2,
\cr} \nn
&&\hspace{-1em}\hbox{second sheet}\nn
&& \rho=
\cases{
 \bar{\rho}_1=\tau_1-i\,\sigma_1
                &for $\tau\leq0$,\ \ $-\pi\leq\sigma\leq\pi$ : region
$1^*$ \cr
 -\bar{\rho}_2=-\tau_2+i\,\sigma_2
                &for $\tau\geq0$,\ \ $-\pi\leq\sigma\leq\pi$ : region
$2^*$. \cr}.
\end{eqnarray}
Then, again two cuts are placed from $\sigma=-\sigma_0$ to 
$\sigma=+\sigma_0$ at $\tau=0$ on the first sheet and on the 
second as shown in Fig.~\ref{fig:cl-cl}, and again the left hand side points 
on the one cut are identified with the right hand side points on the 
other cut.
The resulting $\rho$ plane is drawn in Fig.~\ref{fig:cl-clnew}. 
The holomorphic fields are defined by
\begin{equation}
  \phi(\rho)=
\cases{
\phi^{(1)}(\rho) 
                &in region 1\cr
\bar{\phi}^{(1)}(\rho) 
                &in region $1^*$\cr
(-1)^{-d_\phi}\,\phi^{(2)}(-\rho) 
                &in region 2\cr
(-1)^{-d_\phi}\,\bar{\phi}^{(2)}(-\rho) 
                &in region $2^*$.\cr}
\label{eq:connect-cc}
\end{equation}
As in the above case, we obtain the connection equations
on the closed intersection vertex $\bra{v_\rc}$,
\begin{eqnarray}
&&\cases{
\bra{v_\rc(1,2)}\bigl[\phi^{(1)}(i\sigma)
          -(-1)^{-d_\phi}\,\phi^{(2)}(-i\sigma)\bigr]=0 \cr
\noalign{\vskip.5ex}
\bra{v_\rc(1,2)}\bigl[\bar{\phi}^{(1)}(-i\sigma)
          -(-1)^{-d_\phi}\,\bar{\phi}^{(2)}(+i\sigma)\bigr]=0 \cr}
\qquad (\abs{\sigma}\geq\sigma_0), \nn
&&\cases{
\bra{v_\rc(1,2)}\bigl[\phi^{(1)}(i\sigma)
          -(-1)^{-d_\phi}\,\bar{\phi}^{(2)}(-i\sigma)\bigr]=0 \cr
\noalign{\vskip.5ex}
\bra{v_\rc(1,2)}\bigl[\bar{\phi}^{(1)}(-i\sigma)
          -(-1)^{-d_\phi}\,{\phi}^{(2)}(i\sigma)\bigr]=0 \cr}
\qquad (\abs{\sigma}\leq\sigma_0),
\end{eqnarray}
which implies that twisting of a closed string is made in 
the region $\abs{\sigma}<\sigma_0$.

With these conditions of gluing, the corresponding LPP vertices are 
uniquely specified including their normalization, and they satisfy 
the following BRS invariance:\cite{rf:LPP}
\begin{eqnarray}
&&\bra{u(1,1^\c)}(\QB^{(1)}+\QB^{(1^\c)})=0, \nn
&&\bra{v_\ro(1,2)}(\QB^{(1)}+\QB^{(2)})=0, \nn
&&\bra{v_\rc(1^\c,2^\c)}(\QB^{(1^\c)}+\QB^{(2^\c)})=0.
\end{eqnarray}

But they still do not give the final form of the desired 
vertices which reproduce the correct Polyakov amplitudes:
firstly, for each closed string, we should multiply 
the projection operator ${\cal P}$, which projects only the 
$L_0-\bar L_0=0$ modes out, 
\begin{equation}
{\cal P}\equiv\int^{2\pi}_0
\frac{d\theta}{2\pi}\exp i\theta(L_0-\bar{L}_0),
\end{equation}
and the corresponding antighost zero-mode factor $b_0^-=(b_0-\bbar_0)/2$. 
Secondly, since the intersection vertices have moduli parameters 
$\sigma_0, \sigma_1$ and $\sigma_2$,
we must insert antighost factors associated with the quasi-conformal 
deformations of the Riemann surface corresponding to the changes of 
those moduli parameters.\cite{rf:AGMV} \ 
Finally, unoriented projection operators $\uoprj$ 
have to be multiplied:
\begin{equation}
\uoprj=\half(1+\Omega),
\label{eq:unorientedprj}
\end{equation}
where $\Omega$ denotes a twist operator:
\begin{eqnarray}
\Omega \phi(\rho) \Omega^{-1}
=\cases{
\phi(i\pi+\bar \rho)& for open string\cr
\bar \phi(\bar \rho)& for closed string.\cr}
\end{eqnarray}
Let us understand that, for open string case, the twist operator 
$\Omega$ in Eq.~(\ref{eq:unorientedprj}) also implies to taking the 
transposition of the matrix index, so that the operator $\Pi$ projects 
out the unoriented O($n$) type open string field satisfying 
$\ket{\Psi}^{\rm T}=\Omega\ket{\Psi}$. As a result, we obtain the final 
forms of vertices for those three types of interactions:

\ \ 1.\ open-closed transition vertex
\begin{eqnarray}
\bra{U(1,2^\c)}=\bra{u(1,2^\c)}{b^-_0}^{(2^\c)}{\cal P}^{(2^\c)}
\prod_{r=1,2^\c}\uoprj^{(r)}, 
\end{eqnarray}

\ \ 2.\ open intersection vertex
\begin{eqnarray}
\bra{V_\ro(1,2)}=\int_{0\leq\sigma_1\leq\sigma_2\leq\pi}
d\sigma_1 d\sigma_2\bra{v_\ro(1,2;\sigma_1,\sigma_2)}
b_{\sigma_1}b_{\sigma_2}
\prod_{r=1,2}\uoprj^{(r)},
\end{eqnarray}

\ \ 3.\ closed intersection vertex
\begin{eqnarray}
\bra{V_\rc(1^\c,2^\c)}=\int_0^{\frac{\pi}{2}} \,d\sigma_0
\bra{v_\rc(1^\c,2^\c;\sigma_0)}
b_{\sigma_0}\prod_{r=1^\c,2^\c}{b_0^-}^{(r)}{\cal P}^{(r)}
\uoprj^{(r)}.
\end{eqnarray}
Note that the product of unoriented projections $\prod_{r}\Pi^{(r)}$ 
in these three vertices may be replaced by a single $\Pi$ of either 
one of the two strings since the two $\Pi$'s are the same on these 
vertices. 
As for the matrix index property, the vertex $\bra{U(1,2^\c)}$ 
is proportional to the unit matrix ${\bf 1}_{ij}$ for the matrix 
index $i,j$ of the open string 1, and the vertex $\bra{V_\ro(1,2)}$ 
is proportional to 
$\delta(i_1,j_1;i_2,j_2)\equiv\delta_{i_1,j_2}\delta_{i_2,j_1}$
just like the open reflector. 
As a convention here and henceforth, the order of the 
ghost factors $b_0^{-(r)}$ in $\prod_{r=1^\c,2^\c}b_0^{-(r)}$ is 
defined as the same as that appearing in the arguments of 
the vertex $\bra{v_\rc(1^\c,2^\c;\sigma_0)}$; so here 
$\prod_{r=1^\c,2^\c}{b_0^-}^{(r)}= {b_0^-}^{(1^\c)}{b_0^-}^{(2^\c)}$. 

\begin{wrapfigure}[10]{r}{\halftext}
    \epsfxsize=4cm
    \centerline{\epsfbox{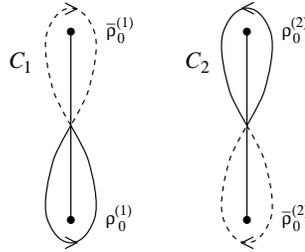}}
\vspace{1ex}
\caption{The contours $C_j$ in Eq.~(\protect\ref{eq:bsigma}). 
}
\label{fig:contour}
\end{wrapfigure}

$b_{\sigma_j}$ ($j=0,1,2$) are the anti-ghost factors for the 
quasi-conformal deformations corresponding to the change of 
the moduli parameters $\sigma_j$. 
They are given by\cite{rf:AGMV,rf:GM,rf:KugoSuehiro}
\begin{eqnarray}
&&b_{\sigma_j}=\left({d\rho_0^{(j)}\over d\sigma_j}\right)
b_{\rho_0^{(j)}}, \ \ {\rm for}\ \ j=1, 2, \nn
&&b_{\sigma_0}=
\left({d\rho_0^{(1)}\over d\sigma_0}\right)b_{\rho_0^{(1)}}
+\left({d\rho_0^{(2)}\over d\sigma_0}\right)b_{\rho_0^{(2)}}, \nn
&&b_{\rho_0^{(j)}}=
 \oint_{C_j}\frac{d\rho}{2\pi i}b(\rho),
\ \ {\rm for}\ \ j=1, 2,
\label{eq:bsigma}
\end{eqnarray}
where the contour $C_j$ is a closed path depicted in
Fig.~\ref{fig:contour} which goes around the 
interaction point $\rho_0^{(j)}$ on the first sheet and then around $\bar
\rho_0^{(j)}$ on the second sheet of the $\rho$ planes in 
Figs.~\ref{fig:op-opnew} and \ref{fig:cl-clnew}.

A comment may be in order on the integration region for $\sigma_0$ 
in the closed intersection vertex $V_\rc$. 
At first sight it seems that the moduli parameter $\sigma_0$ should be 
integrated from 0 to $\pi$. However, since 
the configuration with $\pi/2\leq\sigma_0\leq\pi$ becomes the same as 
that with $0\leq\sigma_0\leq\pi/2$ if the twist operator $\Omega$ 
is acted on either one of the two strings 1 or 2, we have an identity 
for the LPP vertex:
\begin{eqnarray}
\bra{v_\rc(1^\c,2^\c;\pi-\sigma_0)}\Omega^{(2)} \prod_{r=1^\c,2^\c}{\cal
P}^{(r)}&=&
-\bra{v_\rc(1^\c,2^\c;\sigma_0)}\prod_{r=1^\c,2^\c}{\cal P}^{(r)} 
\nn
&& \hspace{5em} (0\leq\sigma_0\leq\pi/2).
\label{eq:Vcidentity}
\end{eqnarray}
But there is an unoriented projection $\uoprj$ acting on the vertex 
and $\Omega\uoprj=\uoprj$, the contribution of 
the configurations with $\pi/2\leq\sigma_0\leq\pi$ is actually the same as
that with $\pi/2\leq\sigma_0\leq\pi$ in our final vertex. So we have 
restricted the integration region to $0\leq\sigma_0\leq\pi/2$ to 
correctly cover the moduli space only once. [The minus sign in 
Eq.~(\ref{eq:Vcidentity}) is important here and later. It should 
appear here since otherwise the integral 
$\int d\sigma_0 \bra{v_\rc(1^\c,2^\c;\sigma_0)}b_{\sigma_0}$, 
if extended over $0\leq\sigma_0\leq\pi$, would vanish because of the 
fact $b_{\sigma_0}=(d\rho_0/d\sigma_0)b_{\rho_0}=-b_{\pi-\sigma_0}$. 
More directly, as we shall see below, the LPP vertex 
$\bra{v_\rc(1^\c,2^\c;\sigma_0)}$ contains two ghost factors at 
the two interaction points $\sigma=\pi\pm\sigma_0$ if rewritten 
in terms of the HIKKO representation. The multiplication order of these 
two is opposite between the two configurations with $\sigma_0$ and 
$\pi-\sigma_0$, which yields the minus sign in 
Eq.~(\ref{eq:Vcidentity}).] 

\section{Action}

We shall see later that there appears a divergence owing to the tachyon 
at order $g^2$ and it is canceled by the shift of the zero intercept 
parameter in $L_0$. Taking this into account in advance, we define here 
tilded (or, `bare') BRS charges by shifting the zero intercept 
parameters in the usual (`renormalized') BRS charges $\QB^\o$ and 
$\QB^\c$ for open and closed strings, respectively, as follows:
\begin{eqnarray}
\tQB^\o &=& \QB^\o + \lambda_\o g^2 c_0, \nn
\tQB^\c &=& \QB^\c + \lambda_\c g^2 c_0^+,
\end{eqnarray}
where $\lambda_\o g^2$ and $\lambda_\c g^2$ are the amounts of the zero 
intercept shift for open and closed strings, respectively, which will be
determined later by the gauge invariance requirement. 

We can now give the more precise form for the action
Eq.~(\ref{eq:action}) and the gauge transformation
Eq.~(\ref{eq:gauge}) using the bra-ket notation:
\begin{eqnarray}
\label{eq:braketAction}
  S &=& -\half \bra{R^\o(2,1)}\ket{\Psi}_1
\tQB^{(2)}\Pi^{(2)}\ket{\Psi}_2
-\half \bra{R^\c(2,1)}\ket{\Phi}_1
\tQB^{(2)}{b^-_0}^{(2)}\bPP{2} \nn
&& \qquad {}+\aa g\,\bra{U(2,1)}\ket{\Phi}_1\ket{\Psi}_{2}
+\bb\,\half{g^2}\bra{V_\ro(2,1)}\ket{\Psi}_1\ket{\Psi}_2 \nn
&& \qquad\qquad {}+\cc\,\half{g^2}
\bra{V_\rc(2,1)}\ket{\Phi}_1\ket{\Phi}_2,
\end{eqnarray}
\begin{eqnarray}
\label{eq:braketgauge}
\delta({b_0^-}^{(1)}\ket{\Phi}_1) &=&
    -\tQB^{(1)}\bPP{1}\ket{\Lambda^\c}_1
    +\cc g^2 \bra{V_\rc(3,2)}\ket{R^\c(1,2)}\ket{\Lambda^\c}_3, \nn
  \delta\ket{\Psi}_1 &=&
    \aa g\bra{U(2,1^\c)}\ket{R^\o(1,2)}\ket{\Lambda^\c}_{1^\c}.
\end{eqnarray}
where $\bPP{r}$ is a shorthand notation for 
${b_0^-}^{(r)}{\cal P}^{(r)}\uoP{r}$. 
Since the gauge transformation $\delta$ is the usual derivation, 
the gauge transformation parameter $\ket{\Lambda^\c}$ is Grassmann even. 
The change of the action under this gauge transformation is given by 
\begin{eqnarray}
\label{eq:dS-bk}
 &&\hspace{-1em} \delta S = 
  \delta S_1 + \delta S_2 + \delta S_3 + \delta S_4, \nn
 &&\delta S_1=
    \aa g\bra{U(1,1^\c)}(\tQB^{(1)}+\tQB^{(1^\c)})
          \ket{\Psi}_1\ket{\Lambda^\c}_{1^\c} \nn
 &&\delta S_2=
    +g^2\left[\aa^2\bra{U(1,1^\c)}\bra{U(2,2^\c)}\ket{R^\o(1,2)}\right.\nn
    && \hspace{5.5em}
          -\cc\bra{V_\rc(2^\c,1^\c)}(\tQB^{(1^\c)}+\tQB^{(2^\c)})\nn
    && \hspace{5.5em}\left. 
         +\lambda_\c\bra{R^\c(2^\c,1^\c)}\{\QB^{(1^\c)},\,
          {c_0^+}^{(1^\c)}\}{b_0^-}^{(1^\c)}\right]
          \ket{\Phi}_{1^\c}\ket{\Lambda^\c}_{2^\c} \nn
 &&\delta S_3=
     +\aa g^3\left[-\cc\bra{\check{U}(1,2^\c)}\bra{V_\rc(3^\c,1^\c)}
          \ket{R^\c(2^\c,3^\c)} \right.\nn
    && \hspace{7em}
         +\bb\bra{V_\ro(1,2)}\bra{U(3,1^\c)}\ket{R^\o(2,3)}\big]
        \ket{\Psi}_1\ket{\Lambda^\c}_{1^\c} \nn
 &&\delta S_4=
     -\cc^2g^4 \bra{\check{V}_\rc(1^\c,3^\c)}
          \bra{V_\rc(4^\c,2^\c)}\ket{R^\c(3^\c,4^\c)}
         \ket{\Phi}_{1^\c}\ket{\Lambda^\c}_{2^\c}.
\end{eqnarray}
Here the checked vertices $\bra{\check{U}}$ and $\bra{\check{V}_\rc}$ 
are those with a factor $\bzm$ removed:
\begin{eqnarray}
  \bra{U(1,2^\c)}=\bra{\check{U}(1,2^\c)}{b_0^-}^{(2^\c)},
\hspace{2em}
  \bra{V_\rc(1,2)}=\bra{\check{V}_\rc(1,2)}{b_0^-}^{(2)}.
\end{eqnarray}
In this calculation, it is useful to remember the symmetry properties 
$\bra{V_\ro(1,2)}=\bra{V_\ro(2,1)}$ and 
$\bra{V_\rc(1,2)}=-\bra{V_\rc(2,1)}$, and also the following Grassmann 
even-odd properties of the vertices: the open-closed transition vertex 
$\bra{U}$ and open reflector $\bra{R^{\o}}$ are {\em odd} and all the 
others $\bra{R^{\c}},\ \bra{V_\ro},\ \bra{V_\rc}$ are {\rm even}. These 
properties can easily be seen from the form of each term in the action 
(which is even, of course) and the fact that $\kPsi$ is even and $\kPhi$ 
odd.

\section{Gauge invariance}

In this section, we show that the parameters 
$\lambda_\c, \lambda_\o, \aa, \bb$ and $\cc$ in the action
(\ref{eq:braketAction}) should satisfy
\begin{eqnarray}
&&\lambda_\c=2\lambda_\o = -\lim_{a\rightarrow\infty}{2n\aa^2\over a^2},
\nn
&&\cc = n\aa^2, \nn
&&\cc = -4\pi i \bb,
\label{eq:relations}
\end{eqnarray}
in order for the theory to be gauge invariant. 
To show this, let us now examine each term in 
Eq.~(\ref{eq:dS-bk}) order by order.

\subsection{Order $g$ invariance}

It follows from the definition of the LPP vertex that
  \begin{eqnarray}
    \bra{U(1,1^\c)}(\QB^{(1)}+\QB^{(1^\c)})=0,
  \end{eqnarray}
which has been proven first in Ref.~\citen{rf:ShapThorn,rf:Hata-Nojiri}.
This finishes the proof of order $g$ invariance. But the first
term $\delta S_1$ in Eq.~(\ref{eq:dS-bk}) contains also the order $g^3$ 
terms coming from the shift of the zero intercept parameter. They have 
to cancel out by themselves since no other counterparts appear to cancel
them. So we here consider the condition for them to cancel, although 
they are of order $g^3$.

{}From the connection equation Eq.~(\ref{eq:OCconnect}) for $\phi=c$, 
we have
\begin{eqnarray}
&&\bra{U(1,1^\c)}\bigl[ c^{(1^\c)}(\sigma)
   +2 c^{(1)}(\pi-\frac{\sigma}{2})\bigr]=0, 
    \nn
&&\bra{U(1,1^\c)}\bigl[ \bar{c}^{(1^\c)}(-\sigma)
    +2 c^{(1)}(\pi+\frac{\sigma}{2})\bigr]=0.
\end{eqnarray}
Summing these two and performing integration $\int_0^{2\pi}d\sigma/2\pi$, 
we find
\begin{equation}
\bra{U(1,1^\c)} \bigl[ 
   {c_0}^{(1^\c)}+{\bar{c}_0}^{(1^\c)} + 4 {c_0}^{(1)} \bigr]
=\bra{U(1,1^\c)}\  2 \bigl[ 
   {c_0^+}^{(1^\c)} + 2{c_0}^{(1)} \bigr]
   = 0
\end{equation}
Therefore, if we take 
\begin{eqnarray}
\label{eq:lambda}
  \lambda_\c=2\lambda_\o,
\end{eqnarray}
it follows that
\begin{eqnarray}
  \bra{U(1,1^\c)}(\tQB^{(1)}+\tQB^{(1^\c)})=0.
\end{eqnarray}
The requirement (\ref{eq:lambda}) is natural since 
the zeroth order zero intercepts $\alpha_0^\o$ and $\alpha_0^\c$ for 
open and closed strings also satisfy the relation 
$\alpha_0^\c=2\alpha_0^\o=2$. Hereafter, we accept the relation of 
Eq.~(\ref{eq:lambda}) and then the `order $g$' gauge invariance 
$\delta S_1=0$ is realized.

\subsection{Order $g^2$ invariance}

\begin{figure}[tb]
 \parbox{\halftext}{\hspace{-1em}
    \epsfxsize=6cm
    \centerline{\epsfbox{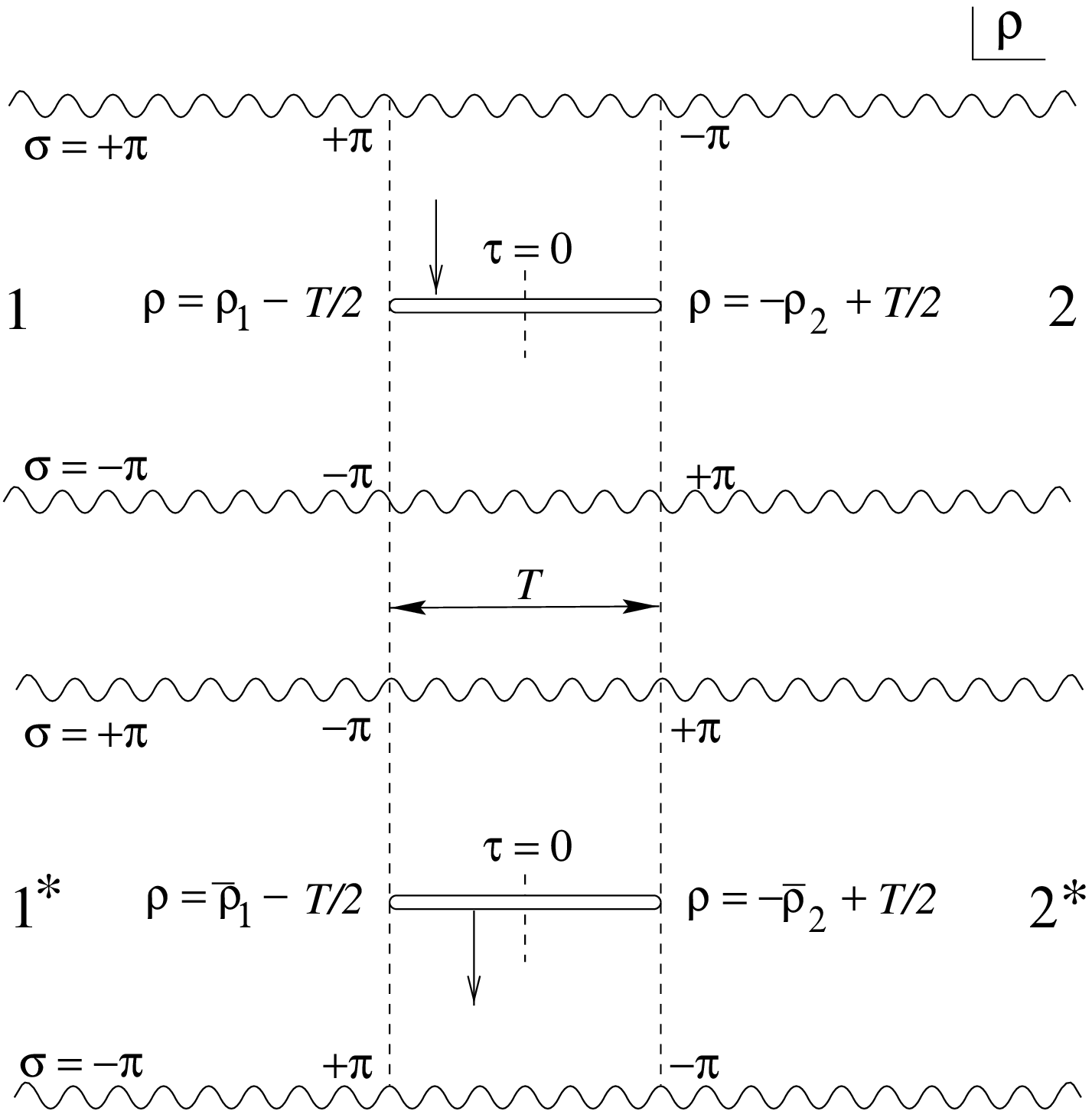}}}
 \hspace{8mm}
 \parbox{\halftext}{
    \vspace{3ex}\hspace{-1.3em}
    \epsfxsize= 7cm
    \centerline{\epsfbox{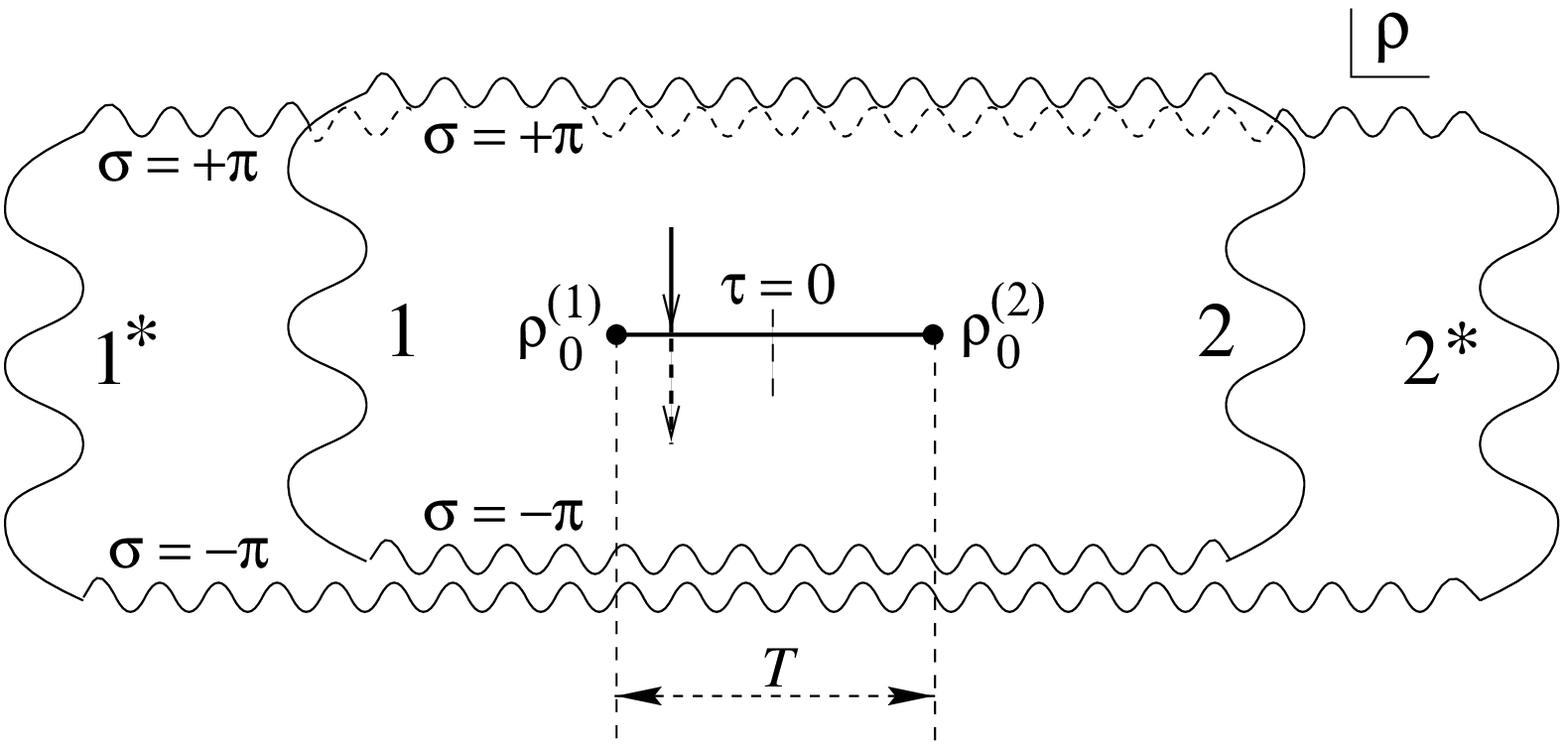}}
    }
\caption{The $\rho$ plane for the vertex $\bra{v_{D_2}}$.}
\label{fig:cl-cl-D2}
\end{figure}

In the order $g^2$ terms $\delta S_2$ in Eq.~(\ref{eq:dS-bk}), 
the first $UU$ and the second $V_\rc$ terms are divergent for these 
correspond to singular configurations in moduli space, so that we need 
regularization for their evaluation. First, we rewrite the first terms 
as follows,
\begin{eqnarray}
  \bra{U(1,1^\c)}\bra{U(2,2^\c)}\ket{R^\o(1,2)}
   &=& \lim_{T\rightarrow0}
\bra{u(1,1^\c)}\bra{u(2,2^\c)}
e^{-(1/2)\Lzero2 T}\ket{R^\o(1,2)} \nn
&& \qquad {}\times e^{+(L^{(1^\c)}+L^{(2^\c)})T/2}\!\!
      \prod_{r=1^\c,2^\c}\!\!\bPP{r},
\label{eq:regularize}
\end{eqnarray}
where $L\equiv L_0 + \bar L_0$ for closed string. 
Here the unoriented projection operators for the intermediate open 
strings 1 and 2 have been dropped since 
$\bra{u(1,1^\c)}\uoP1{\cal P}^{(1^\c)} = 
\bra{u(1,1^\c)}\uoP{1^\c}{\cal P}^{(1^\c)}$. 
$T$ is introduced as a regularization parameter and 
has the meaning of propagation time of the intermediate open string. 
The factor 1/2 in $e^{-(1/2)\Lzero2 T}$ is the factor $1/\alpha_2$ of the 
string length parameter $\alpha_2=2$ for the intermediate open string 2. 
We can replace the operator $e^{-(1/2)\Lzero2 T}$ by a more 
symmetric one $e^{-(\Lzero1+\Lzero2)T/4}$ if necessary since 
$\Lzero1=\Lzero2$ in front of the reflector $\ket{R^\o(1,2)}$. 
Note that we also have added the operator $e^{+(L^{(1^\c)}+L^{(2^\c)})T/2}$:
this is because it almost cancels 
$e^{-(1/2)\Lzero2 T}=e^{-(\Lzero1+\Lzero2)T/4}$ even for 
finite $T$ since the energy momentum tensor $T(\rho)$ is continuous 
on the vertices $\bra{u(r,r^\c)}$ aside from the interaction point,  
although the right hand side in Eq.~(\ref{eq:regularize}), in any case, 
reduces to the left hand side as $T\rightarrow0$, formally.
The part $\bra{u(1,1^\c)}\bra{u(2,2^\c)}e^{-(1/2)\Lzero2 T}
\ket{R^\o(1,2)}$ can be evaluated according to the LPP's 
``Generalized gluing and resmoothing" (GGR) theorem,\cite{rf:LPP} 
and so we obtain
\begin{eqnarray}
\label{eq:D2vertex}
&& \bra{U(1,1^\c)}\bra{U(2,2^\c)}\ket{R^\o(1,2)}\nn
&&\qquad {} = n\,\lim_{T\rightarrow 0}
\bra{v_{D_2}(1^\c,2^\c;T)}e^{+(L^{(1^\c)}+L^{(2^\c)})T/2} 
       \!\!\!\prod_{r=1^\c,2^\c}\!\!\bPP{r},
\end{eqnarray}
where $\bra{v_{D_2}(T)}$ is the LPP vertex, which 
we call D2 vertex, corresponding to 
the $\rho$ plane represented in Fig.~\ref{fig:cl-cl-D2}, 
and the factor $n$ came from the 
trace of Chan-Paton factor of the intermediate open string. 

The second term in the order $g^2$ terms $\delta S_2$
in Eq.~(\ref{eq:dS-bk}) is rewritten as 
\begin{eqnarray}
  &&\bra{V_\rc(1^\c,2^\c)}(\tQB^{(1^\c)}+\tQB^{(2^\c)})
    = \int^{\frac{\pi}{2}}_0 d\sigma_0 
    \bra{v_\rc(1^\c,2^\c;\sigma_0)}\{\QB,\,b_{\sigma_0}\}
      \prod_{r=1^\c,2^\c}\bPP{r} \nn
&&
\hspace{4em} {}+\lambda_\c \int^{\frac{\pi}{2}}_0 d\sigma_0
     \bra{v_\rc(1^\c,2^\c;\sigma_0)}b_{\sigma_0}
      ({c_0^+}^{(1^\c)}+{c_0^+}^{(2^\c)})
      \prod_{r=1^\c,2^\c}\bPP{r}.
\end{eqnarray}
We now use the identity,\cite{rf:AGMV,rf:FMS,rf:KugoSuehiro}
\begin{eqnarray}
  \bra{v_\rc(1^\c,2^\c;\sigma_0)}\{\QB,\,b_{\sigma_0}\}
&=&
  \bra{v_\rc(1^\c,2^\c;\sigma_0)}T_{\sigma_0} \nn
&=& {d\over d\sigma_0}\bigl\{\bra{v_\rc(1^\c,2^\c;\sigma_0)}\bigr\},
\end{eqnarray}
which follows from the fact that 
\begin{eqnarray}
  T_{\sigma_0}=\{\QB,\,b_{\sigma_0}\}
   = \left[
\left({d\rho_0^{(1)}\over d\sigma_0}\right)\oint_{C_1}
+\left({d\rho_0^{(2)}\over d\sigma_0}\right)\oint_{C_2}\right]
\frac{d\rho}{2\pi i}T(\rho),
\end{eqnarray}
is the generator for the change of the moduli parameter $\sigma_0$ 
(since, generally, the contour integration of energy momentum tensor 
$\oint_{C_Z}(dz/2\pi i)T(z)$ along a closed path $C_Z$ 
encircling the point $Z$ is known\cite{rf:GM} to give a generator 
for $\partial/\partial Z$).
Then, the integral over $\sigma_0$ in the first term leaves only 
the surface terms:
\begin{eqnarray}
\label{eq:P2vertex}
\hspace{-2em}  \bra{V_\rc(1^\c,2^\c)}(\tQB^{(1^\c)}+\tQB^{(2^\c)})
    &=& -\lim_{\sigma_0\rightarrow0}
   \bra{v_\rc(1^\c,2^\c;\sigma_0)}
      \prod_{r=1^\c,2^\c}\bPP{r} \nn
&&
\hspace{-9em} {}+\lambda_\c \int^{\frac{\pi}{2}}_0 d\sigma_0
     \bra{v_\rc(1^\c,2^\c;\sigma_0)}b_{\rho_0}
      ({c_0^+}^{(1^\c)}\!+{c_0^+}^{(2^\c)})
      \!\!\prod_{r=1^\c,2^\c}\!\!\bPP{r} ,
\end{eqnarray}
where the surface term at the other end point $\sigma_0=\pi/2$ has dropped
out since it vanishes by itself owing to the identity 
Eq.~(\ref{eq:Vcidentity}). We call this vertex $\bra{v_\rc(\sigma_0)}$ 
P2 vertex for short.

\begin{wrapfigure}{r}{6.6cm}
   \epsfxsize= 5.5cm  
   \centerline{\epsfbox{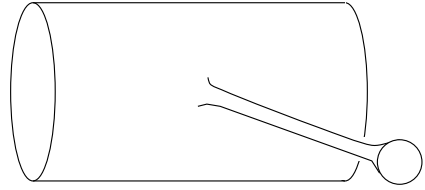}}
 \caption{The configuration yielding the singularities in the D2 and P2 vertices.}
 \label{fig:rappa}
\end{wrapfigure}
The quantities Eq.~(\ref{eq:D2vertex}) for D2 vertex and 
the first term of Eq.~(\ref{eq:P2vertex}) for P2 vertex
correspond to a common configuration as depicted in the Fig.~\ref{fig:rappa} 
in the limit $T,\sigma_0\rightarrow0$, and so have the well-known
singularities owing to the tachyon and dilaton contributions. 
We want to evaluate this singularity in Eqs.~(\ref{eq:D2vertex}) and 
(\ref{eq:P2vertex}) explicitly. For that purpose, it is convenient to 
map the $\rho$ plane to the whole complex $z$ plane. This map is 
constructed as follows: first, consider the former D2 case, 
corresponding to the diagram Fig.~\ref{fig:cl-cl-D2} with open string 
propagating as an intermediate state. By the usual Mandelstam mapping,
\begin{eqnarray}
\label{eq:Mandelmap}
  \rho=\ln(2-\zeta)-\ln(\zeta+2),
\end{eqnarray}
we can first map the $\rho$ plane of Fig.~\ref{fig:cl-cl-D2} to the Riemann 
surface, $\zeta$ plane. The cut between $\rho_0^{(1)}=-T/2$ and 
$\rho_0^{(2)}=T/2$ on the $\rho$ plane, which corresponds to the open string
boundary, is mapped to the cut between $\zeta=-2a$ and $\zeta=2a$ on $\zeta$
plane:
\begin{equation}
  {T\over2} = \ln{1+a\over1-a} \ \Rightarrow\ a = \tanh{T\over4}.
\label{eq:interval}
\end{equation}
So $a$ is a real positive number in $0\leq a\leq1$, corresponding 
the moduli $0\leq T\leq\infty$. This $\zeta$ plane, having two sheets as
$\rho$ plane, is then mapped to the whole complex $z$ plane by the 
Joukowski transformation:
\begin{eqnarray}
\label{eq:Joukowmap}
  \zeta=z+\frac{a^2}{z}.
\end{eqnarray}
\begin{figure}[htb]
 \parbox{\halftext}{
    \epsfxsize=65mm
    \centerline{\epsfbox{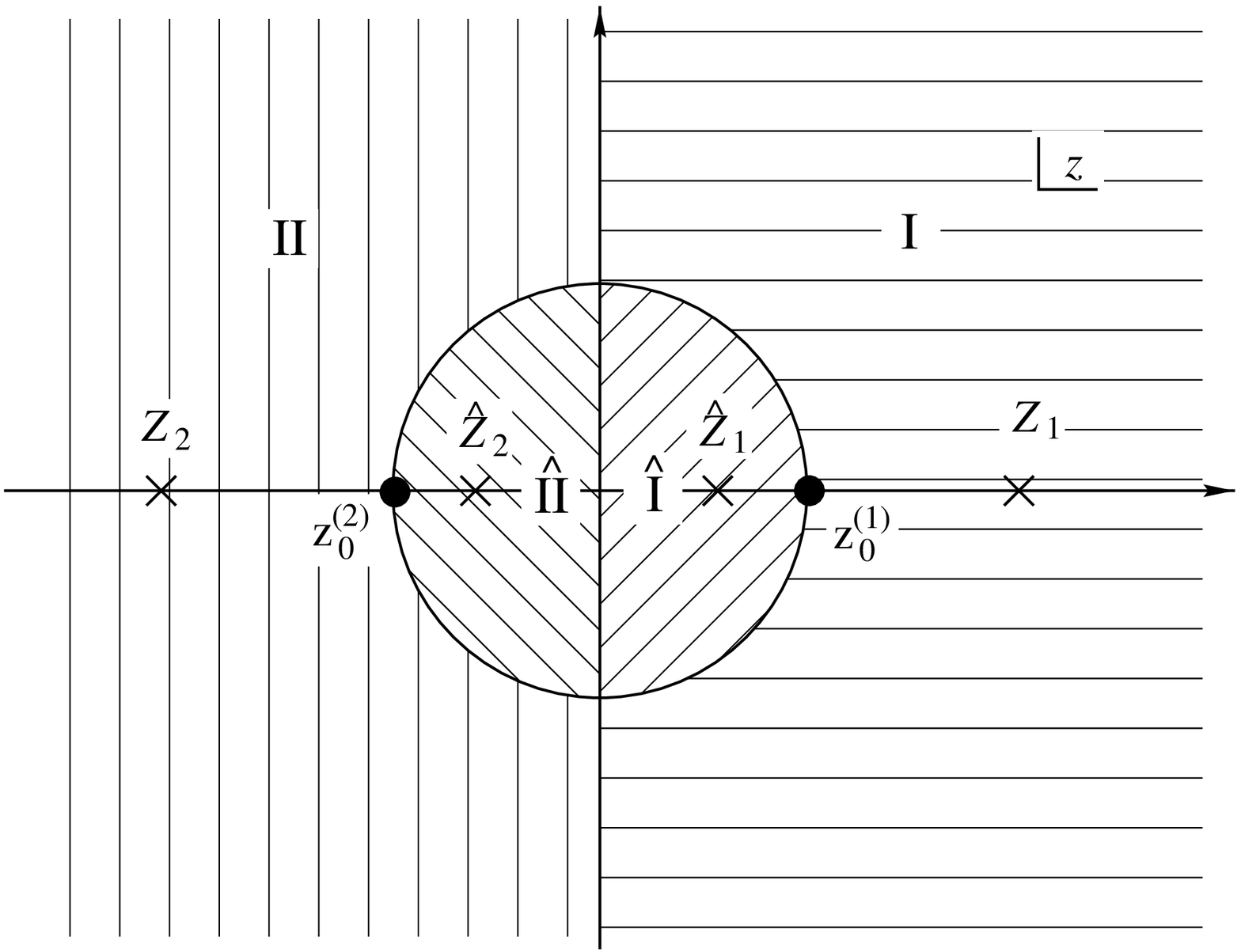}}
    \caption{The $z$ plane of the D2 vertex. 
Regions I and II ($\,\hat{\rm I}$ and $\hat{\rm II}\,$) correspond to
the $\tau{<}0$ and $\tau{>}0$ regions on the first (second) sheet of
the $\rho$ plane in Fig.~8, respectively.}
    \label{fig:z-D2}}
 \hspace{8mm}
 \parbox{\halftext}{
    \epsfxsize= 65mm
    \centerline{\epsfbox{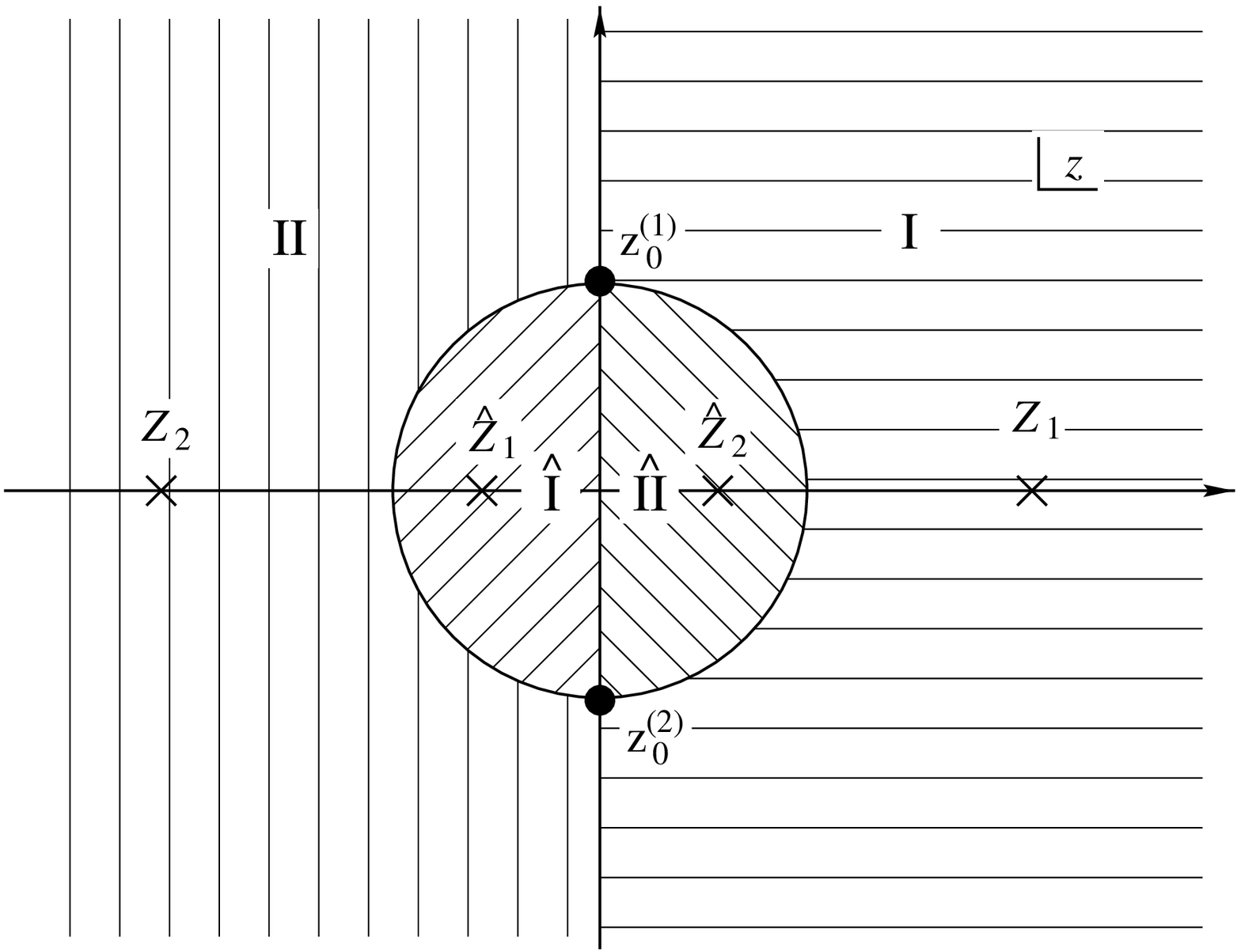}}
    \caption{The $z$ plane of the P2 vertex.
Regions I and II \,($\,\hat{\rm I}$ and $\hat{\rm II}\,$) correspond to
the regions 1 and 2 ($\,1^*$ and $2^*\,$) on the first (second) sheet of
the $\rho$ plane in Fig.~6.}
    \label{fig:z-P2}}
\end{figure}
The correspondence between the each string domain of the $\rho$ plane
and the $z$ plane is represented in Fig.~\ref{fig:z-D2}.
Note that the $z$ plane is now of a single sheet. 
The first sheet of $\rho$ (or $\zeta$) plane is mapped to the 
outer region of the circle with radius $|z|=a$ and the second sheet to 
the inner region. 
If $z$ is a root of
Eq.~(\ref{eq:Joukowmap}) for a given $\zeta$ (or $\rho$), so is
$a^2/z$. When a point $\rho$ on the first sheet is mapped to $z$ in
the outer region, $|z|\geq a$, $\bar\rho$ on the second sheet
corresponds to $a^2/\zbar$ in the inner region. Therefore, in
particular, we have the following correspondence for the real string
coordinate:
\begin{equation}
X^{\mu(1)}(\sigma_1,\tau_1)
=\half\big(X^{\mu(1)}(\rho)+\bar X^{\mu(1)}(\bar\rho)\big)
\quad \Leftrightarrow\quad 
\half\big(X^\mu(z)+ X^\mu({a^2\over\zbar})\big).
\label{eq:realX}
\end{equation}

Consider next the $\rho$ plane of Fig.~\ref{fig:cl-clnew}, for the 
P2 vertex $V_\rc$. We should notice that this plane
differs from the previous $\rho$ plane of Fig.~\ref{fig:cl-cl-D2}, only by 
the cut structure which in this case corresponds to a cross cap instead 
of a open string boundary, and it is well-known that the mapping $\rho$ to
$z$ in this case is given by the same one as above simply by changing 
the moduli from real $a$ to imaginary $ia$. The resultant $z$ plane is 
depicted in Fig.~\ref{fig:z-P2}.

Therefore the evaluation of 
Eq.~(\ref{eq:D2vertex}) and the first term of Eq.~(\ref{eq:P2vertex})
can be carried out in a unified manner. 
The $\rho$ planes are mapped to the $z$ planes by the mapping
\begin{eqnarray}
\label{eq:rho-z map}
  \rho=\ln\left(2-\bigl(z+\frac{q^2}{z}\bigr)\right)
         -\ln\left(z+\frac{q^2}{z}+2\right),
\end{eqnarray}
with real $q=a$ or imaginary $q=ia$ for D2 and P2 vertex cases, 
respectively. 
It is important for later consideration that on the $z$ plane
in Fig.~\ref{fig:z-D2} or in \ref{fig:z-P2}, 
only one complex $z$ plane is needed for our analysis, and 
all of the insertion points of strings
are on the real axis, the situation which is very similar 
to purely open strings system.

The interaction points are given by
\begin{eqnarray}
\label{eq:intpt}
  \frac{d\rho}{dz}=0\ \ \ \Rightarrow\ \ \ 
  z_0^{(1)}=+q,\ z_0^{(2)}=-q.
\end{eqnarray}
The point corresponding to the insertion of string fields
are given by
\begin{eqnarray}
&&  Z_1=1+\sqrt{1-q^2},\hspace{2em}
  \hat{Z}_1=1-\sqrt{1-q^2}, \\
&&  Z_2=-1-\sqrt{1-q^2},\hspace{2em}
  \hat{Z}_2=-1+\sqrt{1-q^2}.
\label{eq:inspt}
\end{eqnarray}

To discuss the singularity of the D2 and P2 vertices, 
it is convenient to use the HIKKO's OSp symmetric expression for the 
vertex. In the Appendix A, we show the following 
expression for the LPP vertex a la HIKKO:
\begin{eqnarray}
\label{eq:LPP2}
\hspace{-2.5em}
\bra{v(1^\c,2^\c;q)}=
  N(q)\,\bidx{2^\c}\bra{\tilde{1},\tilde{1}} 
  \bidx{1^\c}\bra{\tilde{1},\tilde{1}}
  \bigl(\sum_r \frac{b_0^{(r)}}{\alpha_r}\bigr)
  \exp(E_{\rH}(1^\c,2^\c))\,c(\rho_0^{(1)})c(\rho_0^{(2)}).
\end{eqnarray}
where the normalization factor $N(q)$ is given explicitly in 
Eq.~(\ref{eq:N}) and the exponent $E_{\rH}$ is the HIKKO's OSP 
invariant quadratic operator given by\cite{rf:HIKKO1,rf:HIKKO2}
\begin{eqnarray}
&& E_{\rH}=\!\!\sum_{\stackrel{\scriptstyle m,n\geq0}
{\ \ \ r,s=1,\bar{1},2,\bar{2}}}\!\!\!\!
    \Nmn{rs}{mn}\left(\half{\alpha_m^\mu}^{(r)}\eta_{\mu\nu} 
    {\alpha_n^\nu}^{(s)}+i\,{\gamma_m}^{(r)}
    {\beta_n}^{(s)}\right)
    \equiv \sum \Nmn{rs}{mn}\half a_m^{(r)}\cdot a_n^{(s)}, \nn
&& 
{\gamma_n}^{(r)}=i\,n\alpha_r {c_n}^{(r)},\qquad 
{\beta_n}^{(r)}={b_n}^{(r)}/{\alpha_r},
\label{eq:EHIKKO}
\end{eqnarray}
with the Fourier coefficients $\Nmn{rs}{mn}$ of the Neumann function 
for the $X^\mu$ operators.
Here we have used the following notation for OSp$(d,2|2)$ vector 
and invariant metric:
\begin{eqnarray}
 &&a_m^M=(\alpha_m^\mu,\,\gamma_m,\,\beta_m), \nn
 && a \cdot b = a^M \eta_{MN}\, b^N, \hspace{2em}
 \eta_{MN}=\pmatrix{
      \eta_{\mu\nu} &   &   \cr
                    & 0 & i \cr
                    & -i& 0 \cr}.
\label{eq:OSPinv}
\end{eqnarray}

{}From Eqs.~(\ref{eq:rho-z map}) and (\ref{eq:intpt}), we see that
the two interaction points $\rho_0^{(1,2)}$ on the $\rho$ plane are related 
with the parameter $q$ as
\begin{eqnarray}
\label{eq:intpt on rho}
 {\rho_0}^{(2)}=-{\rho_0}^{(1)}=\ln\left(\frac{1+q}{1-q}\right).
\end{eqnarray}
${\rho_0}^{(r)}$ is real if $q=a$ and purely imaginary if $q=ia$.
For real $q=a$ (D2 vertex), the interaction points $\rho_0^{(1,2)}$ 
are on the open string boundary as in Fig.~\ref{fig:cl-cl-D2}, so that 
the holomorphic and anti-holomorphic parts are equal to each other. Then
we find
\begin{eqnarray}
 c(\rho_0^{(1)}) &=& c^{(1)}({\rho_0}^{(1)}-T/2)
    = \half\left(c^{(1)}(0)+\bar{c}^{(1)}(0)\right) = C^{(1)}(0),\nn
 c(\rho_0^{(2)})
    &=& -c^{(2)}(-{\rho_0}^{(2)}+T/2) 
    = -\half\left(c^{(2)}(0)+\bar{c}^{(2)}(0)\right)= -C^{(2)}(0),
\label{eq:D2ghost}
\end{eqnarray}
where we have defined
\begin{eqnarray}
 C(\rho)\equiv\half\bigl(c(\rho)+\bar{c}(\rho)\bigr).
\end{eqnarray}
(Note that the argument of $\bar c$ is not $\bar\rho$ but $\rho$ here.) \ 
If $q=ia$ (imaginary), on the other hand, 
from the connection condition of Eq.~(\ref{eq:connect-cc}),
we have $c^{(1)}(i\sigma)=-\cbar^{(2)}(-i\sigma)= +\cbar^{(1)}(i\sigma)$ 
at both interaction points $i\sigma=\rho_0^{(1)}=-i\sigma_0$ 
and $i\sigma=\rho_0^{(2)}=+i\sigma_0$,
so that
\begin{eqnarray}
 c(\rho_0^{(1)})
    &=& \half\left(c^{(1)}(-i{\sigma_0})
                  +\bar{c}^{(1)}({-i\sigma_0})\right) 
    = C^{(1)}(\rho_0^{(1)}), \nn
 c(\rho_0^{(2)})
    &=& -\half\left(c^{(2)}(-i{\sigma_0})
                  +\bar{c}^{(2)}(-{i\sigma_0})\right)
    = -C^{(2)}(\rho_0^{(1)}).
\label{eq:P2ghost}
\end{eqnarray}
These ghost factors Eqs.~(\ref{eq:D2ghost}) and Eq.~(\ref{eq:P2ghost}) 
for D2 and P2 cases, respectively, look very different. 
We, however, note that, for the former D2 case, what should be evaluated 
is not the LPP vertex $\bra{v_{D_2}(1^\c,2^\c;T)}$ itself but  
that multiplied by the operator $e^{+(L^{(1^\c)}+L^{(2^\c)})T/2}$, 
as seen in Eq.~(\ref{eq:D2vertex}). So, using Eqs.~(\ref{eq:LPP2}) and 
(\ref{eq:D2ghost}), we have 
\begin{eqnarray}
&&\hspace{-2em}\bra{v(1^\c,2^\c;q=a)}e^{+(L^{(1^\c)}+L^{(2^\c)})T/2} \nn
&& = N(q)\,\bidx{2^\c}\bra{\tilde{1},\tilde{1}} 
  \bidx{1^\c}\bra{\tilde{1},\tilde{1}}
  \bigl(\sum_r \frac{b_0^{(r)}}{\alpha_r}\bigr)
  \exp(\hat E_{\rH}(1^\c,2^\c))
    \left(-\hat C^{(1)}(0)\hat C^{(2)}(0)\right),
\label{eq:4.28}
\end{eqnarray}
with hatted operators denoting
$\hat \calO = 
e^{-(L^{(1^\c)}+L^{(2^\c)})T/2} \calO e^{+(L^{(1^\c)}+L^{(2^\c)})T/2}$. 
This hat operation is equivalent to replacing all the oscillators by
\begin{equation}
(\hat \alpha_n^\mu,\ \hat c_n,\ \hat b_n) =
(e^{nT/2}\alpha_n^\mu,\ e^{nT/2}c_n,\ e^{nT/2}b_n).
\label{eq:replace}
\end{equation}
So, now the ghost factors become
\begin{equation}
-\hat C^{(1)}(0)\hat C^{(2)}(0) =
- C^{(1)}(-T/2) C^{(2)}(-T/2) =
- C^{(1)}(\rho_0^{(1)}) C^{(2)}(\rho_0^{(1)}),
\end{equation}
since $\rho_0^{(1)} = -T/2$ in this D2 case. Therefore the ghost factor 
$c(\rho_0^{(1)})c(\rho_0^{(2)})$ in Eq.~(\ref{eq:LPP2}) can now be written 
as $- C^{(1)}(\rho_0^{(1)}) C^{(2)}(\rho_0^{(1)})$ commonly for the D2 and 
P2 cases. 

The exponent $\hat E_{\rH}(1^\c,2^\c)$ for D2 case is 
given by the same expression as $E_{\rH}(1^\c,2^\c)$ in 
Eq.~(\ref{eq:EHIKKO}) with the replacement Eq.~(\ref{eq:replace}) done 
for all the oscillators, while no such is done for the P2 case. 
But we shall see shortly that this also helps to make the expressions 
common for both cases.

Since we want to know the singular structure of the LPP vertex in the 
limit $T,\ \sigma_0 \rightarrow0$, we have only to know the behavior of the
exponent $E_{\rH}$ around $q \sim0$, which is written in terms 
of the Neumann coefficients $\Nmn{rs}{mn}$. 
We can do this task by expanding the Neumann 
function directly. The Neumann function on the $z$ plane is given 
by\footnote{
This Neumann function is for the real coordinate 
$X^\mu=\half\bigl(X^\mu(\rho)+\bar X^\mu(\bar\rho)\bigr)$. 
The 2-point function for the holomorphic coordinate 
$X^\mu(z)$ is of course given by 
$\VEV{X^\mu(z)\,X^\nu(z')}=-\eta^{\mu\nu}\ln(z-z')$. 
If we note the relation Eq.~(\ref{eq:realX}), we can derive the Neumann 
function $N(z,z')$ in Eq.~(\ref{eq:Neumann}) as follows:
omitting the indices $\mu,\nu$, we have 
\begin{eqnarray}
-\VEV{\big(X(z)+ X({q^2\over\zbar})\big)
\big(X(z')+ X({q^2\over\zbar'})\big)} &=& 
\ln(z-z')+\ln\bigl(z-{q^2\over\zbar'}\bigr)
+\ln\bigl({q^2\over\zbar}-z'\bigr)
+\ln\bigl({q^2\over\zbar}-{q^2\over\zbar'}\bigr)  \nn
&=& 2N(z,z')+ \ln\bigl({q^2zz'\over\zbar\zbar'}\bigr) .
\end{eqnarray} 
The last term is irrelevant since it gives functions depending only on 
a single variable $z$ or $z'$ or $\zbar$ or $\zbar'$. 
}
\begin{eqnarray}
 N(z,z')&=&\ln\abs{z-z'}+\ln\abs{1-\frac{q^2}{z\bar{z}'}} \nn
        &=&\ln\abs{z-z'}-\ln\abs{z-\bar{z}'}
           +\ln\abs{z+\frac{q^2}{z}
                   -\bar{z}'-\frac{q^2}{\bar{z}'}}.
\label{eq:Neumann}
\end{eqnarray}
Since $\Nmn{rs}{mn}$ appear as Fourier coefficients of the Neumann 
function Eq.~(\ref{eq:Neumann}), namely the coefficients of the terms 
$w_r^n w'^{\,m}_s= e^{n(\tau_r+i\sigma_r)}e^{m(\tau'_s+i\sigma'_s)}$, 
we rewrite it by substituting 
\begin{eqnarray}
z &=& \alpha_rW_r+\sqrt{W_r^2-q^2} 
= 2\alpha_r\left(W_r-{q^2\over4W_r}\right) + \odr(q^4) \nn
&&{\rm with}\quad W_r \equiv{1-w_r\over1+w_r}, \quad \alpha_1=+1, \ \alpha 
_2=-1, 
\label{eq:4.32}
\end{eqnarray} 
which follows from the mapping of Eq.~(\ref{eq:rho-z map}) and 
the relation $\rho=\alpha_r\rho_r= \alpha_r\ln w_r$ in the string region
$r=1$ and 2. 

Actually, however, the last relation 
$\rho=\alpha_r\rho_r= \alpha_r\ln w_r$ applies only to the P2 vertex 
case, and the correct one for the D2 vertex case is 
$\rho=\alpha_r(\rho_r - T/2) = \alpha_r\ln(w_re^{-T/2})$. So the variable
$w_r$ in Eq.~(\ref{eq:4.32}) should be understood as standing for 
$w_re^{-T/2}$ for D2 case. This means that if we find the terms of the 
form $\Amn{rs}{mn}w_r^n w'^{\,m}_s$ by writing the Neumann function in 
terms of the variable $w_r$ in Eq.~(\ref{eq:4.32}), the true Neumann 
coefficients $\Nmn{rs}{mn}$ in D2 case is given by 
$\Nmn{rs}{mn}=\Amn{rs}{mn}e^{-nT/2}e^{-mT/2}$. But recall that what we 
need for the D2 case is not the exponent $E_{\rH}$ but 
$\hat E_{\rH}$, for which, since the replacement 
Eq.~(\ref{eq:replace}) is done, the coefficients of the oscillators are 
given by $\Nmn{rs}{mn}e^{nT/2}e^{mT/2}$ and just equal the coefficients 
$\Amn{rs}{mn}$. So, for both D2 and P2 cases, we can unifiedly find the 
desired Neumann coefficients simply by using Eq.~(\ref{eq:4.32}). 

Now, substituting Eq.~(\ref{eq:4.32}) into Eq.~(\ref{eq:Neumann}), 
we easily find 
\begin{eqnarray}
 N(\rho_r,\rho'_s) 
&=&  \ln\abs{w_r-w'_s} 
    +{q^2\over2}\sum_{m,n\geq1}\left[
        (w_r^m-\wbar_r^m)(w'^{\,n}_s-\wbar'^{\,n}_s)\right] \nn
&&\qquad \qquad     +(\hbox{irrelevant terms}) 
           \qquad \hbox{for} \ \ (r,s)=(1,1),\,(2,2) \nn
&=& -\sum_{n\geq1}{1\over2n}\left(w_r^n
w'^{\,n}_s+\wbar_r^n\wbar'^{\,n}_s\right) 
    -{q^2\over2}\sum_{m,n\geq1}\left[
        (w_r^m-\wbar_r^m)(w'^{\,n}_s-\wbar'^{\,n}_s)\right] \nn
&&\qquad \qquad     +(\hbox{irrelevant terms}) 
           \qquad \hbox{for} \ \ (r,s)=(1,2),\,(2,1). \nn
\end{eqnarray}
The (irrelevant terms) stands for $2\ln2-\ln\abs{(1+w_r)(1+w'_s)}$ 
which yields functions of a single variable $w_r$ or $w'_s$, not 
contributing to the exponent $E_{\rH}$ in the presence of the 
momentum conservation. 
The exponent $E_{\rH}$ can now be obtained from this 
by formally performing replacements 
$(w_r^m,\ \wbar_r^m,\ w'^{\,n}_s,\ \wbar'^{\,n}_s) \rightarrow 
(a^{(r)}_m,\ \bar a^{(r)}_m,\ a^{(s)}_n,\ \bar a^{(s)}_n)$,
discarding the free propagation term $\ln\abs{w_r-w'_s}$ as well as 
irrelevant terms:
\begin{eqnarray}
\label{eq:LPPexp}
 E_{\rH}(1^\c,2^\c)&=&E^\c_{12}
   +\frac{q^2}{2}\sum_{m\geq1}({a_m}^{(1)}-{\bar{a}_m}^{(1)})\cdot
                 \sum_{n\geq1}({a_n}^{(1)}-{\bar{a}_n}^{(1)}) \nn
 &&    +\frac{q^2}{2}\sum_{m\geq1}({a_m}^{(2)}-{\bar{a}_m}^{(2)})\cdot
                 \sum_{n\geq1}({a_n}^{(2)}-{\bar{a}_n}^{(2)}) \nn
 &&   -q^2\sum_{m\geq1}({a_m}^{(1)}-{\bar{a}_m}^{(1)})\cdot
                 \sum_{n\geq1}({a_n}^{(2)}-{\bar{a}_n}^{(2)})
      +\odr(q^4),
\end{eqnarray}
where $E^\c_{12}$ is the exponent of the closed reflector defined
in Eq.~(\ref{eq:refexp}), and we have used an OSp invariant 
notation $a_m^M=(\alpha_m^\mu,\,\gamma_m,\,\beta_m)$ and 
$a\cdot b = a^M\eta_{MN}\,b^N$ introduced in Eq.~(\ref{eq:OSPinv}).
The appearance of the reflector is natural since the 
configurations of D2 and P2 vertices both reduce to a mere 
connection of two closed strings when $T,\ \sigma_0$ go to zero.
Substitute this expanded from for $E_{\rH}(1^\c,2^\c)$ 
(which is in fact $\hat E_{\rH}(1^\c,2^\c)$ for D2 case) into 
Eq.~(\ref{eq:LPP2}) (or Eq.~(\ref{eq:4.28}) for D2 case) 
and multiply ${b_0^-}^{(1)}{b_0^-}^{(2)}$ to it. Then, noting 
\begin{eqnarray}
&&\bidx{2^\c}\bra{\tilde{1},\tilde{1}} 
  \bidx{1^\c}\bra{\tilde{1},\tilde{1}}
  \bigl(\sum_r \frac{b_0^{(r)}}{\alpha_r}\bigr)
{b_0^-}^{(1)}{b_0^-}^{(2)} \nn
&&\qquad \quad =
-\bidx{2^\c}\bra{1,1}\bidx{1^\c}\bra{1,1}
(c_0^{+(1)}+c_0^{+(2)})(c_0^{-(1)}+c_0^{-(2)}){b_0^-}^{(1)}.
\end{eqnarray}
we find 
\begin{eqnarray}
\label{eq:LPP3}
\bra{v(1^\c,2^\c;q)}{b_0^-}^{(1)}{b_0^-}^{(2)}
&=&
  N(q)\,\bra{R^\c(1^\c,2^\c)}C^{(1)}(\rho_0^{(1)}) C^{(2)}(\rho_0^{(1)}) \nn
&&\hspace{3em} \times 
  \left\{1+q^2 {\cal D}^{(1)}(0)+\odr(q^4)\right\}{b_0^-}^{(1)},
\end{eqnarray}
where we have defined
\begin{eqnarray}
\hspace{-3em} {\cal D}(\rho)= -:\Bigl\{\half \Bigl(\frac{dX}{d\rho}(\rho)
                       -\frac{d\bar{X}}{d\bar \rho}(\bar\rho)\!\Bigr)^2
 \! +\left({dc\over d\rho}(\rho)-{d\cbar\over d\bar\rho}(\bar\rho)\right)
   \bigl(b(\rho)-\bar{b}(\bar\rho)\bigr)\Bigr\}: \, .
\end{eqnarray}
Note that this ${\cal D}(0)$ is commutative with the ghost factor
$C(\rho_0^{(1)})$ since $b(0)-\bar{b}(0)$ contained in the former 
anti-commutes with $C(\rho_0^{(1)})=\half
\bigl(c(\rho_0^{(1)})+\cbar(\rho_0^{(1)})\bigr)$.
[This explains why we have rewritten the original ghost factor 
$c(\rho_0^{(1)})$ into $C(\rho_0^{(1)})$ as done in 
in Eqs.~(\ref{eq:D2ghost}) and Eq.~(\ref{eq:P2ghost}).
If we had kept using $c(\rho_0^{(1)})$, we would have encountered a
divergent expression like $b(0)c(\rho_0^{(1)}) \sim1/\rho_0^{(1)}$ 
as $\rho_0^{(1)}\rightarrow0$.]
We replace $C^{(2)}(\rho_0^{(1)})$ by $-C^{(1)}(-\rho_0^{(1)})$ since 
they are equal on the reflector $\bra{R^\c(1^\c,2^\c)}$, and 
expand the two ghost factors in powers of $q$ using 
Eq.~(\ref{eq:intpt on rho}):
\begin{eqnarray}
\label{eq:CC}
 C^{(1)}(-{\rho_0}^{(1)})C^{(1)}({\rho_0}^{(1)})
  = q\,C^{(1)}\frac{dC^{(1)}}{d\rho}(0)
    +q^3\,{\cal A}(0)+\odr(q^5),
\end{eqnarray}
where ${\cal A}$ is defined by
\begin{eqnarray}
 {\cal A}(\rho)=\frac13 C\frac{dC}{d\rho}
                 -2\frac{dC}{d\rho}\frac{d^2C}{d\rho^2}
                 +\frac23 C\frac{d^3C}{d\rho^3}.
\end{eqnarray}
Finally, substituting Eqs.~(\ref{eq:N}) and (\ref{eq:CC}) to
Eq.~(\ref{eq:LPP3}), we obtain
\begin{eqnarray}
\label{eq:LPP4}
&& \bra{v(1^\c,2^\c;q)}{b_0^-}^{(1)}{b_0^-}^{(2)}\nn
&& \qquad \quad =
\bra{R^\c(1^\c,2^\c)}\left\{
   \frac2{q^2}C^{(1)}\frac{dC^{(1)}}{d\rho}(0)
   +{\cal B}^{(1)}(0)
   +\odr(q^2) \right\}{b_0^-}^{(1)},
\end{eqnarray}
with
\begin{eqnarray}
 {\cal B}(\rho)=\half\left({\cal A}-3C\frac{dC}{d\rho}
               +C\frac{dC}{d\rho}{\cal D} \right).
\end{eqnarray}
This Eq.~(\ref{eq:LPP4}) is a result which applies both to 
D2 and P2 vertices by putting $q=a$ and $q=ia$, respectively.
In the limit $q\rightarrow0$ (i.e., $T\rightarrow0$ for D2 and $\sigma 
_0\rightarrow0$ for P2
cases), this has two contributions: the first divergent term $\propto1/q^2$ 
corresponds to the tachyon contribution and the second to the dilaton, 
as anticipated. The powers of $q$ in these are reduced by 1 since 
a power $q^1$ came from the product of two ghost factors 
$C^{(1)}(-{\rho_0}^{(1)})C^{(1)}({\rho_0}^{(1)})$ at the interaction points
which vanish at the coincident limit.

At last, we can determine the relation for the parameters of 
$\aa$, $\cc$ and $\lambda_c$ to make the action gauge invariant. 
Since the result Eq.~(\ref{eq:LPP4}) is common to D2 and P2 vertices, 
the ${\cal B}^{(1)}(0)$ term can be canceled between 
the first two terms in $\delta S_2$ in Eq.~(\ref{eq:dS-bk}),
if we take $n\aa^2=\cc$:
\begin{eqnarray}
&&\hspace{-2em} \aa^2\bra{U(1,1^\c)}\bra{U(2,2^\c)}\ket{R^\o(1,2)}
 +\cc\bra{V_\rc(1^\c,2^\c)}(\tQB^{(1)}+\tQB^{(2)}) \nn
&&  = \lim_{a\rightarrow0} 
    \frac{4n\aa^2}{a^2}
    \bra{R^\c(1^\c,2^\c)}
    C^{(1)}\frac{dC^{(1)}}{d\rho}(0)\,
    {b_0^-}^{(1)}\prod_{r=1^\c,2^\c}\PP \nn
&& \ \ {}+\cc \lambda_\c \int^{\frac{\pi}{2}}_0 d\sigma_0
     \bra{v_\rc(1^\c,2^\c;\sigma_0)}b_{\sigma_0}
      ({c_0^+}^{(1^\c)}\!{+}{c_0^+}^{(2^\c)})
      \!\!\!\prod_{r=1^\c,2^\c}\!\!\bPP{r}\!,
\label{eq:g2final}
\end{eqnarray}
where we have used $\bra{V_\rc(2^\c,1^\c)}=-\bra{V_\rc(1^\c,2^\c)}$, 
and note the minus sign in the first term in Eq.~(\ref{eq:P2vertex}).
The first singular term, due to the existence of the tachyon 
in bosonic strings, did not cancel but added up between the two since 
$q^2=a^2$ for D2 and $q^2=-a^2$ for P2.
However, since it can be rewritten into the form of a BRS transform as
\begin{eqnarray}
 && \bra{R^\c(1^\c,2^\c)}
    C^{(1)}\frac{dC^{(1)}}{d\rho}(0)\,
    {b_0^-}^{(1)}{\cal P}^{(1)}{\cal P}^{(2)}\nn
&&\qquad \quad = \bra{R^\c(1^\c,2^\c)}
    \{\QB^{(1^\c)},\,\half{c_0^+}^{(1^\c)}\}\,
    {b_0^-}^{(1)}{\cal P}_1{\cal P}_2,
\end{eqnarray}
it turns out that
it can be canceled by the third term in $\delta S_2$ 
in Eq.~(\ref{eq:dS-bk}) if we set 
\begin{equation}
\lambda_\c= -\lim_{a\rightarrow0}{2n\aa^2\over a^2}.
\end{equation}
This corresponds to an infinite renormalization of the zero intercept
parameter as mentioned before, and 
explains the reason why we have had to define the `bare' BRS charge 
$\tQB$ by shifting the zero intercept parameter. 

Finally, let us show that the remaining second term 
in Eq.~(\ref{eq:g2final}) also vanishes. First we note that
\begin{eqnarray}
\bra{v_\rc(1^\c,2^\c;\sigma_0)}b_{\sigma_0}
&=& \bra{v_\rc(1^\c,2^\c;\sigma_0)} 
\left(-i\oint_{C_1}+i\oint_{C_2}\right)
\frac{d\rho}{2\pi i}b(\rho) \nn
&=& \bra{v_\rc(1^\c,2^\c;\sigma_0)} 
\left(-i\oint_{\Czzero{1}}+i\oint_{\Czzero{2}}\right)
\frac{dz}{2\pi i}\left({dz\over d\rho}\right)b(\rho) \nn
&=& A(q)\bra{v_\rc(1^\c,2^\c;\sigma_0)} 
\bigl(b(z_0^{(1)})+b(z_0^{(2)})\bigr),
\label{eq:deformA}
\end{eqnarray}
with $A(q)=iq(1-q^2)/2$, where we have used Eq.~(\ref{eq:bsigma})
and $\rho_0^{(1)}=-\rho_0^{(2)}=-i\sigma_0$ for the definition of $b_{\sigma 
_0}$. The contour $C_1-C_2$ on the $\rho$ plane is deformed as drawn
\begin{wrapfigure}[14]{r}{\halftext}
    \epsfxsize=5.95cm
    \centerline{\epsfbox{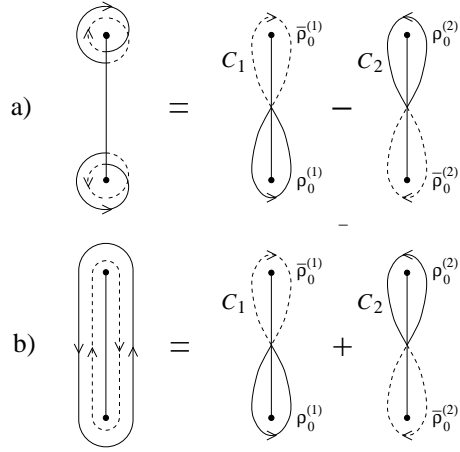}}
    \caption{Contour deformations a) and b) used in 
Eqs.~(\protect\ref{eq:deformA}) and (\protect\ref{eq:deformB}), 
respectively.}
    \label{fig:C1C2}
\end{wrapfigure}
in Fig.~\ref{fig:C1C2} a), and then mapped to $\Czzero{1}-\Czzero{2}$
on the $z$ plane where $\Czzero{i}$ are closed contours encircling
the interaction points $z_0^{(i)}$ in Fig.~\ref{fig:z-P2}. 
Eqs.~(\ref{eq:dzdr}), (\ref{eq:intpt}) and (\ref{eq:inspt}) have been
used for the evaluation of the contour integration on the $z$ plane. 
Next, the ghost zero-mode factor 
$({c_0^+}^{(1^\c)}+{c_0^+}^{(2^\c)})=\half
({c_0}^{(1^\c)}+{\cbar_0}^{(1^\c)}+
{c_0}^{(2^\c)}+{\cbar_0}^{(2^\c)})$ 
in the second term, which is first written as an integral over $\sigma$, 
can also be rewritten as an integral over $z$ again going to the $z$ plane. 
The integral contour can be deformed to the closed paths 
$\Czzero{1}+\Czzero{2}$ around the interaction points, and we find:
\begin{eqnarray}
&&\bra{v_\rc(1^\c,2^\c;\sigma_0)}b_{\sigma_0}
      ({c_0^+}^{(1^\c)}+{c_0^+}^{(2^\c)}) \nn
 &&\ = \half A(q)\bra{v_\rc(1^\c,2^\c;\sigma_0)}
     \Bigl(\oint_{\Czzero1}\!\!\!+\oint_{\Czzero2}\Bigr)
     \frac{dz}{2\pi i}\left(\frac{d\rho}{dz}\right)^2
     \!\bigl(b(z_0^{(1)})\!+b(z_0^{(2)})\bigr)\,c(z).\hspace{3em} 
\end{eqnarray}
But, although $\langle{b(z_0^{(i)})c(z)}\rangle = 1/(z_0^{(i)}-z)$, 
since the factor $(d\rho/dz)^2$ contains double zero $(z_0^{(i)}-z)^2$, 
the integrand is regular around the interaction points and 
the contour integral vanishes. Thus we have finished the proof of the 
O($g^2$) gauge invariance $\delta S_2=0$ and found the first two 
relations in Eq.~(\ref{eq:relations}) for the parameters.

\subsection{Order $g^3$ and $g^4$ invariance}

\begin{figure}[tb]
 \parbox{\halftext}{
    \vspace{4.3ex}
    \epsfxsize=60mm
    \centerline{\epsfbox{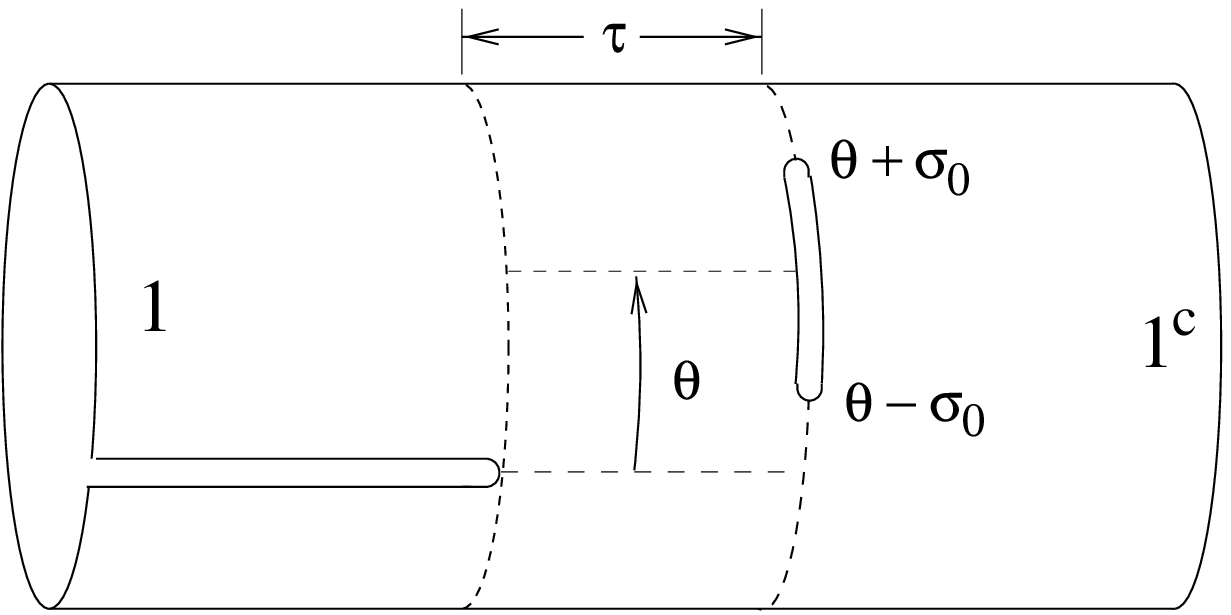}}
    \vspace{3.2ex}
    \caption{The string diagram corresponding to the vertex 
$\bra{v_1(\sigma_0,\theta,\tau)}$.}
    \label{fig:UVc}}
 \hspace{5mm}
 \parbox{\halftext}{
    \epsfxsize= 60mm \hspace{1mm}
    \centerline{\epsfbox{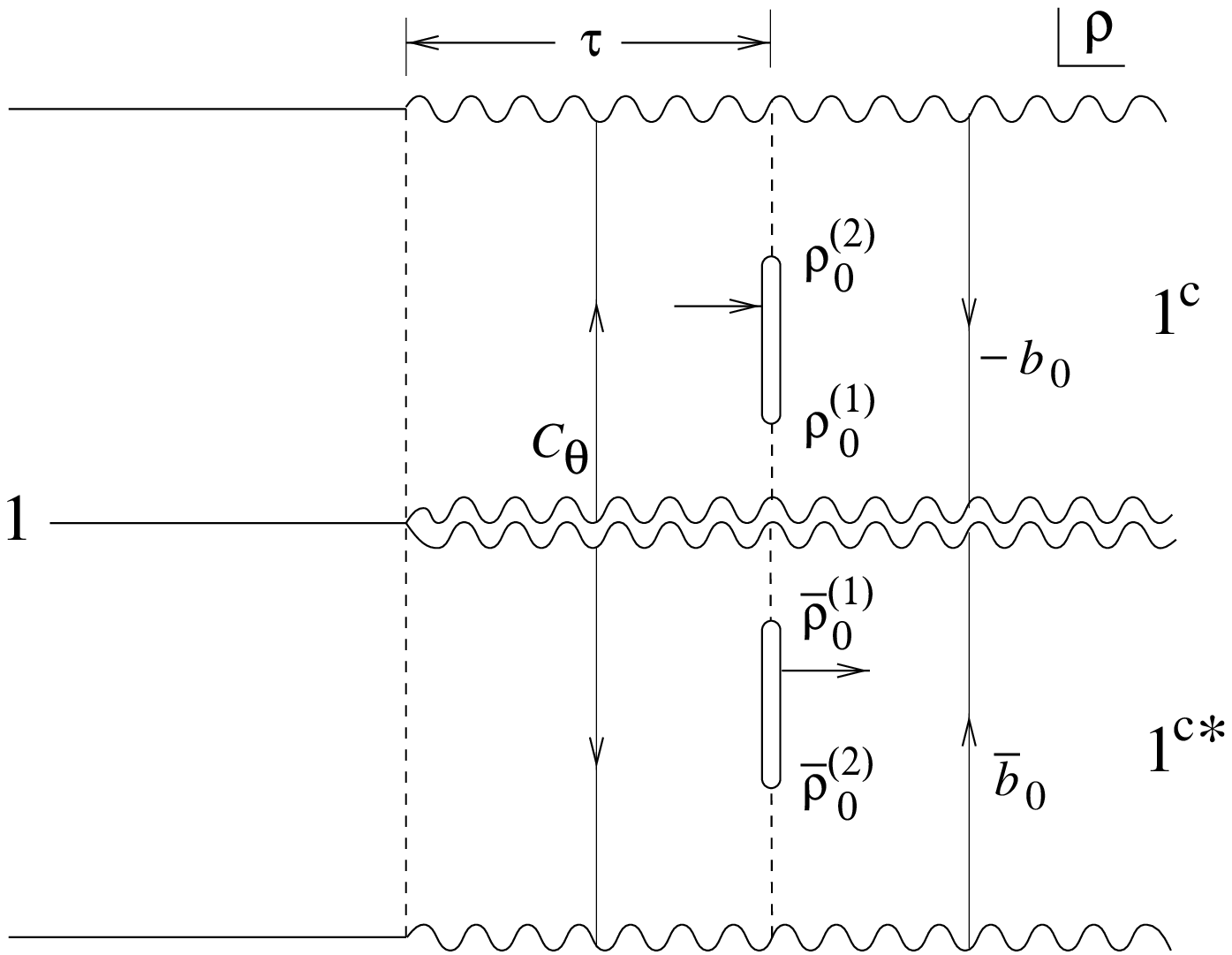}}
    \caption{The $\rho$ plane of the vertex 
    $\bra{v_1(\sigma_0,\theta,\tau)}$.}
    \label{fig:UVc-rho}}
\end{figure}

Now we turn to the consideration of the 
order $g^3$ terms $\delta S_3$ in Eq.~(\ref{eq:dS-bk}).
Again, inserting a propagator of the intermediate state, 
we rewrite the first term of the two $O(g^3)$ terms into the form
\begin{eqnarray}
\label{eq:v1}
&& \hspace*{-3em}
\bra{\check{U}(1,2^\c)}\bra{V_\rc(3^\c,1^\c)}\ket{R^\c(2^\c,3^\c)} \nn
&=& \lim_{\tau\rightarrow0}
\int_0^{\frac{\pi}{2}}\,d\sigma_0
\int_0^{2\pi}\frac{d\theta}{2\pi}
\bra{u(1,2^\c)}\bra{v_\rc(3^\c,1^\c;\sigma_0)}b_{\sigma_0}{b_0^-}^{(3^\c)} 
\nn
&& \hspace{8em}\times e^{-\tau(L_0^{(3^\c)}+\overline{L}_0^{(3^\c)})}
e^{i\theta(L_0^{(3^\c)}-\overline{L}_0^{(3^\c)})}
\ket{R^\c(2^\c,3^\c)}\left(b_0^-{\cal P}\Pi\right)^{(1^\c)} \nn
&=& \lim_{\tau\rightarrow0}
\int_0^{\frac{\pi}{2}}\,d\sigma_0
\int_0^{2\pi}\frac{d\theta}{2\pi}
\bra{v_1(1,1^\c;\sigma_0,\theta,\tau)}
b_{\sigma_0}b_{C_\theta}\left(b_0^-{\cal P}\Pi\right)^{(1^\c)},
\end{eqnarray}
where $\tau$ is the propagation time interval, 
the explicit expression for 
$\calP=\int_0^{2\pi}(d\theta/2\pi) \newline \exp(i\theta(L_0-\bar L_0))$ 
was inserted for $\calP^{(3)}$, 
and $\bra{v_1(\sigma_0,\theta,\tau)}$ denotes the effective LPP vertex 
following from the GGR theorem, which corresponds to the string diagram 
depicted in Fig.~\ref{fig:UVc}. (We introduced $\tau$ for clarity for 
drawing the figure, although there occurs actually no singularity here 
even at $\tau=0$.)\ \ $b_{C_\theta} \equiv(1/2)\oint_{C_\theta} 
d\rho/(2\pi i)\, b(\rho)$ is a rewriting of 
${b_0^-}^{(3^\c)}=(1/2)(b_0-\bbar_0)^{(3^\c)}$ and the contour 
$C_\theta$ is the path traversing the intermediate closed string as 
drawn in Fig.~\ref{fig:UVc}.

Since ${b_0^-}^{(1^\c)}$ factor for the external closed string $1^\c$ 
is also multiplied in Eq.~(\ref{eq:v1}), 
we can replace $b_{C_\theta}$ by $b_{C_\theta}- {b_0^-}^{(1^\c)}$, and
the latter can be seen to reduce to the following expression by making
a deformation of the integration contour as drawn in
Fig.~\ref{fig:C1C2} b): 
\begin{eqnarray}
\label{eq:b_C}
 b_{C_\theta}- {b_0^-}^{(1^\c)}= -\half\left(
           \oint_{C_1}+\oint_{C_2}
           \right)\frac{d\rho}{2\pi i}b(\rho)
    = -\half(b_{\rho_0^{(1)}}+b_{\rho_0^{(2)}}),
\label{eq:deformB}
\end{eqnarray}
where $C_i$ ($i=1,2$) are the closed contours given before in
Fig.~\ref{fig:contour}. This expression can also be understood from
the meaning of $b_{C_\theta}$: it is originally the anti-ghost factor
$b_0^{-(3)}$ corresponding to the quasi-conformal deformation for the
change of $\theta$, whereas the $\theta$ change is now equivalent to
shifting the interaction points $\rho_0^{(1)}=
\tau_0+i(\theta-\sigma_0)$ and $\rho_0^{(2)}=
\tau_0+i(\theta+\sigma_0)$ simultaneously by a common amount, and
$b_{\rho_0^{(i)}}$ are clearly the anti-ghost factors corresponding to
the shift of $\rho_0^{(i)}$. 

On the other hand, the other anti-ghost factor $b_{\sigma_0}$ is 
\begin{eqnarray}
\label{eq:b_rho}
 b_{\sigma_0}
         = -i(b_{\rho_0^{(1)}}-b_{\rho_0^{(2)}})
\end{eqnarray}
by the definition Eq.~(\ref{eq:bsigma}). 
Substituting this and Eq.~(\ref{eq:b_C}) into
Eq.~(\ref{eq:v1}), we find
\begin{eqnarray}
&& \bra{\check{U}(1,2^\c)}\bra{V_\rc(3^\c,1^\c)}\ket{R^\c(2^\c,3^\c)} \nn
&& \qquad {}= i \lim_{\tau\rightarrow0}
\int_0^{\frac{\pi}{2}}\,d\sigma_0
\int_0^{2\pi}\frac{d\theta}{2\pi}
\bra{v_1(1,1^\c;\sigma_0,\theta,\tau)}
b_{\rho_0^{(1)}}b_{\rho_0^{(2)}}(b_0^-{\cal P}\Pi)^{1^\c}.
\end{eqnarray}

\begin{figure}[tb]
 \parbox{\halftext}{
\vspace{4ex}
    \epsfxsize=60mm
    \centerline{\epsfbox{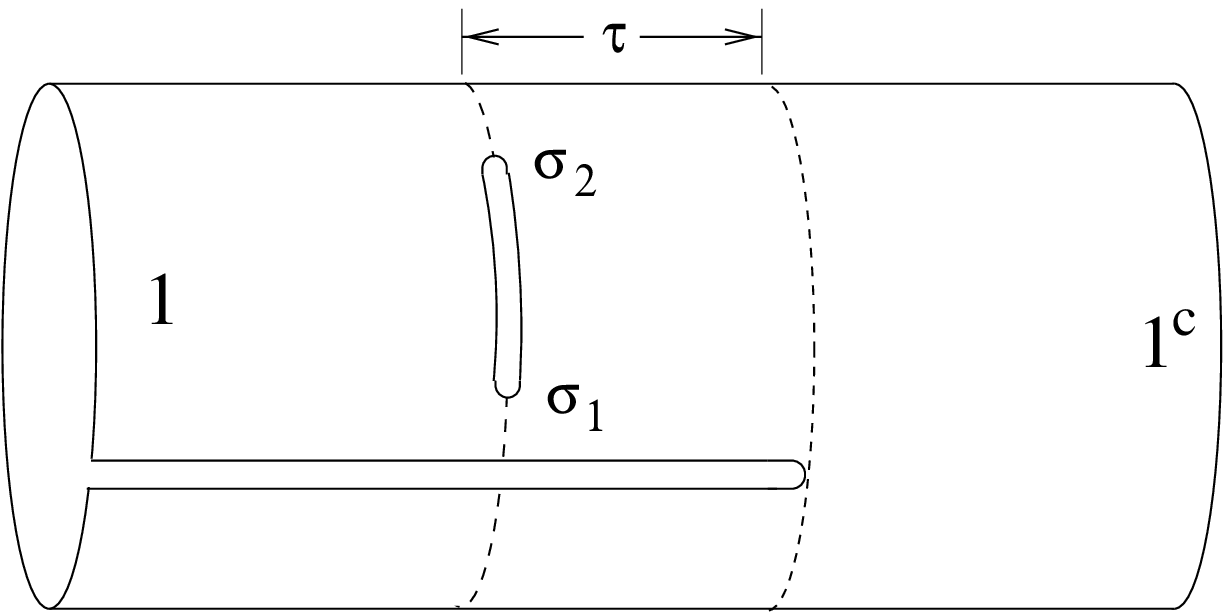}}
\vspace{2.5ex}
    \caption{The string diagram corresponding to the vertex 
$\bra{v_2(\sigma_1,\sigma_2,\tau)}$.}
    \label{fig:VoU}}
 \hspace{4mm}
 \parbox{\halftext}{
    \epsfxsize= 60mm 
\hspace{1mm}
    \centerline{\epsfbox{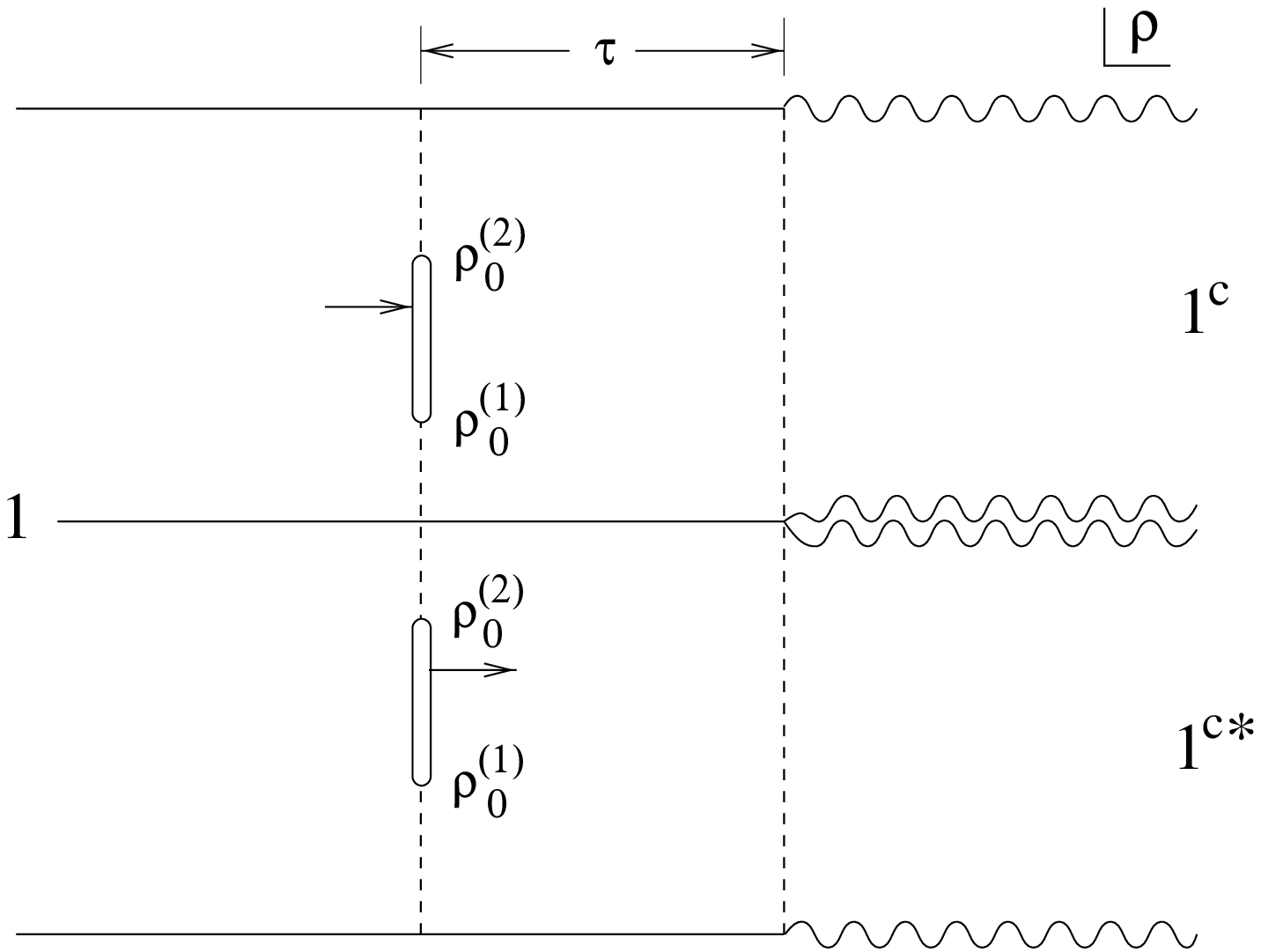}}
    \caption{\hspace{-1mm}The $\rho$ plane of the vertex 
    $\bra{v_2(\sigma_1,\sigma_2,\tau)}$.}
    \label{fig:VoU-rho}}
\end{figure}

The second one of the two $O(g^3)$ terms $\delta S_3$ 
in Eq.~(\ref{eq:dS-bk}) is rewritten as 
\begin{eqnarray}
&&\hspace{-2em}\bra{V_\ro(1,2)}\bra{U(3,1^\c)}\ket{R^\o(2,3)} \nn
&=& -\lim_{\tau\rightarrow0}
    \int_{0\leq\sigma_1 \leq\sigma_2 \leq\pi}\hspace{-4em}
    d\sigma_1 d\sigma_2 \bra{v_\ro(1,2;\sigma_1,\sigma_2)}
    \bra{u(3,1^\c)}b_{\sigma_1}b_{\sigma_2}e^{-\tau L_0^{(3)}}\ket{R^\o(2,3)}
    \left(b_0^-{\cal P}\Pi\right)^{(1^\c)} \nn
&=& +4\lim_{\tau\rightarrow0}
        \int_{0\leq\sigma_1 \leq\sigma_2 \leq\pi}
    d\sigma_1 d\sigma_2 \bra{v_2(1,1^\c;\sigma_1,\sigma_2,\tau)}
        b_{\rho_0^{(1)}}b_{\rho_0^{(2)}}\left(b_0^-{\cal P}\Pi\right)^{(1^\c)},
\end{eqnarray}
where the minus in front comes from the order exchange of 
$b_0^{-(1^\c)}$ and $\ket{R^\o(2,3)}$, and the factor $-4$ 
from $b_{\sigma_i}=(d\rho_0^{(i)}/d\sigma_i)b_{\rho_0^{(i)}}
=2ib_{\rho_0^{(i)}}$ since $\rho_0^{(i)}= \alpha_1(\tau_0+i\sigma_i)$ and $\alpha_1=2$ 
(Note that the interaction points are now on the open string side).
$\bra{v_2(\sigma_1,\sigma_2,\tau)}$ represents the effective LPP vertex 
corresponding to the string diagram depicted in Fig.\ref{fig:VoU},

Since the two string diagrams Figs.~\ref{fig:UVc} and \ref{fig:VoU} 
clearly reduce to the same configuration as $\tau\rightarrow0$ and there are 
no singularities at $\tau=0$, 
the corresponding effective LPP vertices must also 
become the same at $\tau=0$:
\begin{eqnarray}
 \bra{v_1(1,1^\c;\sigma_0,\theta,\tau=0)}
 =\bra{v_2(1,1^\c;\sigma_1,\sigma_2,\tau=0)}.
\end{eqnarray}
Therefore, if 
\begin{eqnarray}
 -i\cc\int_0^{\frac{\pi}{2}}d\sigma_0 \int_0^{2\pi}
 \frac{d\theta}{2\pi}
 +4\bb\int_{0\leq\sigma_1 \leq\sigma_2 \leq\pi} d\sigma_1d\sigma_2 = 0,
\end{eqnarray}
holds for a common integrand,
the two $O(g^3)$ terms in $\delta S_3$ in Eq.~(\ref{eq:dS-bk}) 
cancel each other. Noting the relations 
$\alpha_1\sigma_1=\theta-\sigma_0, \ \alpha_1\sigma_2=\theta+\sigma_0$ 
with $\alpha_1=2$ and so $d\sigma_0d\theta = 2d\sigma_1d\sigma_2$, and 
confirming that the both terms cover the same integration region once, 
we obtain
\begin{eqnarray}
 \cc = -4\pi i\bb,
\end{eqnarray}
as a sufficient condition for the $O(g^3)$ gauge invariance.

\begin{figure}[tb]
    \epsfxsize=65mm
    \centerline{\epsfbox{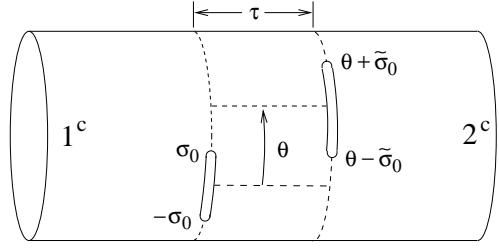}}
    \caption{The string diagram corresponding to the vertex 
$\bra{v_3(\sigma_0,\tilde{\sigma_0},\theta,\tau)}$.}
    \label{fig:VcVc}
\end{figure}

Finally consider the $O(g^4)$ term $\delta S_4$ in Eq.~(\ref{eq:dS-bk}). 
It is is rewritten as
\begin{eqnarray}
\label{eq:VcVc}
&&\hspace{-2em} \bra{\check{V}_\rc(1^\c,3^\c)}
          \bra{V_\rc(4^\c,2^\c)}\ket{R^\c(3^\c,4^\c)} 
          \nn
&=& 
\lim_{\tau\rightarrow0}\int_0^{\frac{\pi}{2}}d\sigma_0
\int_0^{\frac{\pi}{2}}d\tilde{\sigma}_0\,
\bra{v_\rc(1^\c,3^\c;\sigma_0)}\bra{v_\rc(4^\c,2^\c;\tilde{\sigma}_0)}
b_{\sigma_0}b_{\tilde{\sigma}_0}{b_0^-}^{(4)}\nn
&& \hspace{8em}\times 
e^{-\tau(L_0^{(4^\c)}+\overline{L}_0^{(4^\c)})}
e^{i\theta(L_0^{(4^\c)}-\overline{L}_0^{(4^\c)})}
\ket{R^\c(3^\c,4^\c)}
\!\!\prod_{r=1^\c,2^\c}\!\!\left(b_0^-{\cal P}\Pi\right)^{(r)}\nn
&=& \lim_{\tau\rightarrow0}\int_0^{\frac{\pi}{2}}\!\!d\sigma_0
\!\int_0^{\frac{\pi}{2}}\!\!d\tilde{\sigma}_0\!
\int^{2\pi}_0\!\!\frac{d\theta}{2\pi}
\bra{v_3(1^\c,2^\c;\sigma_0,\tilde{\sigma_0},\theta,\tau)}
b_{C_\theta}b_{\sigma_0}b_{\tilde{\sigma}_0}
\!\!\!\!\prod_{r=1^\c,2^\c}\!\!\!\bPP{r}\!,
\end{eqnarray}
where $\bra{v_3(\sigma_0,\tilde{\sigma_0},\theta,\tau)}$ 
is the effective LPP vertex 
corresponding to the string diagram depicted in Fig.~\ref{fig:VcVc}.
Again, there appears no singularity even at $\tau=0$.

Consider string configurations corresponding to the 
LPP vertex $\bra{v_3(\sigma_0,\tilde{\sigma_0},\theta,\tau)}$.
We note that, for each possible configuration at $\tau=0$, 
there are always a pair of configurations with $\tau\not=0$ 
which reduce to the same configuration at $\tau=0$; they are clearly the 
two configurations with $(\sigma_0{=}x,\,\tilde{\sigma}_0{=}y,\,\theta{=}\phi)$ 
$(\sigma_0{=}y,\,\tilde{\sigma}_0{=}x,\,\theta{=}{-}\phi)$.
Though the pair reduce to the same configuration, the anti-ghost factors
$b_{\sigma_0}b_{\tilde{\sigma}_0}$ appear in the reverse order. Since 
anti-ghost factors anti-commute, the pair cancel each other and 
Eq.~(\ref{eq:VcVc}) becomes zero. 
We thus have proved that the order $g^4$ gauge invariance is realized.

\section{String 2-point Amplitudes}

In this section, we calculate closed string 2-point amplitudes 
corresponding to a disk and a projective plane using the Feynman rule 
which follows from the gauge invariant action determined in the previous 
section. 

We must fix the gauge first. 
From the gauge invariance under the transformation of Eq.~(\ref{eq:gauge}),
it is possible to impose the usual gauge fixing condition for
the closed string field;
\begin{eqnarray}
 b_0^+\ket{\Phi}=0.
\label{eq:GFclosed}
\end{eqnarray}
The present open-closed mixed system is naturally expected to possess 
also the gauge symmetry with an open string field parameter $\Lambda^\o$, 
although we have not discussed it. 
(Actually, it is violated at the tree level but is realized at the loop 
level, as we shall see in later publication.\cite{rf:KT})
So here we simply assume that this is the case and we 
impose the same type of gauge condition also for the 
open string field:
\begin{eqnarray}
 b_0\ket{\Psi}=0.
\label{eq:GFopen}
\end{eqnarray}
Once the gauge is fixed, we can derive the gauge fixed 
action and the Feynman rules following 
by now the standard procedure: 
the gauge fixed action takes the same form as the above 
gauge invariant one (\ref{eq:braketAction}), 
but here $\ket\Psi$ and $\ket\Phi$ are now subject to the gauge conditions 
(\ref{eq:GFopen}) and (\ref{eq:GFclosed}) and contain only the 
$\ket\phi$ component defined in Eq.~(\ref{eq:component}). (However, the 
$\ket\phi$ here in the gauge fixed action contains {\it much more} 
component fields than the original $\ket\phi$ (or even the whole fields 
$\ket\Psi$ and $\ket\Phi$) of the gauge invariant action, since now in 
the gauge fixed action the restriction that the component fields carry 
vanishing ghost number is no longer imposed. The infinite tower of 
component fields carrying non-zero ghost number here correspond to the 
FP ghost fields, ghosts' ghost fields, and so on.) \ The Feynman 
propagators are given by
\begin{eqnarray}
&&\wick{2}{<1{\ket{\Psi}}_1>1{\ket{\Psi}}_2} 
= \left({b_0\over L_0}\right)^{(1)}\ket{R^\o(1,2)}, \nn
&&\wick{2}{{b_0^{-(1)}}<1{\ket\Phi}_1{b_0^{-(2)}}>1{\ket\Phi}_2} 
= \left({b_0\bbar_0\over L_0+\bar L_0}\right)^{(1)}\ket{R^\c(1,2)}.
\end{eqnarray}
Note that these propagators indicate that only the $\ket\phi$ components 
are propagating. Therefore, in the Feynman diagrams, we can use the same 
form of vertices as appearing in the gauge invariant action 
Eq.~(\ref{eq:braketAction}). [Precisely speaking, we should put 
the projections $\Pi^{(1)}$ and $(\calP \Pi)^{(1)}$ for the open and 
closed propagators, respectively, since our fields are subject to 
the constraints $\Pi^{(1)}\ket\Psi=\ket\Psi$ and 
$(\calP \Pi)^{(1)}\ket\Phi=\ket\Phi$. 
But, since all the vertices are constructed to contain those projections,
we can omit them from the propagators.]

Now, we consider the disk amplitude for the closed tachyon 2-point 
function which comes from using the open-closed transition vertex twice. 
Noting that normalized closed tachyon state is given by 
\begin{eqnarray}
&&\ket{\Phi(p)} = \sqrt{2} c_0^-\ket{\varphi(p)}, \nn
&&\ket{\varphi (p)}=\left[c(w)\cbar(\wbar)
\exp{\bigl(ip\cdot \half(X(w)+\bar X(\wbar))\bigr)}
\right]_{\scriptstyle w=0 \atop \scriptstyle \wbar=0}\ket0,
\end{eqnarray}
and satisfies $\calP\Pi\ket{\varphi(p)}=\ket{\varphi(p)}$, 
we find the disk amplitude as follows:
\begin{eqnarray}
&&\hspace*{-2em} (2\pi)^d\delta^d(p_1+p_2){\cal A}_{D_2}\nn
&=& 
(\aa g)^2 
\wick{2}{
\bra{U(4,2^\c)}{\ket{\Phi(p_2)}}_{2^\c}
<1 {\ket{\Psi}}_4 \bra{U(3,1^\c)}
{\ket{\Phi(p_1)}}_{1^\c}
>1 {\ket{\Psi}}_3 
} \nn
&=&
2\aa^2\,g^2\bra{u(4,2^\c)}\bra{u(3,1^\c)}
 \frac{{b_0}^{(4)}}{L_0^{(4)}}\ket{R^\o(3,4)}
 \ket{\varphi(p_1)}_{1^\c}\ket{\varphi(p_2)}_{2^\c} \nn
 &=& 2\aa^2\,g^2\int_0^\infty d\tau 
\bra{u(3,1^\c)} \bra{u(4,2^\c)} \nn
 && \qquad \times \oint_C\frac{d\rho}{2\pi i}\left({d\rho\over d\rho_4}\right)b(\rho)
 e^{-L_0^{(4)}\tau}\ket{R^\o(3,4)}
 \ket{\varphi(p_1)}_{1^\c}\ket{\varphi(p_2)}_{2^\c} \nn
 &=& -4n\aa^2\,g^2\int_0^\infty d\tau 
 \bra{v_{D_2}(1^\c,2^\c;\,T{=}2\tau)}
 \oint_C\frac{d\rho}{2\pi i}b(\rho)
 \ket{\varphi(p_1)}_{1^\c}\ket{\varphi(p_2)}_{2^\c}\ ,
\end{eqnarray}
where $\bra{v_{D_2}(1^\c,2^\c;\,T{=}2\tau)}$ is the 
LPP vertex introduced in Eq.~(\ref{eq:D2vertex}) corresponding to 
Fig.~\ref{fig:cl-cl-D2} and the contour $C$ is the path 
traversing the intermediate open string strip (from bottom to top).
Note that $d\rho/d\rho_4=\alpha_4=2$ was used since string 4 is an open string.
The front minus sign in the last 
expression comes from the exchange of the order of $b(\rho)$ and 
$\ket{R^\o(3,4)}$. 
By the definition of the LPP vertex and the mapping (\ref{eq:rho-z map}), 
we can evaluate the amplitude as follows by the conformal field theory 
on the $z$ plane in Fig.~\ref{fig:z-D2}:
\begin{eqnarray}
 &=& -4n\,\aa^2\,g^2\, \delta^d(p_1+p_2)\int_0^\infty d\tau 
 \oint_{C'}\frac{dz}{2\pi i}\left(\frac{d\rho}{dz}\right)^{-1}
 \left<b(z)c(Z_1)c(\hat{Z}_1)c(Z_2)c(\hat{Z}_2)\right> \nn
 && \qquad \qquad \times 
 \left<e^{ip_1\cdot X(Z_1)/2}e^{ip_1\cdot X(\hat{Z}_1)/2}
             e^{ip_2\cdot X(Z_2)/2}e^{ip_2\cdot X(\hat{Z}_2)/2}\right>,
\end{eqnarray}
where $d=26$ and the contour $C'$ is the image of $C$ on the $z$ plane 
which is just the line from $i\infty$ to $-i\infty$ along the imaginary 
axis of the $z$ plane in Fig.~\ref{fig:z-D2}. The correlation function 
for the ghost part is calculated in Appendix A, and the matter part is 
easily evaluated by using 
$\VEV{X^\mu(z)\,X^\nu(\tilde z)}=-\eta^{\mu\nu}\ln(z-\tilde z) $ as 
\begin{eqnarray}
&& \left<e^{ip_1\cdot X(Z_1)/2}e^{ip_1\cdot X(\hat{Z}_1)/2}
             e^{ip_2\cdot X(Z_2)/2}e^{ip_2\cdot X(\hat{Z}_2)/2}\right> \nn
&&\qquad \qquad = \prod_{\stackrel{\scriptstyle r\neq s}
{r,s=1,2,\hat{1},\hat{2}}}
    (Z_r-Z_s)^{p_r\cdot p_s/4} 
= \frac{(1-q^2)^2}{2^4q^4},
\end{eqnarray}
where we have used the momentum conservation and the on-shell condition
$p_1^2=p_2^2=8$ together with Eq.~(\ref{eq:inspt}).
Thus, using Eqs.~(\ref{eq:ghost5}) and (\ref{eq:dzdr}) in the Appendix A, 
we find the disk amplitude as follows,
\begin{eqnarray}
\label{eq:AD2}
 {\cal A}_{D_2}
&=& 
-4n\,\aa^2\,g^2\int_0^\infty d\tau \oint_{C'}\frac{dz}{2\pi i}
 {\prod_{r=1,\hat1,2,\hat2}(z-Z_r)\over4(z-z_0^{(1)})(z-z_0^{(2)})}
 {2^6q^2(1-q^2)\over\prod_{r=1,\hat1,2,\hat2}(z-Z_r)}
 \frac{(1-q^2)^2}{2^4q^4} \nn
&=& 
-4n\,\aa^2\,g^2\int_0^\infty d\tau \frac{(1-q^2)^3}{2q^3} 
= -4n\,\aa^2\,g^2
  \int_0^1 dq\frac{1}{q^3}(1-q^2)^2,
\end{eqnarray}
where we have evaluated the $\oint_{C'} dz/2\pi i$ integration simply by 
evaluating the residue at $z_0^{(1)}(=q)$, and then used 
$d\tau= d(T/2) = (2/(1-q^2))dq$ which follows from Eq.~(\ref{eq:interval}).

Similarly, we can calculate the projective plane amplitude for the 
tachyon 2-point function, which comes from the closed intersection 
vertex as follows:
\begin{eqnarray}
\label{eq:AP2}
&&\hspace*{-3em} (2\pi)^d\delta^d(p_1+p_2){\cal A}_{P_2} \nn
&=& -\cc\,g^2 \bra{V_\rc(1^\c,2^\c)}
\sqrt{2}{c_0^-}^{(1^\c)}\ket{\varphi(p_1)}_{1^\c}
\sqrt{2}{c_0^-}^{(2^\c)}\ket{\varphi(p_2)}_{2^\c} \nn
&=& -\cc\,g^2\,\int_0^{\pi/2}d\sigma_0
  \bra{v_\rc(1^\c,2^\c;\sigma_0)}b_{\sigma_0}\sqrt{2}\ket{\varphi(p_1)}_{1^\c}
\sqrt{2}\ket{\varphi(p_2)}_{2^\c} \nn
  &=& -2\cc\,g^2\,\delta^d(p_1+p_2)\int_0^{\pi/2} d\sigma_0
    \left(-i\oint_{\Czzero1}+i\oint_{\Czzero2}\right)\frac{dz}{2\pi i}
      \left(\frac{d\rho}{dz}\right)^{-1} \nn
 && \qquad \times 
 \left<b(z)c(Z_1)c(\hat{Z}_1)c(Z_2)c(\hat{Z}_2)\right> \nn 
 && \qquad \times
 \left<e^{ip_1\cdot X(Z_1)/2}e^{ip_1\cdot X(\hat{Z}_1)/2}
       e^{ip_2\cdot X(Z_2)/2}e^{ip_2\cdot X(\hat{Z}_2)/2}\right> 
\end{eqnarray}
so that
\begin{eqnarray}
 {\cal A}_{P_2} &=& -2\cc\,g^2\int_0^{\pi/2} d\sigma_0
    \left(-i\oint_{\Czzero1}+i\oint_{\Czzero2}\right)\frac{dz}{2\pi i}
 {1\over(z-z_0^{(1)})(z-z_0^{(2)})} {(1-q^2)^3\over q^2} \nn
  &=& -2\cc\,g^2\int_0^{\pi/2} d\sigma_0
 \left(-i{1\over2q}+i{1\over-2q}\right) {(1-q^2)^3\over q^2} \nn
 &=& +2\cc\,g^2\int_0^{\pi/2} id\sigma_0\frac{1}{q^3}(1-q^2)^3
 = +2\cc\,g^2\int_0^{+i} dq \frac{2}{q^3}(1-q^2)^2\nn
 &=& -4\cc\,g^2\int_0^1 dq\frac{1}{q^3}(1+q^2)^2.
\end{eqnarray}
where use has been made of 
$i\sigma_0= \ln[(1+q)/(1-q)]$ and $d(i\sigma_0)= (2/(1-q^2))dq$, 
and, at the final step, we have changed the variable, $q \rightarrow iq$ so 
that the new $q$ takes real values in $[0,1]$ as in the preceding case.

The amplitudes of Eqs.~(\ref{eq:AD2}) and (\ref{eq:AP2})
are singular at the point $q=0$, and so we introduce
a cut-off parameter $a$:
\begin{eqnarray}
 {\cal A}_{D_2 \atop P_2} 
&=& -4g^2\left({n\aa^2 \atop \cc}\right)
   \lim_{a\rightarrow0}
   \int_a^1dq\frac{1}{q^3}(1\mp q^2)^2 \nn
&=& -4g^2\left({n\aa^2 \atop \cc}\right)
       \lim_{a\rightarrow0} \left(\frac{1}{2a^2}\pm2\ln a\right).
\label{eq:cutoffa}
\end{eqnarray}
The quadratic divergences correspond to an emission of the closed tachyon 
to the vacuum, and the logarithmic divergence to the dilaton. The two 
dilaton divergences are combined to cancel each other 
if $\cc=n\aa^2$, which is the same condition as required
by the gauge invariance in the previous section. So we are left with
\begin{eqnarray}
\label{eq:ADP}
 {\cal A}_{D_2}+{\cal A}_{P_2}=
  -\lim_{a\rightarrow0}\frac{4n\aa^2}{a^2}\,g^2 .
 \end{eqnarray}

We still have another amplitude contribution to the same order, coming 
from the counter term which was introduced as a renormalization of the 
intercept. The counterterm is contained in 
$\tQB^\c= \QB^\c + \lambda_\c g^2 {c_0^+}$, and hence 
contributes to the tachyon amplitude as
\begin{eqnarray}
&&\hspace*{-3em} (2\pi)^d\delta^d(p_1+p_2){\cal A}_{\rm count} \nn
&=&
 - \bra{R^\c(2^\c,1^\c)}
\sqrt{2}{c_0^-}^{(1^\c)}\ket{\varphi(p_1)}_{1^\c}
\lambda_\c g^2 {c_0^+}^{(2^\c)}{b_0^-}^{(2^\c)}
\sqrt{2}{c_0^-}^{(2^\c)}\ket{\varphi(p_2)}_{2^\c}\nn
&=&
 -2\lambda_\c g^2 \bra{R^\c(2^\c,1^\c)}
{c_0^-}^{(1^\c)}\ket{\varphi(p_1)}_{1^\c}
{c_0^+}^{(2^\c)}\ket{\varphi(p_2)}_{2^\c} .
\end{eqnarray}
This is easily evaluated using the oscillator expression directly and 
yield
\begin{equation}
\label{eq:Acount}
{\cal A}_{\rm count} 
= -2\lambda_\c g^2 = +\lim_{a\rightarrow0}\frac{4n\aa^2}{a^2}g^2.
\end{equation}
The cut-off parameter $a$ here is the one introduced in the previous 
section and has obviously the same physical meaning as that introduced 
in Eq.~(\ref{eq:cutoffa}), and so we can identify these parameters. 
Comparing Eq.~(\ref{eq:ADP}) and (\ref{eq:Acount}), it turns out that 
also the divergences originated in the tachyon cancel after summation of
the three amplitudes if the action has the gauge symmetry, and we 
conclude that the net tachyon amplitude vanishes in this order:
\begin{equation}
 {\cal A}_{D_2}+{\cal A}_{P_2} + {\cal A}_{\rm count} = 0.
\end{equation}

\section{Summary and discussion}

In this paper we have considered the mixed system of open and closed 
strings and constructed the string field theory action explicitly 
up to the terms quadratic in the fields. We have proved in detail 
the invariance of the action under the gauge transformation with 
closed string field parameter $\Lambda^\c$. 

It is not difficult, in principle, to construct the full action which is
invariant under both type gauge transformations with open and 
closed string field parameters $\Lambda^\o$ and $\Lambda^\c$. But, to do this, we 
need to consider one-loop amplitudes with open-string external states. 
This is because of anomaly: the gauge invariance under the open string 
transformation parameter $\Lambda^\o$ {\it is} violated at one-loop diagrams 
for open string 2-point function constructed with 3-open-string-vertex 
twice. And those violations can be canceled by the tree diagrams 
with the open-closed transition interaction $U$ twice and with open 
intersection interaction $V_\ro$ once. 
[As for the orientable diagram parts, this anomaly is essentially the 
same as the situation for the Lorentz invariance known already in the 
light-cone gauge string field theory.\cite{rf:ST1,rf:ST2,rf:KikkawaSawada}]
The string diagrams between which these cancellations occur for 
non-orientable diagram parts are locally 
the same as the D2 and P2 diagrams in Figs.~\ref{fig:cl-cl-D2} and 
\ref{fig:cl-clnew} considered in this paper, and the 
only difference is that there the external states are also open strings 
(so D2 becomes a one-loop diagram). 
It is there that the relation between the coupling constants $\bb$ and  
$\aa$ is fixed and the condition 
\begin{equation}
n= 2^{26/2}= 8192
\end{equation}
is required. Because of the complications due to loop-amplitude 
calculation, we deferred these discussions to the forthcoming 
paper in which we shall present the full gauge invariant action.

The main motive for this work is to develop the string field theory 
which can describe the D-branes. Since T-dual transformation of the 
usual open string gives the open string with Dirichlet boundary, the 
D-brane dynamics can be studied in detail in our string field theory 
for open-closed mixed system. But, to do this fully we have to wait 
for the completion of the full gauge invariant action. 
Even at this stage, however, we can see some connection of our string 
field theory and the Dirac-Born-Infeld action for D-branes. Let us 
explain this here briefly. 

We consider the gauge transformations for some 
massless fields. If the T-duality transformation is done, the 
unoriented closed string is known to look like an oriented closed string 
apart from the orientifold plane,\cite{rf:DLP}
and so it becomes to contain an anti-symmetric 
tensor field $B_{\mu\nu}(x)$ in the form
\begin{eqnarray}
 b_0^-\ket{\Phi}&=& \frac{1}{2} B_{\mu\nu}(x)
   \left(\alpha_{-1}^{\mu}\overline{\alpha}_{-1}^{\nu}
   -\alpha_{-1}^{\nu}\overline{\alpha}_{-1}^{\mu}\right)
   \ket{1,1} +\cdots\nn
    &=& -\frac{1}{2} B_{\mu\nu}(x)\left[
     c\partial X^{\mu}(w)\cbar\bar{\partial}\overline{X}^{\nu}(\wbar)
       -(\mu\leftrightarrow \nu)
        \right]_{\scriptstyle w=0\atop \scriptstyle \wbar=0} \ket{0} + \cdots.
\end{eqnarray}
The transformation parameter $\Lambda^\c$ has the following mode:
\begin{eqnarray}
\label{eq:zeta}
 b_0^-\ket{\Lambda^\c} &=& -2i\zeta_\mu(x)
 \left(b_{-1}\overline{\alpha}_{-1}^\mu 
       -\overline{b}_{-1}\alpha_{-1}^\mu\right) 
       \ket{1,1}+\cdots                          \nn
    &=& 2\zeta_\mu(x)
        \left[c\partial X^\mu(w)
        +\cbar\overline{\partial}\overline{X}^\mu(\wbar)
        \right]_{\scriptstyle w=0\atop \scriptstyle \wbar=0} \ket{0} + \cdots.
\end{eqnarray}
Noting that the last expression can also be written as
\begin{equation}
     2\left[
    \zeta_\mu\bigl(\half(X(w)+\bar X(\wbar))\bigr)
        \left( c\partial X^\mu(w)
        +\cbar\overline{\partial}\overline{X}^\mu(\wbar)\right)
        \right]_{\scriptstyle w=0\atop \scriptstyle \wbar=0} \ket{0}
\end{equation}
(with normal ordering implied) and that $[\QB,\,X]=c\partial X$ 
and $[\QB,\,\overline{X}]=\cbar\bar{\partial}\overline{X}$,
we find the BRS transform of this as 
\begin{eqnarray}
 -\QB b_0^-\ket{\Lambda^\c} &=& -\frac{1}{2}
     \left(\partial_{\mu} \zeta_{\nu}(x)
     -\partial_{\nu} \zeta_{\mu}(x)\right)\nn
   &&\qquad \times\left[
     c\partial X^{\mu}(w)\cbar\bar{\partial}\overline{X}^{\nu}(\wbar)
       -(\mu\leftrightarrow \nu)
        \right]_{\scriptstyle w=0\atop\scriptstyle \wbar=0} \ket{0} + \cdots.
\end{eqnarray}
Then the anti-symmetric tensor field is transformed
under the gauge transformation of Eq.~(\ref{eq:gauge}) 
at the leading order as follows,
\begin{eqnarray}
\label{eq:dB}
 \delta B_{\mu\nu}=\partial_\mu\zeta_\nu-\partial_\nu\zeta_\mu.
\end{eqnarray}
Next consider the gauge transformation for an open string massless gauge 
mode generated by the same parameter of Eq.~(\ref{eq:zeta}). The massless 
gauge field $A_\mu$ is contained in the open string field as
\begin{eqnarray}
 \ket{\Psi} &=& -i A_\mu(x)\alpha_{-1}^\mu\ket{1}+\cdots\nn
  &=& A_\mu(x)[c\partial X^\mu(w)]_{w=0}\ket{0}+\cdots.
\end{eqnarray}
{}From this equation, we can find the gauge transformation for
the massless mode by
\begin{eqnarray}
 \delta A^\mu&=& i\bra{1}c_0\alpha^\mu_1\left(\delta\ket{\Psi}\right).
\end{eqnarray}
On a (group) symmetric background, only the trace part of the 
open string field gets transformed under the gauge transformation 
Eq.~(\ref{eq:braketgauge}). The gauge modes belong to the adjoint 
representation, which is anti-symmetric and has no trace part 
in the present O($n$) case, and hence receives no transformation 
under the closed parameter gauge transformation. But, 
if a non-trivial Wilson line background appears (which gives rise to the
coordinates of the D-branes), then, the gauge field components 
corresponding to the Cartan subalgebra become to receive the 
same form of gauge transformation as in Eq.~(\ref{eq:braketgauge}) 
without taking trace.\footnote{Indeed, if $A_\mu$ develops a VEV 
$\VEV{A_\mu}=\theta$ with $\theta$ taking a value in the Cartan subalgebra, 
the open-open-closed string interaction term will induces an 
open-closed transition vertex with the present $\bra{U}$ replaced by 
$\bra{U}\theta$. So the gauge transformation Eq.~(\ref{eq:braketgauge}) 
get additional contribution of the form 
$\delta\ket\Psi_1\propto \bra{U(2,1^\c)}\theta\ket{R^\o(1,2)}
\ket{\Lambda^\c}_{1^\c}$, which contains the transformation for the 
gauge field components of the Cartan subalgebra.} 
Assuming this, the gauge transformation of a Cartan subalgebra component 
of the gauge field is given by
\begin{equation}
\label{eq:dA1}
 \delta A^\mu \propto \bidx{1}\bra{1}{c_0}^{(1)}{\alpha_1^\mu}^{(1)} 
    \bra{U(2,1^\c)}\ket{R^\o(1,2)}\ket{\Lambda^\c}_{1^\c},
\end{equation}
with the understanding that $\bra{U(2,1^\c)}\ket{R^\o(1,2)}$ now 
does not imply taking the trace. 

\begin{wrapfigure}{r}{6.6cm}
   \epsfxsize= 5.5cm   
   \centerline{\epsfbox{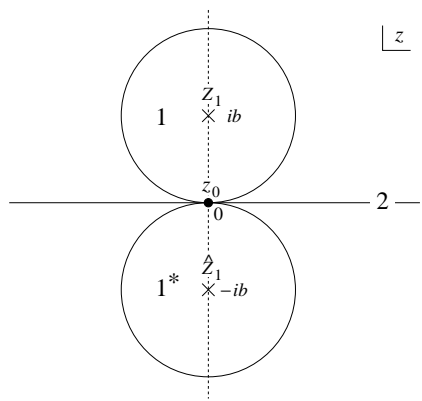}}
 \caption{$z$ plane of the open-closed transition vertex $\bra{U(2,1^\c)}$.}
 \label{fig:z-opcl}
\end{wrapfigure}
In order to evaluate this, it is convenient to map
the $\rho$ plane in Fig.~\ref{fig:op-cl} of the open-closed transition 
vertex to the $z$ plane in Fig.~\ref{fig:z-opcl} by the following
Mandelstam mapping:
\begin{eqnarray}
\label{eq:Mandel o-c}
 \hspace*{-1cm}\rho=\ln(z-ib)+\ln(z+ib).
\end{eqnarray}
On the $z$ plane, the real axis corresponds to the open string boundary.
$Z_1=+ib$ and $\hat Z_{1}=-ib$ correspond to the closed string insertion
points, and $Z_2=\infty$ to the open string insertion point. 
We can evaluate the gauge transformation as follows:\footnote{It can be 
evaluated also by using the oscillator representation, but we need the 
Neumann coefficients and very bored calculation. In Appendix B, we give 
the explicit Neumann coefficients for the open-closed transition vertex.
They were partially given in Ref.~\citen{rf:KK}, and completely in 
Refs.~\citen{rf:GreenSchwarz2} and \citen{rf:ShapThorn}. 
The latter authors' derivations are more complicated than ours.} 
Substituting Eq.~(\ref{eq:zeta}) into 
Eq.~(\ref{eq:dA1}), and going to the $z$ plane, we find
\begin{eqnarray}
\label{eq:dA2}
 \delta A^\mu 
 &\propto& \half
   \bigl\langle \zeta_\nu\bigl(\half(X(Z_1)+X(\hat Z_1))\bigr) 
\left(c\partial X^\nu(Z_1)
         +c\partial X^\nu(\hat Z_1)\right)
        c\partial c\partial X^\mu(Z_2)\bigr\rangle \nn
 &=& -\zeta^\mu(x).
\end{eqnarray}
The gauge transformations  (\ref{eq:dB}) and (\ref{eq:dA2})
are of the same form as those in the Dirac-Born-Infeld action, and so
the open-closed string field theory possesses the symmetry of the 
Dirac-Born-Infeld action as a part of its gauge symmetry.

\section*{Acknowledgements}
The authors would like to express their sincere thanks to K.\
Hashimoto, H.\ Hata, Y.\ Imamura, K.\ Kikkawa, H.\ Kunitomo, M.\
Maeno, S.\ Sawada, K.\ Suehiro and S.\ Yahikozawa for the valuable and
helpful discussions. They also acknowledge the hospitality in the
Summer Institute Kashikojima '96, in which this work was motivated,
and Kyoto '97, where it was completed. T.~K. is supported in part by
the Grant-in-Aid for Scientific Research (\#08640367) and T.~T. by the
Grant-in-Aid (\#6844) from the Ministry of Education, Science, Sports
and Culture.

\appendix
\section{Relation between the LPP, KS and HIKKO vertices
\label{sec:relation}}

We prove Eq.~(\ref{eq:LPP2}) in this appendix. So the discussion is 
performed on the concrete case of two closed string vertices, but 
it will also clarifies general relations between the three apparently 
different expressions for the vertices given by LPP, KS and HIKKO. 

Let us denote the LPP's D2 vertex 
$\bra{v_{D_2}(1^\c,2^\c;T)}$ and P2 vertex $\bra{V_\rc(1^\c,2^\c;\sigma_0)}$ 
unifiedly by 
\begin{eqnarray}
  \bra{v(1^\c,2^\c;q)}.
\end{eqnarray}
Namely, this corresponds to the configuration 
Fig.~\ref{fig:cl-cl-D2} for real $q=a$ and Fig.~\ref{fig:cl-clnew} 
for imaginary $q=ia$, respectively. 
These $\rho$ planes are mapped to the $z$ planes in 
Fig.~\ref{fig:z-D2} and Fig.~\ref{fig:z-P2}, respectively, via the 
mapping Eq.~(\ref{eq:rho-z map}). 


By the definition of the LPP vertex, we have 
\begin{eqnarray}
\label{eq:LPP*vac}
&&  \bra{v(1^\c,2^\c;q)} {\cal O}(z) \cdots\,
\ket{1,1}_{1^\c}\ket{1,1}_{2^\c}
=\left(\frac{dZ_1}{dw_1}\right)^{-1}
     \left(\frac{d\hat{Z}_1}{d\bar{w}_1}\right)^{-1}
     \left(\frac{dZ_2}{dw_2}\right)^{-1}
     \left(\frac{d\hat{Z}_2}{d\bar{w}_2}\right)^{-1} \nn
&&
\hspace{12em}\times\left< 
\,{\cal O}(z) \cdots\,c(Z_1)c(\hat{Z}_1)c(Z_2)c(\hat{Z}_2)\right>,
\end{eqnarray}
for any operators $\calO(z) \cdots$, 
where $\langle\ \ \rangle$ denotes the correlation function of the 
conformal field theory on the $z$ plane and the factors 
\begin{eqnarray}
  \left(\frac{dZ_r}{dw_r}\right)^{-1}
  =\lim_{z\rightarrow Z_r}\left(\frac{dz}{dw_r}\right)^{-1},
\hspace{1em}\pmatrix{Z_r = Z_1,\,\hat{Z}_1,\,Z_2,\,\hat{Z}_2 \cr
w_r=w_1,\,\bar{w}_1,\,w_2,\,\bar{w}_2 \cr},
\end{eqnarray}
appear from the conformal mapping of the ghost factors contained 
in the Fock vacuum; $\ket{1,1}= c(w{=}0) \cbar(\wbar{=}0)\ket{0,0}$.
Henceforth, we use the notation $\ket{\bf1}_{1^\c,2^\c}$ to denote 
$\ket{1,1}_{1^\c}\ket{1,1}_{2^\c}$, for brevity.

On the other hand, Kunitomo and Suehiro have defined in Ref.~\citen{rf:KS}
the vertex by the condition:
\begin{eqnarray}
\label{eq:bcKS}
  \bra{V_{\rm KS}(1^\c,2^\c;z_0^{(s)})}
    c(z)b(\tilde{z})\ket{\bf1}_{1^\c,2^\c}
  &=& \frac{\left<
       b(z_0^{(s)})c(z)b(\tilde{z})
      c(Z_1)c(\hat{Z}_1)c(Z_2)c(\hat{Z}_2)\right>}
    {\left<
       b(z_0^{(s)})c(Z_1)c(\hat{Z}_1)c(Z_2)c(\hat{Z}_2)\right>} \nn
  &=& \frac{1}{z-\tilde{z}}\,
      \frac{\tilde{z}-z_0^{(s)}}{z-z_0^{(s)}}
      \prod_{r=1,\hat{1},2,\hat{2}}
        \frac{z-Z_r}{\tilde{z}-Z_r},
\end{eqnarray}
where $z_0^{(s)}$ denotes one of the two interaction points $z_0^{(1)}$ 
and $z_0^{(2)}$ on the $z$ plane, on which the KS vertex depends.
The KS vertex satisfies the normalization condition 
$\bra{V_{\rm KS}(1^\c,2^\c;z_0^{(s)})}\ket{\bf1}_{1^\c,2^\c}=1$ by this
definition.
It is important that, as KS noted, the information of 
the two point function of this type is enough to determine the vertex. 

If we take $b(z_0^{(s)})\,c(z)b(\tilde z)$ for $\calO(z)\cdots$
in Eq.~(\ref{eq:LPP*vac}), we have 
\begin{eqnarray}
&&  \bra{v(1^\c,2^\c;q)} b(z_0^{(s)})\,c(z)b(\tilde z)
\ket{\bf1}_{1^\c,2^\c} \nn
&& \quad= \dZdW \left< 
\,b(z_0^{(s)})\,c(z)b(\tilde z)
\,c(Z_1)c(\hat{Z}_1)c(Z_2)c(\hat{Z}_2)\right>.
\end{eqnarray}
Comparing this with Eq.~(\ref{eq:bcKS}), we see that the KS vertex 
$\bra{V_{\rm KS}(1^\c,2^\c;z_0^{(s)})}$ coincides with 
$\bra{v(1^\c,2^\c;q)} b(z_0^{(s)})$ up to a proportional 
factor depending on the moduli $q$:
\begin{eqnarray}
\label{eq:LPP-KS}
  \bra{v(1^\c,2^\c;q)} b(z_0^{(s)})
    =\calN(q)\bra{V_{\rm KS}(1^\c,2^\c;z_0^{(s)})}, 
\end{eqnarray}
where $\calN(q)$ can be found by multiplying $\ket{\bf1}_{1^\c,2^\c}$ 
on both sides of this as
\begin{eqnarray}
 \calN(q)&=& 
\bra{v(1^\c,2^\c;q)} b(z_0^{(s)})\ket{\bf1}_{1^\c,2^\c} \cr
&=&
\dZdW \left< 
\,b(z_0^{(s)})c(Z_1)c(\hat{Z}_1)c(Z_2)c(\hat{Z}_2)\right>.
\end{eqnarray}
Each factor of this can be calculated as follows: using
\begin{eqnarray}
 \left<(b(z)c(Z_1)c(Z_2)c(Z_3)c(Z_4)\right>
&=&
\det \pmatrix{
 \displaystyle{1\over z-Z_1} & \displaystyle{1\over z-Z_2} &
\displaystyle{1\over z-Z_3}
 & \displaystyle{1\over z-Z_4} \cr
 Z_1^2 & Z_2^2 & Z_3^2 & Z_4^2 \cr
 Z_1   & Z_2   & Z_3   & Z_4   \cr
  1    &   1   &  1    &  1    \cr} \nn
&=& - {\prod_{1\leq i<j\leq4}(Z_i-Z_j)\over\prod_{r=1}^4(z-Z_r)} 
\label{eq:ghost5}
\end{eqnarray}
and Eqs.~(\ref{eq:intpt}) and (\ref{eq:inspt}) for $\zzs$ and 
$Z_1, \hat Z_1, Z_2, \hat Z_2$, we find
\begin{eqnarray}
 \left<(b(\zzs)c(Z_1)c(\hat{Z}_1)c(Z_2)c(\hat{Z}_2)\right>
  = -{-2^6q^2(1-q^2)\over2^2q^2(q^2-1)}=-2^4.
\end{eqnarray}
{}From Eq.~(\ref{eq:rho-z map}) and $\rho=\alpha_r\ln w_r$ with $\alpha_r=
+1$ $(-1)$ 
for $r=1,\hat1$ ($2, \hat2$), we have 
\begin{equation}
{dw_r\over dz} = {-4\alpha_rz^2\over(z^2+2\alpha_rz+q^2)^2}
\left( 1-{q^2\over z^2}\right).
\end{equation}
Putting $z=Z_1, \hat Z_1, Z_2, \hat Z_2$ and using (\ref{eq:inspt}), 
we find
\begin{eqnarray}
 \dZdW = {1\over4^4}\prod_{r=1,\hat1,2,\hat2}\left( 1-{q^2\over Z_r^2}\right)
  = {(1-q^2)^2\over2^4q^4}.
\end{eqnarray}
So we obtain
\begin{eqnarray}
  \calN(q)= -{(1-q^2)^2\over q^4}.
\end{eqnarray}

Now that the relation between the LPP and KS vertices is established, 
we can next clarify the relation between the KS and HIKKO vertices. 
Guided by the OSP($d,2|2$) symmetry, HIKKO defined the following ghost 
fields on the $z$ plane in their papers \citen{rf:HIKKO1,rf:HIKKO2}:
\begin{eqnarray}
\label{eq:cbHIKKO}
  c_{\rH}(z)=\left(\frac{dz}{d\rho}\right)^{-1}c(z),
\qquad 
  b_{\rH}(z)=\left(\frac{dz}{d\rho}\right)b(z),
\end{eqnarray}
or equivalently, in terms of the original fields on the $\rho$ plane, by 
\begin{eqnarray}
  c_{\rH}(z)= c(\rho),
\qquad 
  b_{\rH}(z)= \left({d\rho\over dz}\right)b(\rho).
\end{eqnarray}
Namely, HIKKO treated the ghost fields as if $c$ and $b$ carry 
the weights 0 and +1, respectively, with which they could obtain 
OSp symmetric expression for the vertex.


We first note that the KS vertex realizes the following 2-point 
function for the HIKKO's ghost fields:
\begin{eqnarray}
&& \bra{V_{\rm KS}(1^\c,2^\c;z_0^{(1)})}
 {\rm T}c_{\rH}(z)b_{\rH}(\tilde{z})
  \ket{\bf1}_{1^\c,2^\c} \nn
&&\qquad = \bra{V_{\rm KS}(1^\c,2^\c;z_0^{(1)})}
 {\rm T}c(z)b(\tilde{z})
  \ket{\bf1}_{1^\c,2^\c}\left({dz\over d\rho}\right)^{-1}
  \left({d\tilde z\over d\tilde \rho}\right)\nn
&& \qquad =\frac{1}{z-\tilde{z}}\,\frac{z-z_0^{(2)}}
  {\tilde{z}-z_0^{(2)}},
\label{eq:KScb}
\end{eqnarray}
as well as the same form of equation with the two interaction points 
$z_0^{(1)}$ and $z_0^{(2)}$ being interchanged, 
where use has been made of 
the definitions of KS vertex and HIKKO's field, Eqs.~(\ref{eq:bcKS}) and
(\ref{eq:cbHIKKO}), and the relation:
\begin{equation}
{d\rho\over dz} = 
 {4(z-z_0^{(1)})(z-z_0^{(2)})\over 
\prod_{r=1,\hat1,2,\hat2}(z-Z_r)}.
\label{eq:dzdr}
\end{equation}
Looking at this 2-point function in Eq.~(\ref{eq:KScb}), we can guess 
that the KS vertex is given in the form:
\begin{eqnarray}
\label{eq:KS-HIKKO}
\hspace*{-0.5cm}  \bra{V_{\rm KS}(1^\c,2^\c;z_0^{(1)})}
 =\bidx{2^\c,1^\c}\bra{\tilde{\bf1}}
  \left(\sum_{r=1,\hat1,2,\hat2} \frac{b_0^{(r)}}{\alpha_r}\right)
  \exp(E_{\rH}(1^\c,2^\c))\,c_{\rH}(z_0^{(2)}) 
\end{eqnarray}
(and the same form equation with $z_0^{(1)}$ and $z_0^{(2)}$ 
interchanged),
where $E_{\rH}$ is the HIKKO's OSp symmetric quadratic operator 
written by annihilation operators alone, the explicit form of which will
be given later in Eq.~(\ref{eq:EHIKKO}). 
The vacuum $\bidx{2^\c,1^\c}\bra{\tilde{\bf1}}$ 
here is the abbreviation for 
$\bidx{2^\c}\bra{\tilde{1},\tilde{1}}\bidx{1^\c}\bra{\tilde{1},\tilde{1}}$,
 and the bra vacuum $\bra{\tilde{1},\tilde{1}}$ denotes 
$\bra{\tilde{1},\tilde{1}}=\bra{1,1}c_0\cbar_0$ conjugate to the
ket Fock vacuum $\ket{1,1}$: $\langle\tilde{1},\tilde{1}|1,1\rangle=1$. Since 
the bra vacuum $\bra{\tilde{1},\tilde{1}}$ is annihilated by $c_0$ and 
$\cbar_0$ and the ket vacuum $\ket{1,1}$ by $b_0$ and $\bbar_0$, so we 
call $c_0, \cbar_0$ creation operators and $b_0,\bbar_0$ annihilation 
operators, henceforth. 

{}From now on we prove Eq.~(\ref{eq:KS-HIKKO}); 
namely, the right-hand side of (\ref{eq:KS-HIKKO}) also realizes 
the same 2-point function as in Eq.~(\ref{eq:KScb}).

Noting that $E_{\rH}$ is a quadratic form of annihilation 
operators alone, we have the following equation generally for operators 
${\cal O}$ linear in the creation and annihilation operators:
\begin{eqnarray}
 e^{E_{\rH}}\left\{
  {\cal O}_1{\cal O}_2\cdots\right\}=
  \bigl\{ \left({\cal O}_1+[E_{\rH},\,{\cal O}_1]\right)
        \left({\cal O}_2+[E_{\rH},\,{\cal O}_1]\right)\cdots 
       \bigr\} e^{E_{\rH}}.
\end{eqnarray}
Now introduce the following notations,
\begin{eqnarray}
&& \hat{{\cal O}}\equiv{\cal O}+[E_{\rH},\,{\cal O}]
= \hat{{\cal O}}^{(-)}+\hat{{\cal O}}^{(+)}, \nn
 &&\hat{{\cal O}}^{(-)}\equiv{\cal O}^{(-)}, \qquad 
 \hat{{\cal O}}^{(+)}\equiv{\cal O}^{(+)}+[E_{\rH},\,{\cal O}^{(-)}],
\end{eqnarray}
where superscripts $(-)$ and $(+)$ denote the creation and 
annihilation operator parts, respectively. The hatted operators are 
also linear in the creation and annihilation operators. 
Applying the Wick theorem to the hatted operators $\hat\calO$, we have
\begin{eqnarray}
 &&e^{E_{\rH}}{\rm T}\left(
   {\cal O}_1{\cal O}_2{\cal O}_3\cdots\right)
  = {\rm T}\left(\hat{{\cal O}_1}\hat{{\cal O}_2}\hat{{\cal O}_3}\cdots 
\right)
   e^{E_{\rH}} \nn
  &&\qquad = \left(:\hat{{\cal O}_1}\hat{{\cal O}_2}\hat{{\cal O}_3}\cdots:
     + :\wick{2}{<1{\hat{\cal O}}_1 >1{\hat{\cal O}}_2}{\hat{\cal O}_3}
        \cdots :
     + \ \cdots\  \right)e^{E_{\rH}},
\label{eq:hatWick}
\end{eqnarray}
with the contraction of the $\hat\calO$ operators given by
\begin{equation}
 \wick{2}{<1{\hat{\cal O}}_1 >1{\hat{\cal O}}_2}
  =\bidx{2^\c,1^\c}\bra{\tilde{\bf1}}{\rm T}
\hat{{\cal O}_1}\hat{{\cal O}_2}
\ket{\bf1}_{1^\c,2^\c}
  =\wick{1}{<1{{\cal O}}_1>1{{\cal O}}_2}
   +\bigl[[E_{\rH},\,{\cal O}_1^{(-)}],\,{\cal O}_2^{(-)}\bigr].
\end{equation}
This contraction of the hatted operators is of just the same form 
as defined by HIKKO. Then, we obtain
\begin{eqnarray}
\label{eq:ccb}
&& \bidx{2^\c,1^\c}\bra{\tilde{\bf1}}
  \left(\sum_r \frac{b_0^{(r)}}{\alpha_r}\right)
  e^{E_{\rH}}\,{\rm T}\,c_{\rH}(z_0^{(2)})
  c_{\rH}(z)b_{\rH}(\tilde{z}) \nn
&&\hspace{2em}=
\bidx{2^\c,1^\c}\bra{\tilde{\bf1}}
  \left(\sum_r \frac{b_0^{(r)}}{\alpha_r}\right)
\bigl[
\wick{2}{<1{\hat c}_{\rH}(z_0^{(2)})\hat{c}_{\rH}(z)
>1{\hat b}_{\rH}(\tilde{z})}
+\wick{2}{\hat{c}_{\rH}(z_0^{(2)})
<1{\hat c}_{\rH}(z)>1{\hat b}_{\rH}(\tilde{z})} \nn
&&\hspace{13em}
+ :{\hat c}_{\rH}(z_0^{(2)})
{\hat c}_{\rH}(z){\hat b}_{\rH}(\tilde{z}): 
        \bigr]
    e^{E_{\rH}}.
\end{eqnarray}
If $E_{\rH}$ is such a quadratic form that it realizes
\begin{eqnarray}
\label{eq:EH}
 \wick{2}{<1{\hat c}_{\rH}(z)>1{\hat b}_{\rH}(\tilde{z})}
 =\frac{1}{z-\tilde{z}},
\end{eqnarray}
then, with the help of the formula (\ref{eq:hatWick}), 
Eq.~(\ref{eq:ccb}) becomes
\begin{eqnarray}
 = \bidx{2^\c,1^\c}\bra{\tilde{\bf1}}
  \left(\sum_r \frac{b_0^{(r)}}{\alpha_r}\right)
\left[-\frac{1}{z_0^{(2)}-\tilde z}\hat{c}_{\rH}(z)
     +\frac{1}{z-\tilde{z}}\hat{c}_{\rH}(z_0^{(2)})
+ :{\hat c}_{\rH}(z_0^{(2)})
{\hat c}_{\rH}(z){\hat b}_{\rH}(\tilde{z}): 
        \right]
    e^{E_{\rH}}.\nonumber
\end{eqnarray}
Acting the Fock vacuum $\ket{\bf1}_{1^\c,2^\c}$ on this equation,
we have
\begin{eqnarray}
\label{eq:ccb2}
&&  
\bidx{2^\c,1^\c}\bra{\tilde{\bf1}}
  \left(\sum_r \frac{b_0^{(r)}}{\alpha_r}\right)
  e^{E_{\rH}}c_{\rH}(z_0^{(2)})
  \,{\rm T}\,c_{\rH}(z)b_{\rH}(\tilde{z})
  \ket{\bf1}_{1^\c,2^\c} \nn
&&\quad 
 = \bidx{2^\c,1^\c}\bra{\tilde{\bf1}}
  \left(\sum_r \frac{b_0^{(r)}}{\alpha_r}\right)
\left[-\frac{1}{z_0^{(2)}-\tilde z}\hat{c}_{\rH}(z)
     +\frac{1}{z-\tilde{z}}\hat{c}_{\rH}(z_0^{(2)})
        \right]
    \ket{\bf1}_{1^\c,2^\c}.
\end{eqnarray}
Since $b_0^{(r)}$ are the annihilation operators, 
only the $c_0$ mode part in $\hat{c}_{\rH}(z)$ and 
$\hat{c}_{\rH}(z_0^{(2)})$ can survive the VEV in the the right hand side.
Noting
\begin{eqnarray}
 c_{\rH}(z)=\alpha_r\sum_n {c_n}^{(r)}e^{in\rho_r} = 
\alpha_rc_0^{(r)} + (\hbox{non-zero modes}),
\end{eqnarray}
we have 
\begin{eqnarray}
\hspace*{-1cm}\bidx{2^\c,1^\c}\bra{\tilde{\bf1}}
  \left(\sum_r \frac{b_0^{(r)}}{\alpha_r}\right)
\hat{c}_{\rH}(z)
  \ket{\bf1}_{1^\c,2^\c}
= \bidx{2^\c,1^\c}\bra{\tilde{\bf1}}
  \left(\sum_r \frac{b_0^{(r)}}{\alpha_r}\right)
\hat{c}_{\rH}(z_0^{(2)})\ket{\bf1}_{1^\c,2^\c}
 =1,
\end{eqnarray}
so that Eq.~(\ref{eq:ccb2}) finally becomes
\begin{eqnarray}
 =-\frac{1}{z_0^{(2)}-\tilde{z}}+\frac{1}{z-\tilde{z}}
 =\frac{1}{z-\tilde{z}}\,\frac{z-z_0^{(2)}}{\tilde{z}-z_0^{(2)}}.
\end{eqnarray}
This is just equal to the right hand side of Eq.~(\ref{eq:KScb}) and 
therefore, the Eq.~(\ref{eq:KS-HIKKO}) is proved. $E_H$ is determined by
the requirement of
\begin{eqnarray}
 \wick{1}{<1c_{\rH}(z)>1b_{\rH}(\tilde{z})}
 +\bigl[[E_{\rH},\,c_{\rH}(z)],\,b_{\rH}(\tilde{z})\bigr]
 =\frac{1}{z-\tilde{z}}.
\end{eqnarray}
But HIKKO already showed that such $E_H$ is just given by the OSp invariant 
quadratic form given in
Eq.~(\ref{eq:EHIKKO}).\cite{rf:HIKKO1,rf:HIKKO2} \ 

Combining Eqs.~(\ref{eq:KS-HIKKO}) and (\ref{eq:LPP-KS}), we find 
a direct relation between the LPP and HIKKO vertices:
\begin{equation}
  \bra{v(1^\c,2^\c;q)} b(z_0^{(1)})
    =\calN(q)\bidx{2^\c,1^\c}\bra{\tilde{\bf1}}
  \left(\sum_{r=1,\hat1,2,\hat2} \frac{b_0^{(r)}}{\alpha_r}\right)
  \exp(E_{\rH}(1^\c,2^\c))\,c_{\rH}(z_0^{(2)}).
\label{eq:final}
\end{equation}
The same form equation with $z_0^{(1)}$ and $z_0^{(2)}$ 
interchanged also holds. To convert this into the 
desired Eq.~(\ref{eq:LPP2}), we note that
\begin{eqnarray}
 &&\bra{v(1^\c,2^\c;q)} b(z_0^{(1)})c_{\rH}(z_0^{(1)})
  = \lim_{z\rightarrow z_0^{(1)}}\bra{v(1^\c,2^\c;q)} b(z_0^{(1)})c(z)
\left({dz\over d\rho}\right)^{-1} \nn
  &&\qquad \qquad = \lim_{z\rightarrow z_0^{(1)}}\bra{v(1^\c,2^\c;q)} 
\left({1\over z_0^{(1)}-z} +:b(z_0^{(1)})c(z):\right)
 {4(z-z_0^{(1)})(z-z_0^{(2)})\over 
\displaystyle \prod_{r=1,\hat1,2,\hat2}(z-Z_r)} \nn
&& \qquad \qquad = {2\over q(1-q^2)}\cdot\bra{v(1^\c,2^\c;q)},
\end{eqnarray}
where we have used Eq.~(\ref{eq:dzdr}), and 
Eqs.~(\ref{eq:intpt}) and (\ref{eq:inspt}) for $\zzs$ and 
$Z_1, \hat Z_1, Z_2, \hat Z_2$.
Thus, multiplying both sides of Eq.~(\ref{eq:final}) by 
$-c_{\rH}(z_0^{(1)})$, we can really obtain the desired expression 
Eq.~(\ref{eq:LPP2}) with the coefficient
\begin{equation}
  N(q)= -{(1-q^2)^2\over q^4}\times{q(1-q^2)\over2}\times(-1) = 
   {(1-q^2)^3\over2q^3}.
\label{eq:N}
\end{equation}

\section{Neumann coefficients 
for the open-closed transition vertex}

The $\rho$ plane of Eq.~(\ref{eq:rho o-c})
can be mapped to the whole $z$ plane
by the Mandelstam mapping (\ref{eq:Mandel o-c}).
The interaction point is given by
\begin{eqnarray}
 \frac{d\rho}{dz}=0 \ \ \Longrightarrow \ \ \tau_0=\ln b^2\ \ (z_0=0).
\end{eqnarray}
The Neumann function on the $z$ plane is given by
\begin{eqnarray}
 N(\rho,\rho') &=& \ln\left|z-z'\right|+\ln\left|z-z'^*\right|.
\end{eqnarray}

To define the Neumann coefficients, we need to know some boundary conditions 
for the Neumann function. From the Mandelstam 
mapping (\ref{eq:Mandel o-c}), we have
\begin{eqnarray}
\label{eq:Ntau}
\pderop{\tau}N(\rho,\rho')&=&\left(\sum_{r=1,1^*}
    \frac{1}{z-Z_r}\right)^{-1}
  \frac{1}{2}\left[\frac{1}{z-z'}+\frac{1}{z-z'^*}\right] \nn
&&\hspace{2em} +\left(\sum_{r=1,1^*}\frac{1}{z^*-Z^*_r}\right)^{-1}
  \frac{1}{2}\left[\frac{1}{z^*-z'^*}+\frac{1}{z^*-z'}\right],
\end{eqnarray}
from which we find the following boundary conditions:
\begin{eqnarray}
  \pderop{\tau}N(\rho,\rho')\ \ \longrightarrow \ \ \left\{
    \begin{array}{cl}
        0 & \hspace{2em}(z\longrightarrow Z_1,\,\hat Z_{1}) \\
      {\displaystyle -\frac{2}{\alpha_2}}
          & \hspace{2em}(z \longrightarrow Z_2=\infty) \\
       \infty& \hspace{2em}(z \longrightarrow z_0)
    \end{array} \right.,
\end{eqnarray}
where and hereafter we take $\alpha_1=\alpha_{1^*}=1$
and $\alpha_2=-2$.
Taking these boundary conditions into account, we can define 
the Neumann coefficients using the Fourier expansion in the $\rho$ 
coordinates as follows:
\begin{eqnarray}
\label{eq:Neucoef}
\hspace{-2em}
&&\hspace{-0.5cm}N(\rho_r,\rho'_s) =  
   -\sum_{n\geq1}\frac{1}{n}  e^{-n\left|\tau_1-\tau'_1
 \right|} \cos[n(\sigma_1-\sigma'_1)] +\max(\tau_1,\tau'_1) \nn
  && +\frac{1}{2}\sum_{n,m\geq0} e^{n\tau_1+m\tau'_1}
   \left( \Nnm{11}e^{i(n\sigma_1+m\sigma'_1)}
        +\Nnm{11^*}e^{i(n\sigma_1-m\sigma'_1)}
        +\Nnm{1^*1}e^{i(-n\sigma_1+m\sigma'_1)} \right. \nn
  && \hspace{8em}\left.+\Nnm{1^*1^*}e^{i(-n\sigma_1-m\sigma'_1)}
  \right) \hspace{2em}(r=1,\ s=1)\nn
&&\hspace{-0.5cm}=   -\sum_{n\geq1}\frac{2}{n} e^{-n\left|\tau_2-\tau'_2
 \right|} \cos(n\sigma_2) \cos(n\sigma'_2) +2\max(\tau_2,\tau'_2) \nn
 &&-2\sum_{n,m\geq0} \Nnm{22} e^{n\tau_2+m\tau'_2} \cos(n\sigma_2)
 \cos(m\sigma'_2)-2\tau_2 -2\tau'_2\hspace{2em}(r=2,\ s=2) \nn
&&\hspace{-0.5cm}= \sum_{n,m \geq0} e^{n\tau_1+m\tau'_2} \left(
     \Nnm{12}e^{in\sigma_1}+\Nnm{1^*2}e^{-in\sigma_1}\right)\cos(n\sigma'_2) 
    -2\tau'_2\ \ (r=1,\ s=2) \nn
&&\hspace{-0.5cm}=\sum_{n,m \geq0} e^{n\tau_2+m\tau'_1} \left(
     \Nnm{21}e^{in\sigma'_1}+\Nnm{21^*}e^{-in\sigma'_1}\right)\cos(n\sigma_2) 
    -2\tau_2\ \ (r=2,\ s=1).
\end{eqnarray}

Now, introduce a function,\cite{rf:KK}
\begin{eqnarray}
\label{eq:Mfunc}
 M(\rho,\rho') =\left(\pderop{\tau}+\pderop{\tau'}\right)
 N(\rho,\rho').
\end{eqnarray}
{}From (\ref{eq:Ntau}), this function is rewritten as
\begin{eqnarray}
\label{eq:Mfunc2}
  M(\rho,\rho')=\frac{1}{4\alpha_1}\left[-b^2\left(\frac{1}{z}+\frac{1}{z^*}
  \right)\left(\frac{1}{z'}+\frac{1}{z'^*}\right)+4\right].
\end{eqnarray}
Using the Mandelstam mapping Eq.~(\ref{eq:Mandel o-c}), we can expand
$z$ by the $\rho$-plane coordinates:
\begin{eqnarray}
&&\frac{b}{2}\left(\frac{1}{z}+\frac{1}{z^*}\right)=
\sum_{n\geq1}\frac{(2n-1)!!}{(2n)!!} e^{n\tau_1} \sin{n\sigma_1} \nn
&&\hspace{5.5em}= \sum_{n\geq0}\frac{(2n-1)!!}{(2n)!!} e^{(2n+1)\tau_2}
\cos[(2n+1)\sigma_2],
\end{eqnarray}
where use has been made of the formula 
\begin{eqnarray}
  \left(1-x\right)^{-\frac{1}{2}}=\sum_{n=0}^{\infty}
 \frac{(2n-1)!!}{(2n)!!}x^n,\hspace{2em}(\abs{x}<1).
\end{eqnarray}
Substituting this into Eq.~(\ref{eq:Mfunc2}), we obtain
\begin{eqnarray}
\label{eq:Mfuncexp}
  M(\rho,\rho')&=&\frac{1}{\alpha_1}+\sum_{n,m\geq 
    0}e^{n\tau_1+m\tau'_1} \frac{1}{4\alpha_1}
    \nm \left(e^{i(n\sigma_1+m\sigma'_1)}
           -e^{i(n\sigma_1-m\sigma'_1)} \right.\nn
   && \hspace{5em}   \left.-e^{i(-n\sigma_1+m\sigma'_1)}
           +e^{i(-n\sigma_1-m\sigma'_1)} \right) 
      \hspace{2em}(r=1,\ s=1) \nn
&=& -\frac{2}{\alpha_2} +\sum_{n,m\geq0} \frac{2}{\alpha_2} \nm
      e^{(2n+1)\tau_2+(2m+1)\tau'_2}   \nn
   && \hspace{5em} \times\cos[(2n+1)\sigma_2] \cos[(2m+1)\sigma'_2] 
      \hspace{2em}(r=2,\ s=2) \nn
&=& -\frac{2}{\alpha_2} +\sum_{n,m\geq0} e^{n\tau_1+(2m+1)\tau'_2}
            \frac{i}{2\alpha_1} \nm \nn
   && \hspace{5em} \times\left(e^{in\sigma_1}-e^{-in\sigma_1}\right)
      \cos[(2m+1)\sigma'_2] \hspace{2em}(r=1,\ s=2) \nn
&=& -\frac{2}{\alpha_2} +\sum_{n,m\geq0} e^{(2n+1)\tau_2+m\tau'_1}
            \frac{i}{2\alpha_1} \nm \nn
   && \hspace{5em} \times\left(e^{im\sigma'_1}-e^{-im\sigma'_1}\right)
      \cos[(2n+1)\sigma_2] \hspace{1em}(r=2,\ s=1).
\end{eqnarray}
From Eqs.~(\ref{eq:Neucoef}), (\ref{eq:Mfunc}) and (\ref{eq:Mfuncexp}), 
we find the Neumann coefficients of the open-closed vertex as follows:
\begin{eqnarray}
\label{eq:neucoeff}
  &&\Nnm{11}=\Nnm{1^*1^*}=\frac{1}{2(n+m)}\nm, \nn
  &&\Nnm{11^*}=\Nnm{1^*1}=-\frac{1}{2(n+m)}\nm, \nn
  &&\overline{N}^{22}_{2n+1,2m+1}=\frac{1}{2(n+m+1)}\nm, \nn
  &&\overline{N}^{12}_{n,2m+1}=\overline{N}^{21}_{2m+1,n}
  =\frac{i}{2n-2m-1}\nm, \nn
  &&\overline{N}^{1^*2}_{n,2m+1}=\overline{N}^{21^*}_{2m+1,n}
  =-\frac{i}{2n-2m-1} \nm, \nn
  &&\overline{N}^{22}_{2n,2m}=0.
\end{eqnarray}

This derivation of the Neumann coefficients up to here is more or less 
a review of Ref.~\citen{rf:KK}. Note, however, that 
the components $\overline{N}^{12}_{n,2n}$ are not yet determined 
in this derivation, since they cannot appear in $M(\rho,\rho')$:
\begin{eqnarray}
 \left(\pderop{\tau}+\pderop{\tau'}\right)e^{n(\tau_1+2\tau_2)}
 = \frac{1}{\alpha_1}\left(\pderop{\tau_1}-\half\pderop{\tau_2}\right)
 e^{n(\tau_1+2\tau_2)}=0.
\end{eqnarray}
Substituting the above Neumann coefficients
into Eq.~(\ref{eq:neucoeff}),
we can write the Neumann function with
$\rho$ and $\rho'$ on the strings 1 and 2, respectively, as
\begin{eqnarray}
\label{eq:N1}
 N(\rho,\rho')&=&\sum_{n,m\geq0}e^{n\tau_1+(2m+1)\tau_2}\left(
 \overline{N}^{12}_{n,2m+1} e^{in\sigma_1}+
 \overline{N}^{1^*2}_{n,2m+1} e^{-in\sigma_1}\right)
 \cos[(2m+1)\sigma_2] \nn
 &&+\sum_{n\geq1}e^{n(\tau_1+2\tau'_2)}\left(\N{12}{n,2n}e^{in\sigma_1}
 +\N{1^*2}{n,2n}e^{-in\sigma_1}\right)
 \cos(2n\sigma_2)-2\tau_2.
\end{eqnarray}
Considering the Mandelstam mapping Eq.~(\ref{eq:Mandel o-c}) 
and the reflection symmetry with respect to the imaginary axis of 
the z-plane, we find that the transformation
$z'\rightarrow-z'^*$ corresponds to the transformation $\sigma_2\rightarrow\pi-\sigma_2$. Then,
it follows that
\begin{eqnarray}
\label{eq:N2}
 \tilde{N}(\rho,\rho')&\equiv&\ln\abs{z+z'}+\ln\abs{z+z'^*} \nn
 &=&-\sum_{n,m\geq0}e^{n\tau_1+(2m+1)\tau_2}\left(
 \overline{N}^{12}_{n,2m+1} e^{in\sigma_1}+
 \overline{N}^{1^*2}_{n,2m+1} e^{-in\sigma_1}\right)
 \cos[(2m+1)\sigma_2] \nn
 &&+\sum_{n\geq1}e^{n(\tau_1+2\tau'_2)}\left(\N{12}{n,2n}e^{in\sigma_1}
 +\N{1^*2}{n,2n}e^{-in\sigma_1}\right)
 \cos(2n\sigma_2)-2\tau_2.
\end{eqnarray}
Combining Eqs.~(\ref{eq:N1}) and (\ref{eq:N2}),
we find
\begin{eqnarray}
&& N(\rho,\rho')+\tilde{N}(\rho,\rho') \nn
&&\qquad=
2\sum_{n\geq1}e^{n(\tau_1+2\tau'_2)}\left(\N{12}{n,2n}e^{in\sigma_1}
 +\N{1^*2}{n,2n}e^{-in\sigma_1}\right)
 \cos(2n\sigma_2)-4\tau_2.
\end{eqnarray}
On the other hand, we have another expression for this:
\begin{eqnarray}
 N(\rho,\rho')+\tilde{N}(\rho,\rho')&=&
 \ln\abs{z^2-z'^2}+\ln\abs{z^2-z'^2} \nn
 &&\hspace{-2cm}=\ln\abs{e^{\rho_1+\frac{\tau_0}{\alpha_1}}-
     e^{-2\rho_2+\frac{\tau_0}{\alpha_1}}}
  + \ln\abs{e^{\rho_1+\frac{\tau_0}{\alpha_1}}-
     e^{-2\rho^*_2+\frac{\tau_0}{\alpha_1}}} \nn
 && \hspace{-2cm}=-2\sum_{n\geq1}\frac{1}{n}e^{n(\tau_1+2\tau_2)}\cos(n\sigma_1)
 \cos(2n\sigma_2)+2\ln b^2-4\tau_2.
\end{eqnarray}
Comparing these two expressions, we find the remaining
Neumann coefficients as follows,\cite{rf:ShapThorn}
\begin{eqnarray}
 \overline{N}^{12}_{n,2n}=\overline{N}^{1^*2}_{n,2n}=-\frac{1}{2n},
 \hspace{3em}\overline{N}^{12}_{00}+\overline{N}^{1^*2}_{00}
 =2\ln b.
\end{eqnarray}

Finally, let us consider the components 
of $n=m=0$.
From Eq.~(\ref{eq:Mandel o-c}) and
(\ref{eq:rho o-c}), we find 
\begin{eqnarray}
  \tau_1&=&\ln\abs{z-Z_1}+\ln\abs{z-Z^*_1}-\ln b^2, \nn
  \tau_2&=&-\frac{1}{2}\ln\abs{z-Z_1}-\frac{1}{2}\ln\abs{z-Z^*_1}
       +\frac{1}{2}\ln b^2.
\end{eqnarray}
Substituting this into Eq.~(\ref{eq:Neucoef}) and taking the limit
$z,z'\longrightarrow Z_r, Z_s$, we obtain
\begin{eqnarray}
  \overline{N}^{rs}_{00}=\ln b,\hspace{2em}(r,s=1,1^*,2).
\end{eqnarray}
They, however, do not appear and are actually irrelevant
to the open-closed transition vertex because of the momentum conservation.


\begin{thebibliography}{99}
\bibitem{rf:pol}
For a review,
J.~Polchinski, ``Tasi Lectures on D-Branes'',
  {\tt hep-th/9611050}.
\bibitem{rf:BFSS}
T.~Banks, W.~Fischler, S.H.~Shenker and L.~Susskind,
  \PR{D55,1997,5112}.
\bibitem{rf:IKKT}
N.~Ishibashi, H.~Kawai, Y.~Kitazawa and A.~Tsuchiya,
  \NP{B498,1997,467}.
\bibitem{rf:HashiHata}
K.~Hashimoto and H.~Hata,
 \PR{D56,1997,5179}.
\bibitem{rf:zwie}
B.~Zwiebach, ``Oriented Open-Closed String Theory Revisited'',
 {\tt  hep-th/9705241}.
\bibitem{rf:GreenSchwarz}
M.B.~Green and J.H.~Schwarz,
  \PL{151B,1985,21}.
\bibitem{rf:ItoyamaMoxhay}
H.~Itoyama and P.~Moxhay,
  \NP{B293,1987,685}.
\bibitem{rf:DougGrin}
M.R.~Dougras and B.~Grinstein,
  \PL{183B,1987,52}.
\bibitem{rf:wendt}
C.~Wendt,
  \NP{B314,1989,209}.
\bibitem{rf:KugoTerao}
T.~Kugo and H.~Terao,
  \PL{208B,1988,416}.
\bibitem{rf:ST1}
Y.~Saitoh and Y.~Tanii,
  \NP{B325,1989,161}.
\bibitem{rf:ST2}
Y.~Saitoh and Y.~Tanii,
  \NP{B331,1990,744}.
\bibitem{rf:KikkawaSawada}
K.~Kikkawa and S.~Sawada,
  \NP{B335,1990,677}.
\bibitem{rf:KT}
T.~Kugo and T.~Takahashi,
  in preparation.
\bibitem{rf:CLNY}
C.G.~Callan, C.~Lovelace, C.R.~Nappi and S.A.~Yost,
  \NP{B288,1987,525}.
\bibitem{rf:LPP}
A.~LeClair, M.E.~Peskin and C.R.~Preitschopf,
  \NP{B317,1989,411}.
\bibitem{rf:KS}
H.~Kunitomo and K.~Suehiro,
  \NP{B289,1987,157}.
\bibitem{rf:HIKKO1}
H.~Hata, K.~Itoh, T.~Kugo, H.~Kunitomo and K.~Ogawa
  \PR{D34,1986,2360}.
\bibitem{rf:HIKKO2}
H.~Hata, K.~Itoh, T.~Kugo, H.~Kunitomo and K.~Ogawa
  \PR{D35,1987,1318}.
\bibitem{rf:kugozwie}
T.~Kugo and B.~Zwiebach,
  \PTP{87,1992,801}.
\bibitem{rf:Siegel}
W.~Siegel,
  \PL{142B,1984,276}.
\bibitem{rf:NW}
A.~Neveu and P.C.~West,
  \NP{B293,1987,266}.
\bibitem{rf:Uehara}
S.~Uehara,
  \PL{190B,1987,76}.
\bibitem{rf:Kugo}
T.~Kugo,
  {\it in} Quantum Mechanics of Fundamental Systems 2,
  ed. by C.~Teitelboim and J.~Zanelli
  (Plenum Publishing Corporation, 1989).
\bibitem{rf:AGMV}
L.~Alvarez-Gaum$\acute{e}$, C.~Gomez,
G.~Moore and C.~Vafa,
  \NP{B303,1988,411}.
\bibitem{rf:GM}
S.B.~Giddings and E.~Martinec,
  \NP{B278,1986,91}.
\bibitem{rf:KugoSuehiro}
T.~Kugo and K.~Suehiro,
  \NP{B337,1990,434}.
\bibitem{rf:ShapThorn}
J.A.~Shapiro and C.B.~Thorn,
  \PR{D36,1987,432}.
\bibitem{rf:Hata-Nojiri}
H.~Hata and M.M.~Nojori,
  \PR{D36,1987,1193}.
\bibitem{rf:FMS}
D.~Friedam, E.~Martinec and S.~Shenker,
  \NP{B271,1986,93}.
\bibitem{rf:DLP}
J.~Dai, R.G.~Leigh and J.~Polchinski,
  \JL{Mod.~Phys.~Lett.,A4,1989,2073}.
\bibitem{rf:KK} 
M.~Kaku and K.~Kikkawa, 
  \PR{D10,1974,1823}.
\bibitem{rf:GreenSchwarz2}
M.B.~Green and J.H.~Schwarz,
  \NP{B243,1984,475}.
\end{thebibliography}
\end{document}